\documentclass[amsmath,amssymb,floatfix,lengthcheck,preprintnumbers,prd,showpacs,superscriptaddress,twocolumn,aps]{revtex4-1}
\pdfoutput=1

\usepackage{graphicx}
\usepackage{slashed}
\usepackage{xcolor}
\usepackage{ytableau}
\usepackage{mathtools}
\newcommand{\beq}{\begin{equation}}
\newcommand{\eeq}{\end{equation}}
\newcommand{\beqs}{\begin{eqnarray}}
\newcommand{\eeqs}{\end{eqnarray}}
\newcommand{\lsim}{\mathrel{\raisebox{-.6ex}{$\stackrel{\textstyle<}{\sim}$}}}

\definecolor{red}{rgb}{1.0, 0, 0}

\begin{document}

\title{Composite bosonic baryon dark matter on the lattice: \\
SU(4) baryon spectrum and the effective Higgs interaction}

\author{T.~Appelquist}
\affiliation{Department of Physics, Sloane Laboratory, Yale University,
             New Haven, Connecticut 06520, USA}
\author{E.~Berkowitz}
\affiliation{Lawrence Livermore National Laboratory, Livermore, California 94550, USA}
\author{R.~C.~Brower}
\affiliation{Department of Physics, Boston University,
             Boston, Massachusetts 02215, USA}
\author{M.~I.~Buchoff}
\affiliation{Institute for Nuclear Theory, Box 351550, Seattle, WA 98195-1550, USA}
\author{G.~T.~Fleming}
\affiliation{Department of Physics, Sloane Laboratory, Yale University,
             New Haven, Connecticut 06520, USA}
\author{J.~Kiskis}
\affiliation{Department of Physics, University of California,
             Davis, California 95616, USA}
\author{M.~F.~Lin}
\affiliation{Argonne Leadership Computing Facility, Argonne National Laboratory,
             Argonne, IL 60439, USA}
\affiliation{Computational Science Center, Brookhaven National Laboratory, 
             Upton, NY 11973, USA}
\author{E.~T.~Neil}
\affiliation{Department of Physics,
             University of Colorado, Boulder, CO 80309, USA}
\affiliation{RIKEN-BNL Research Center, Brookhaven National Laboratory, 
             Upton, NY 11973, USA}
\author{J.~C.~Osborn}
\affiliation{Argonne Leadership Computing Facility, Argonne National Laboratory,
             Argonne, IL 60439, USA}
\author{C.~Rebbi}
\affiliation{Department of Physics, Boston University,
             Boston, Massachusetts 02215, USA}
\author{E.~Rinaldi}
\affiliation{Lawrence Livermore National Laboratory, Livermore, California 94550, USA}
\author{D.~Schaich}
\affiliation{Department of Physics, Syracuse University, Syracuse, NY 13244, USA}
\author{C.~Schroeder}
\affiliation{Lawrence Livermore National Laboratory, Livermore, California 94550, USA}
\author{S.~Syritsyn}
\affiliation{RIKEN-BNL Research Center, Brookhaven National Laboratory, 
             Upton, NY 11973, USA}
\author{G.~Voronov}
\affiliation{Department of Physics, Sloane Laboratory, Yale University,
             New Haven, Connecticut 06520, USA}
\author{P.~Vranas}
\affiliation{Lawrence Livermore National Laboratory, Livermore, California 94550, USA}
\author{E.~Weinberg}
\affiliation{Department of Physics, Boston University,
	Boston, Massachusetts 02215, USA}
\author{O.~Witzel}
\affiliation{Department of Physics, Boston University,
             Boston, Massachusetts 02215, USA}
\author{(Lattice Strong Dynamics (LSD) Collaboration)}
\noaffiliation
\author{G.~D.~Kribs}
\affiliation{School of Natural Sciences, Institute for Advanced Study, 
             Princeton, NJ 08540, USA}
\affiliation{Department of Physics, University of Oregon, Eugene, OR, 97403 USA}

\begin{abstract}

We present the spectrum of baryons in a new SU$(4)$ gauge theory with 
fundamental fermion constituents. The spectrum of these bosonic baryons 
is of significant interest for composite dark matter theories. 
Here, we compare the spectrum and properties of SU$(3)$ and SU$(4)$ baryons,
and then compute the dark-matter direct detection cross section via Higgs boson
exchange for TeV-scale composite dark matter arising from a confining 
SU$(4)$ gauge sector. Comparison with the latest 
LUX results leads to tight bounds on the fraction of the constituent-fermion 
mass that may arise from electroweak symmetry breaking. Lattice calculations of 
the dark matter mass spectrum and the Higgs-dark matter coupling are performed 
on quenched $16^{3} \times 32$,  $32^{3} \times 64$, $48^{3} \times 96$, and 
$64^{3} \times128$ lattices with three different lattice spacings, 
using Wilson fermions with moderate to heavy pseudoscalar meson masses.  
Our results lay a foundation for future analytic and numerical study of 
composite baryonic dark matter.

\end{abstract}

\preprint{INT-PUB-14-005, LLNL-JRNL-650612}

\maketitle

\section{Introduction}

Weakly-interacting massive particles provide an intriguing but 
increasingly constrained model for dark matter.  Weak interactions
play an essential role in obtaining a thermal relic abundance 
and may play a significant role in obtaining an asymmetric abundance 
\cite{Nussinov:1985xr,Chivukula:1989qb,Barr:1990ca,Kaplan:1991ah} of dark matter.  
However, electrically-neutral particles with standard model strength 
couplings to the weak neutral current  
(e.g., a Dirac fermion with the quantum numbers of a neutrino) 
have been ruled out for over two decades.
Now that a particle consistent with the Higgs boson has been
observed \cite{Aad:2012tfa,Chatrchyan:2012ufa}, recent bounds from 
direct detection experiments \cite{Aprile:2012nq,Agnese:2013rvf,Akerib:2013tjd} 
significantly constrain the coupling of dark matter to standard model
particles via exchange of a Higgs boson.

Models of electroweak-neutral dark matter whose constituents carry
electroweak charges are much less constrained by direct detection
experiments, since the neutral composite has only higher-dimensional
interactions suppressed by the confinement scale.
These suppressed interactions---magnetic and electric dipole moments 
(dimension-5), charge radius (dimension-6), polarizability and anapole
(dimension-7), etc.--- are familiar from known composite systems and have 
been studied with respect to dark matter interactions in \cite{Chivukula:1992pn,Bagnasco:1993st,Pospelov:2000bq,Sigurdson:2004zp,Gudnason:2006ug,Alves:2009nf,Kribs:2009fy,Barbieri:2010mn,Banks:2010eh,Chang:2010en,Barger:2010gv,Weiner:2012cb}. 
The bound from direct detection experiments are already sufficient to 
significantly constrain some of these interactions, such as the 
magnetic dipole moment interaction for fermionic 
dark matter~\cite{Banks:2010eh,Chang:2010en,Barger:2010gv,Fortin:2011hv}.  
For example, in the composite SU(3) baryonic dark matter model studied 
in Ref.~\cite{Appelquist:2013ms}, the mass of the dark matter must exceed 
$\simeq 10$~TeV to be safe from XENON100 
constraints \cite{Aprile:2012nq,Akerib:2013tjd}. 
Dimension-6 interactions, for example corresponding to an 
electromagnetic charge radius, also appear to be significantly 
constrained \cite{Bagnasco:1993st,Pospelov:2000bq,Kribs:2009fy,Banks:2010eh}.

There are simple composite dark matter theories, however, that do
not have dimension-5 (or dimension-6) interactions with the standard model. 
Composite bosonic dark matter theories, such as the baryons from an
$SU(N_c)$ with even $N_c$ strongly-coupled theory, 
do not have dipole moment interactions if the dark matter candidate is spin-zero.
The dimension-6 charge radius interaction is also not generated in
composite theories that, for example, preserve a custodial SU$(2)$ symmetry.
The leading interactions that remain include Higgs exchange, 
whose strength depends on the constituent fermion-Higgs couplings, 
and the dimension-7 electromagnetic polarizability interaction 
studied in \cite{Pospelov:2000bq,Kribs:2009fy,Weiner:2012cb}. 

Dark matter stability, for at least the age of the Universe,
can also be an automatic consequence of global symmetries
of the low energy effective theory.  
Within even-$N_c$ gauge theories, we consider an SU$(4)$ dark sector 
with fermions in the 
fundamental representation of the gauge group, where baryonic 
dark matter stability is an automatic consequence of baryon number 
conservation.  This is as opposed to SU$(2)$, where there is no 
dynamical distinction between mesons and baryons, and therefore
requires imposing an additional global or discrete symmetry
\cite{Kribs:2009fy,Lewis:2011zb,Buckley:2012ky}.  

Aside from the implications for dark matter direct detection, 
another strong motivation for composite dark matter with constituents 
that transform under the electroweak group is the possibility of obtaining
the cosmological abundance of dark matter through an asymmetry.  
The observational relation of densities, $\rho_{\rm DM} \simeq 5 \rho_{b}$, 
strongly hints at an asymmetric origin of dark matter, which was recognized 
long ago in the context of technibaryon dark matter 
\cite{Nussinov:1985xr,Chivukula:1989qb,Barr:1990ca,Kaplan:1991ah}
and more recently in other models, e.g.\  \cite{Kitano:2004sv,Farrar:2005zd,Banks:2005hc,Kitano:2008tk,Kaplan:2009ag,Kribs:2009fy,An:2009vq,Alves:2010dd,Dulaney:2010dj,Cohen:2010kn,Shelton:2010ta,Buckley:2010ui,Haba:2010bm,Blennow:2010qp,Hall:2010jx,Falkowski:2011xh,Graesser:2011wi,Kaplan:2011yj,Cui:2011qe,Petraki:2013wwa,Zurek:2013wia}. 
There are proposals to obtain asymmetric
dark matter through electroweak sphalerons 
\cite{Barr:1990ca,Kribs:2009fy,Buckley:2010ui}
as well as various other particle physics models 
(for reviews, see \cite{Petraki:2013wwa,Zurek:2013wia}). 
One of the principal
difficulties in realizing asymmetric dark matter in elementary dark matter
theories is to suppress the thermal relic abundance \cite{Kaplan:2009ag}.
This happens automatically in models of strongly-coupled composite 
dark matter, e.g.~\cite{Barr:1990ca,Gudnason:2006ug,Kribs:2009fy}. 

While na\"ive dimensional analysis can give crude estimates
of the effective couplings of composite dark matter, confronting 
experiment requires much better precision to determine the 
viability of models.  In some special cases, 
e.g.\ \cite{Alves:2009nf,Kaplan:2009de,Kribs:2009fy},
the effective couplings can be estimated using non-relativistic 
effective theory.  
When the fermion masses and the confinement scale are comparable, 
lattice calculations are perfectly suited to determine
the non-perturbative spectrum and observables. 
Unlike lattice gauge theory applied to QCD, where one
goal is to extrapolate to small (light) fermion masses, 
here lattice calculations at relatively heavy fermion masses
provide exactly the computations we are interested
in, \emph{without} the need for mass extrapolations.
In this way, studying strongly-coupled
composite dark matter on the lattice provides access to 
interesting regimes in dark matter ``theory space.''  

We present here the results of lattice simulations of SU$(4)$ gauge theory, with particular focus on the spectrum of baryons and on the baryonic matrix element of the scalar current, a necessary input to calculate the Higgs-exchange scattering cross section for comparison to direct-detection experiment.  To date, there have been only a handful of other lattice calculations with focus on applications in composite dark matter \cite{Lewis:2011zb,Hietanen:2012sz,Appelquist:2013ms,Hietanen:2013fya}.  This work also represents the first calculation on the lattice of the SU$(4)$ baryon spectrum, and one of the first calculations of baryon properties for any $N_c>3$ (the baryon spectrum for $N_c = 3,5,7$ was explored recently in \cite{DeGrand:2012hd,DeGrand:2013nna}.)  Separate from the application to composite dark matter, in this work we will place our simulation results into the larger context of large-$N_c$ calculations, both analytic and numerical.

Our lattice calculations 
span four different volumes ($16^3$, $32^3$, $48^3$, and $64^3$) 
with aspect ratio of 2, and three different lattice spacings. 
In principle a wide range of fermion masses is interesting, though in
this paper we concentrate on moderate to heavy fermion masses corresponding 
to meson mass ratios of $0.5 < m_{PS}/m_{V} < 0.9$.  
Unimproved Wilson fermion propagators are calculated on quenched lattices 
(an approximation that should be reasonable due to a larger $N_c$ 
value with moderate to large fermion masses).  We present results for the 
spin-0, spin-1, and spin-2 baryon masses, as well as pseudoscalar and 
vector masses as reference scales, and the spin-0 baryon scalar matrix element 
(also known as the ``sigma term'').  
Once the spectrum and matrix elements are determined in units of the 
inverse lattice spacing, we set the scale by choosing an appropriate 
physical mass for the spin-0 baryon, our dark matter candidate.  
This allows us to consider a range of dark matter masses.

The paper is organized as follows.  In Sec.~\ref{sec:Models}, we classify the general features of interesting composite models and detail all the pieces in calculating the Higgs exchange cross-section from these composite theories.  In Sec.~\ref{sec:SU4 Baryons}, we discuss the properties of SU(4) baryons and the interpolating operators used to calculate the baryon two point function on the lattice.  Sec.~\ref{sec:3_4_comp} and Sec.~\ref{sec:Higgs_Results} highlight the primary results of this work.  In Sec.~\ref{sec:3_4_comp}, three and four color baryons are compared using the standard large $N_c$ framework.  In Sec.~\ref{sec:Higgs_Results}, cross-sections from direct Higgs exchange are presented along with robust restrictions on allowed values in model space.   Sec.~\ref{sec:Sim_Detail} and Sec.~\ref{sec:Calc_Fit} discuss lattice simulation and fitting details, respectively, while the detailed presentations of the baryon spectra and baryon matrix element for all lattice spacings are contained in Sec.\ref{sec:Baryon_Spectrum} and Sec.~\ref{sec:Baryon_Deriv}.  The final section before the concluding, Sec.~\ref{sec:Lattice_Artifacts} examines the systematic effects from lattice artifacts, namely lattice spacing and finite volume errors.

\section{General Overview of models}
\label{sec:Models}

The ``theory space'' of possible composite dark matter models is quite large, even when limited to SU$(N_c)$ gauge theories with $N_f$ fermions in the fundamental representation.  In order to carry out a lattice calculation, we must specify a particular theory to simulate.  As discussed in the introduction, we choose SU$(4)$ as a minimal example of a composite model with bosonic dark matter candidates (the case of SU$(2)$ is more complicated, due to the presence of an enhanced chiral symmetry; see \cite{Kribs:2009fy,Lewis:2011zb,Buckley:2012ky}).  We require that the constituent fermions carry electroweak charges of some sort, allowing the existence of various interactions between the dark sector and the Standard Model relevant for direct and indirect detection dark matter experiments, as well as for giving the observed dark matter relic density.

There are three regimes for the relative scales between the fermion mass $m_f$
and the confinement scale, $\Lambda_4$, of SU$(4)$:  ``QCD-like''
$m_f \ll \Lambda_4$; ``comparable scales'' $m_f \sim \Lambda_4$;
and ``quarkonia-like'' $m_f \gg \Lambda_4$.  We focus on the ``comparable scales'' regime in this paper, for which lattice calculation is necessary to make progress.  Because $m_f$ is relatively heavy in this regime along with a large $N_c$ value of 4, the effects of fermion loops will be suppressed, justifying the use of the quenched approximation in our lattice simulations.

If the baryonic dark matter is composed of an even number of constituent fermions whose electric charge is plus-minus pairs of equal magnitude, a discrete symmetry of the model forbids the existence of a dimension-6 charge radius operator \cite{Kribs:2009fy}.  Assuming the dark matter is also spin-zero, then it does not carry a magnetic moment either, and the leading interaction relevant for scattering off of ordinary matter mediated by electroweak bosons is the dimension-7 electromagnetic polarizability.  Using na\"ive dimensional analysis, we estimate
that the dark matter mass must be greater than $\sim$ tens of GeV \cite{Chivukula:1992pn} in order to avoid existing direct-detection constraints.  In a future publication we will return to this topic and use lattice simulations to make precise predictions for the spin-independent scattering rate through the polarizability interaction.

An important experimental constraint is the 
non-observation of the mesons of this theory.
Strongly-coupled theories with constituent fermions transforming 
non-trivially under the electroweak group are expected to have 
electrically charged mesons, just like the pions of QCD. 
While the precise constraints are model-dependent, we require that 
the lightest (pseudoscalar) meson masses satisfy $m_{PS} > 100$~GeV\@.
At energies accessible by LEP, the Drell-Yan production cross-section mediated by a photon would otherwise be
quite substantial, and the decay modes are expected to be predominantly to the heaviest standard-model states
kinematically allowed \cite{Buckley:2012ky}, at LEP energies charm-strange and 
$\tau + \nu_\tau$ pairs.  Existing LEP searches
for $\tau^+ \tau^-$ plus missing energy, targeted at pair-production of scalar tau partners in supersymmetric models, place the limit $m_{\tilde{\tau}} > 86$~GeV \cite{Heister:2001nk,Heister:2003zk,Abdallah:2003xe,Abbiendi:2004gf}.  We anticipate this limit applies to the charged pseudoscalars from this dark matter model as well, but we have not attempted detailed collider simulations, and so we have chosen  $m_{PS} > 100$~GeV\@.  Additional collider constraints on meson production and decay could be used to place more stringent constraints on specific models.

If the fermions were to acquire masses purely through the Higgs mechanism, 
then the Higgs coupling to the dark matter baryon would be substantial.
We will show that this case is ruled out
by existing direct detection bounds and LEP exclusion bounds on charged particles.
A viable model has fermions transforming in vector-like representations
of the electroweak group.  This means ``vector-like'' fermion masses are
possible without electroweak symmetry breaking.  
Depending on the particular model,
the fermions can also acquire additional masses through electroweak 
symmetry breaking.  This is unlike the quarks and charged leptons
of the Standard Model that acquire masses solely from 
electroweak symmetry breaking.
We will calculate the bounds on the Higgs couplings 
to these fermions in this paper.  The mixed nature of the fermion masses 
also implies the dynamical breaking of electroweak symmetry by the 
strong dynamics can be controllably suppressed, as well as the contributions 
to the electroweak precision observables.  Detailing these effects will 
be reserved for future work.

\subsection{\textit{Higgs exchange cross-section}}

The calculation of the spin-independent scattering cross section of
the composite dark matter baryon scalar $B$ 
with a nucleus $N$ through Higgs exchange is given by 
\cite{McDonald:1993ex,Kim:2006af,Burgess:2000yq,Essig:2007az,Andreas:2008xy}
\begin{equation}\label{eq:cross_section}
\sigma(B,N) = \frac{\mu(m_B,m_N)^2}{\pi}(Z \mathcal{M}_p + (A-Z) \mathcal{M}_n)^2,
\end{equation} 
where $A$ and $Z$ are the total number of nucleons and protons in the target nucleus, respectively, $\mu(m_1,m_2) = m_1 m_2 / (m_1+m_2)$ is the reduced mass,  and entirety of the interactions with the protons and neutrons within the nucleus 
are contained in $\mathcal{M}_p$ and $\mathcal{M}_n$, respectively.
In order to compare the spin-independent scattering cross section
between experiments, this is conventionally re-written as
\begin{equation}
\sigma_0(B,a) = \sigma(B,N) \frac{\mu(m_B,m_a)^2}{\mu(m_B,m_N)^2 A^2}
\label{eq:pernucleon}
\end{equation}
where $\sigma_0(B,a)$ is the scattering cross section \emph{per nucleon}, $a$, 
at zero momentum transfer.  

In the high energy theory, matrix elements are determined from 
the scattering of the Higgs boson between a fermion within the 
dark matter baryon and a quark within the nucleon:
\begin{equation}
\mathcal{M}_{a} = 
    \frac{y_f y_q}{2 m_h^2} \sum_f \langle B| \bar{f} f | B \rangle 
              \sum_q \langle a| \bar{q} q | a \rangle \, ,
\end{equation}  
where the baryonic matrix elements use the non-relativistic normalization 
of one-particle states, 
$\langle B({\bf p}) | B({\bf q}) \rangle = 
\langle a({\bf p}) | a({\bf q}) \rangle = (2\pi)^3\delta^{(3)}({\bf p} - {\bf q})$
(for a nice discussion of the matching, see Ref.~\cite{Agrawal:2010fh})  
and the label $a=p,n$ for the proton and neutron.  It should be noted 
that $\mathcal{M}_{p,n}$ has units of length squared.
Here $y_f,y_q$ are the effective (Yukawa) couplings of the Higgs with the 
fermions and quarks, respectively.  Our normalization for the quark 
Yukawa coupling is $y_q \equiv \sqrt{2} m_q/v$, where $v \simeq 246$~GeV is the 
Higgs vacuum expectation value.

The matrix elements of the light quarks ($u$,$d$,$s$) in the neutron 
and proton are defined by
\begin{equation}
\langle a | m_q \bar{q} q | a \rangle \equiv m_a f^{(a)}_{q}
\end{equation}   
while the heavy quarks contribute \cite{Shifman:1978zn}
\begin{equation}
\langle a | m_q \bar{q} q | a \rangle = \frac{2}{27} m_a 
\left( 1 - \sum_{q=u,d,s} f^{(a)}_{q} \right) \, .
\end{equation}   
Again, label $a=p,n$ represents the proton and neutron, respectively. 
Chiral perturbation theory as well as lattice techniques allow the
extraction of the nucleon sigma terms 
$\sigma_q^{(p,n)} \equiv \langle p,n | m_q \bar{q} q | p,n \rangle$
that provide the numerical values for the $f^{(p,n)}_{q}$.
We use the values obtained in Ref.~\cite{Hill:2011be}.

For the composite scalar baryon considered here, by analogy with
nucleons we write
\beq
\langle B | m_f \bar{f} f | B \rangle \equiv m_B f^{(B)}_{f} \, ,
\eeq  
that are, like the strange quark content in ordinary baryons, 
determined on the lattice through the analogous sigma term.
The subtlety is that the fermion mass for the composite dark matter is
assumed \emph{not} to be solely EW breaking, and thus $m_f \not= y_f v/\sqrt{2}$.  
To determine the Higgs coupling, we write the constituent fermion
masses as an implicit function of the Higgs, $m_f(h)$.
The Yukawa coupling is thus expressed as an effective Higgs coupling given by
\begin{equation}
\frac{1}{\sqrt{2}} y_f \equiv 
    \left. \frac{\partial \, m_f(h)}{\partial \, h} \right|_{h=v} \, .
\end{equation}
(The $1/\sqrt{2}$ normalization implies a fermion with a mass
solely from electroweak symmetry breaking has a Higgs coupling 
of $y_f/\sqrt{2} = m_f/v$.)

Putting all of this together, we obtain
\begin{equation}\label{eq:f_tilde}
\mathcal{M}_{p,n} = \frac{g_{p,n} g_B}{m_h^2} 
\end{equation}
where 
\begin{equation}
g_{p,n} = \frac{m_{p,n}}{v} \left[ \sum_{q=u,d,s} f^{(p,n)}_{q}
            + \frac{6}{27} \left( 1 - \sum_{q=u,d,s} f^{(p,n)}_{q} \right) \right] 
\end{equation}
\begin{equation}
g_B = \frac{m_B}{v} \sum_f \frac{v}{m_f} 
                      \left. \frac{\partial \, m_f(h)}{\partial \, h} \right|_{h=v} 
                      f^{(B)}_{f} \label{gB-eq} 
\end{equation}
The expression for $g_B$ is determined by three factors.
The first, $m_B/v$ is completely analogous to the $m_{p,n}/v$ factor that 
occurs for the proton and neutron.  The baryon mass itself is extracted
from the lattice.  The second, 
\begin{equation}
\frac{v}{m_f} \frac{\partial \, m_f(h)}{\partial \, h}
\label{moddep-eq}
\end{equation}
is determined completely from the microscopic model.  Specific models
have specific interactions of the Higgs with the constituent fermions.
The third factor, 
\begin{equation}
f^{(B)}_{f} = \frac{\langle B | m_f \bar{f} f | B \rangle}{m_B} = \frac{m_f}{m_B} 
                                       \frac{\partial \, m_B}{\partial \, m_f}
\end{equation}
is extracted from the lattice.  In this way, we have factorized
the Higgs couplings into a model-dependent part (the second factor)
and two dimensionless non-perturbative parts to be extracted from the
lattice.  This is the main focus of the lattice calculation in this paper.

\section{$\mathbf{SU(4)}$ baryons} 
\label{sec:SU4 Baryons}

\begin{figure}[!t]
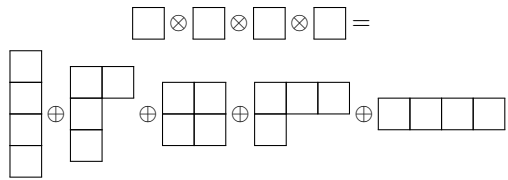
 
\centering
\begin{align}
\ytableausetup{boxsize=1.25em}
\ytableausetup{aligntableaux=center}
&\ydiagram{1} \otimes \ydiagram{1} \otimes \ydiagram{1} \otimes \ydiagram{1} = \nonumber\\
\ydiagram{1,1,1,1} \oplus \ydiagram{2,1,1}&\oplus \ydiagram{2,2} \oplus \ydiagram{3,1} \oplus \ydiagram{4} \nonumber
\end{align}
\caption{Young-tableau representation of SU(4) baryon group theory.}
\label{fig:yt}
\end{figure}

For a four-color theory with a generic number of flavors, there are 5 irreducible representations that emerge (corresponding Young-tableau diagrams are shown in Fig.~\ref{fig:yt}).  When there are four or more degenerate flavors, there exists a totally antisymmetric flavor combination (leftmost Young-tableau in Fig.~\ref{fig:yt}), a totally symmetric flavor combination (rightmost Young-tableau in Fig.~\ref{fig:yt}), and a variety of mixed symmetric-antisymmetric representations in between.  If there are only three degenerate flavors, the totally antisymmetric flavor combination no longer exists (leaving only the other 4 representations).  When there are only two degenerate flavors, only the symmetric and pair-wise antisymmetric states remain (the three rightmost Young-tableau in Fig.~\ref{fig:yt}, which would correspond to spin-0, spin-1, and spin-2 baryons from left to right).  Lastly, when there is only one flavor, only the totally symmetric spin-2 state exists.

\begin{figure}[!t] 
   \centering
   \includegraphics[width=0.20\textwidth]{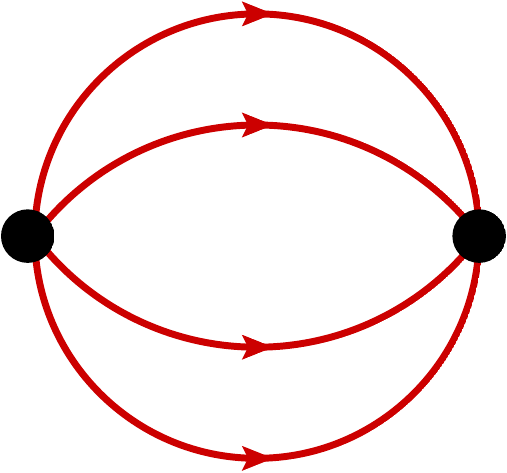}
   \caption{Contraction for 4-color baryons.}
\label{fig:4_color_contract} 
\end{figure}

As in QCD, one would expect the lightest state to be the lowest spin state with the most pairwise antisymmetric combinations.  Here, that would be the middle Young-tableau in Fig.~\ref{fig:yt} for all number of flavors of 2 or greater.  This state is the primary focus for our dark matter search.  The relative mass differences between the spin-0 state and the first spin-1 and spin-2 states are also of interest as they give a sense as to the maximum isospin splitting allowed before inverting the hierarchy.  These mass differences can also play a significant role in determining the remaining thermal relic.

On the lattice, rotational symmetry is broken by the discretization, and states of definite spin get mixed.  What the lattice does preserve is a set of the hypercubic groups, a Clebsch-Gordon decomposition of which is the most optimal way to extract multiple states and disentangle mixing of states of definite spin.
To increase computational efficiency at extracting the ground state in each spin-channel, more simplified operators were employed,
\beq
\mathcal{O}_B = (\psi^T_1 X_1 \psi_2)(\psi^T_3 X_2 \psi_4)
\eeq
where
\beqs
\text{Spin-0}&:& \quad X_1=C\gamma^5 \quad X_2=C\gamma^5\nonumber\\
\text{Spin-1}&:& \quad  X_1=C\gamma^i \quad X_2=C\gamma^5\quad i=1,2,3\nonumber\\
\text{Spin-2}&:& \quad X_1=C\gamma^i \quad X_2=C\gamma^j\quad  i\neq j.\nonumber
\eeqs
Again, these operators allow for mixing with higher angular-momentum baryons (spin-3 and above), but the ground state extracted should nevertheless correspond to the lowest spin state.

The $\psi_i$ notation is used to denote the possibility of different flavors of fermions.  In the degenerate mass limit, for one-flavor ($\psi_i=U$), there is one unique combination
\beq
\mathcal{O}_{B,1}^{N_F=1} = (U^T X_1 U)(U^T X_2 U),
\eeq
for two flavors ($\psi_i=U,D$) there are four unique combinations
\beqs
\mathcal{O}_{B,1}^{N_F=2} &=& (U^T X_1 U)(U^T X_2 D), \nonumber\\
\mathcal{O}_{B,2}^{N_F=2} &=& (U^T X_1 D)(U^T X_2 U), \nonumber\\
\mathcal{O}_{B,3}^{N_F=2} &=& (U^T X_1 U)(D^T X_2 D), \nonumber\\
\mathcal{O}_{B,4}^{N_F=2} &=& (U^T X_1 D)(U^T X_2 D),
\eeqs
for three flavors ($\psi_i=U,D,S$), there are three unique combinations
\beqs
\mathcal{O}_{B,1}^{N_F=3} &=& (U^T X_1 U)(D^T X_2 S), \nonumber\\
\mathcal{O}_{B,2}^{N_F=3} &=& (U^T X_1 D)(U^T X_2 S), \nonumber\\
\mathcal{O}_{B,3}^{N_F=3} &=& (D^T X_1S)(U^T X_2 U)
\eeqs
and for four flavors ($\psi_i=U,D,S,C$), there is only one unique combination
\beq
\mathcal{O}_{B,1}^{N_F=4} = (U^T X_1D)(S^T X_2 C).
\eeq
Since these combinations span over the entirety of the flavor space, one would expect to have overlap with the ground state in each (lattice) spin channel.

\section{Comparison of 3 and 4 color baryons}
\label{sec:3_4_comp}

Our study of SU$(4)$ baryons fits into a larger framework of large-$N_c$ lattice calculations.  Much of the large-$N_c$ lattice effort has focused on gluonic observables and spectra \cite{Lucini:2001ej,Lucini:2010nv,Athenodorou:2010cs}, whose calculations on fermion-quenched lattices yield complete calculations of pure Yang-Mills theories.  However, since fermion loops in the sea are suppressed at large-$N_c$, the quenched approximation is reasonable for fermionic observables as well.  It should be noted, however, that these observables are significantly more expensive to compute than gluonic since fermion operators need to be inverted.  Nonetheless, there have been multiple calculations of meson spectra \cite{DelDebbio:2007wk,Bali:2008an}, with the most complete and comprehensive calculation occurring recently \cite{Bali:2013kia} (for a complete review of large $N_c$ lattice calculations, see Ref.~\cite{Lucini:2012gg}).

While there has been a wealth of literature discussing large $N_c$ baryons for over 30 years \cite{Witten:1979kh,Adkins:1983ya,Jenkins:1993zu,Jenkins:1994md,Dashen:1994qi,Jenkins:1995td,Cohen:2003tb,Cherman:2009fh}, the first large $N_c$ lattice calculations of baryons have only occurred in the last couple of years.  This is mostly due to the fact that baryon contractions are significantly more involved than mesons and grow in computational cost as $N_c!$.  In Ref.~\cite{DeGrand:2012hd}, the baryon spectrum is calculated for $N_c=3,5,7$ on quenched configurations for degenerate fermion masses and in Ref.~\cite{DeGrand:2013nna}, these results are generalized to splitting a third fermion mass akin to the strange quark in QCD.  These results show strong agreement with the large $N_c$ predictions.

At a fixed scale (normalized by some physical quantity such as the mass of the lightest baryon in the chiral limit), the meson and baryon spectra are expected to have significantly different behavior at different values of $N_c$.  Low-lying meson states are not expected to change appreciably as $N_c$ increases ($\mathcal{O}(1)$ in $N_c$-scaling), while baryons, which contain $N_c$ fermions in a color-antisymmetric combination, are expected to have masses that scale linearly with $N_c$ at leading order.  For large $N_c$ baryons made from degenerate-mass fermions, the behavior of the spectrum is contained in a simple relation based on the rotor spectrum \cite{Adkins:1983ya,Jenkins:1995td,DeGrand:2012hd}
\beq\label{eq:rotor}
M(N_c, J)=N_c m_0 + \frac{J(J+1)}{N_c}B  + \mathcal{O}(1/N_c^2),
\eeq
where $J$ is the baryon spin and $m_0$ and $B$ are constants that need to be extracted from two initial input values.  These kinds of large-$N_c$ relations have been seen to work remarkably well for phenomenological extractions of the baryon spectrum and splitting (including strange-light quark mass splittings) \cite{Jenkins:1995td}, in lattice 3-color calculations with a variety of light and strange quark masses \cite{Jenkins:2009wv}, and this rotor spectrum itself has been shown to work to high precision for odd $N_c$ baryons with three, five, and seven colors \cite{DeGrand:2012hd,DeGrand:2013nna}; well better than one would na\"ively expect up to $\mathcal{O}(1/N_c^2)$ corrections.  One common theme among previous spectrum comparisons of this kind is that all the baryons were fermionic as in QCD.  This raises the question of how the large $N_c$ relations fare when comparing fermionic baryons in odd $N_c$ theories to bosonic baryons in even $N_c$.

One point that was emphasized in Ref.~\cite{DeGrand:2013nna} is the fact that each coefficient in Eq.~\eqref{eq:rotor} has corrections that go as $N_c^{-1}$, $N_c^{-2}$, etc.  With that in mind, in the formal large $N_c$ limit, Eq.~\eqref{eq:rotor} should be written as \cite{DeGrand:2013nna}
\beq\label{eq:rotor_c}
M(N_c, J)=N_c m_0^{(0)} +C+ \frac{J(J+1)}{N_c}B  + \mathcal{O}(1/N_c^2),
\eeq
where $m_0^{(0)}$ is the leading $\mathcal{O}(1)$ contribution to $m_0$ and $C$ is the subleading $\mathcal{O}(1/N_c)$ correction to $m_0$.  With two degenerate flavors, 3-color QCD can only provide two points of input for these formulas (the spin-1/2 and spin-3/2 baryon mass).  For that reason, Eq.~\eqref{eq:rotor} with two free parameters is completely determined by 3-color QCD input, while Eq.~\eqref{eq:rotor_c} requires one more state from a different $N_c$ to fix its three free parameters.

\begin{figure}[!t] 
   \centering
   \includegraphics[width=0.46\textwidth]{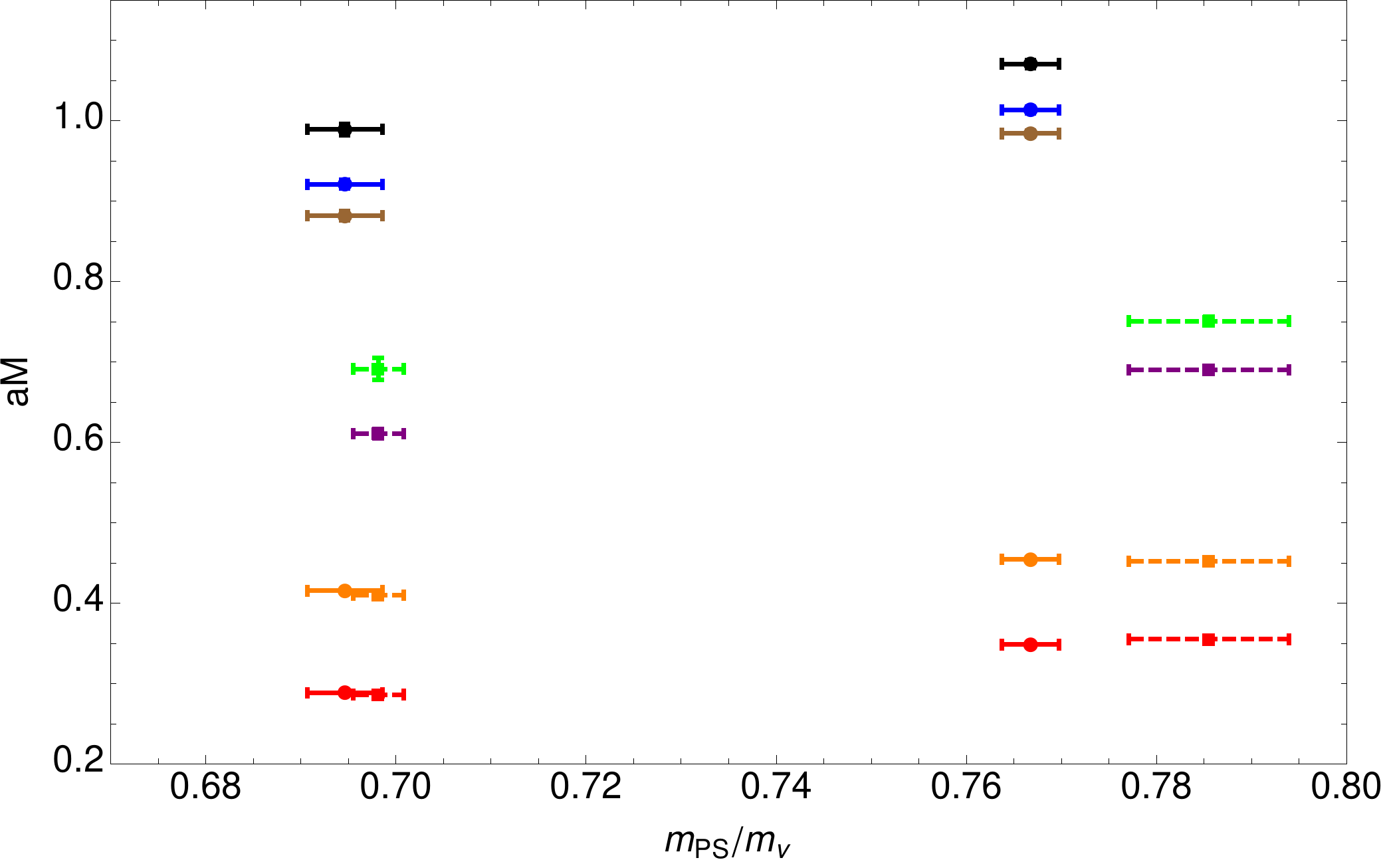}
   \caption{Comparison of SU(3) (square, dashed) and SU(4) (circle, solid) spectrum for the pseudoscalar (red), vector (orange), 3-color spin-1/2 baryon (purple), 3-color spin-3/2 baryon (green), 4-color spin-0 baryon (brown), 4-color spin-1 baryon (blue), and 4-color spin-2 baryon (black).  All calculations shown were for $32^3 \times 64$ lattices. }
\label{fig:3c4c}
\end{figure}

\begin{figure}[!t] 
   \centering
   \includegraphics[width=0.46\textwidth]{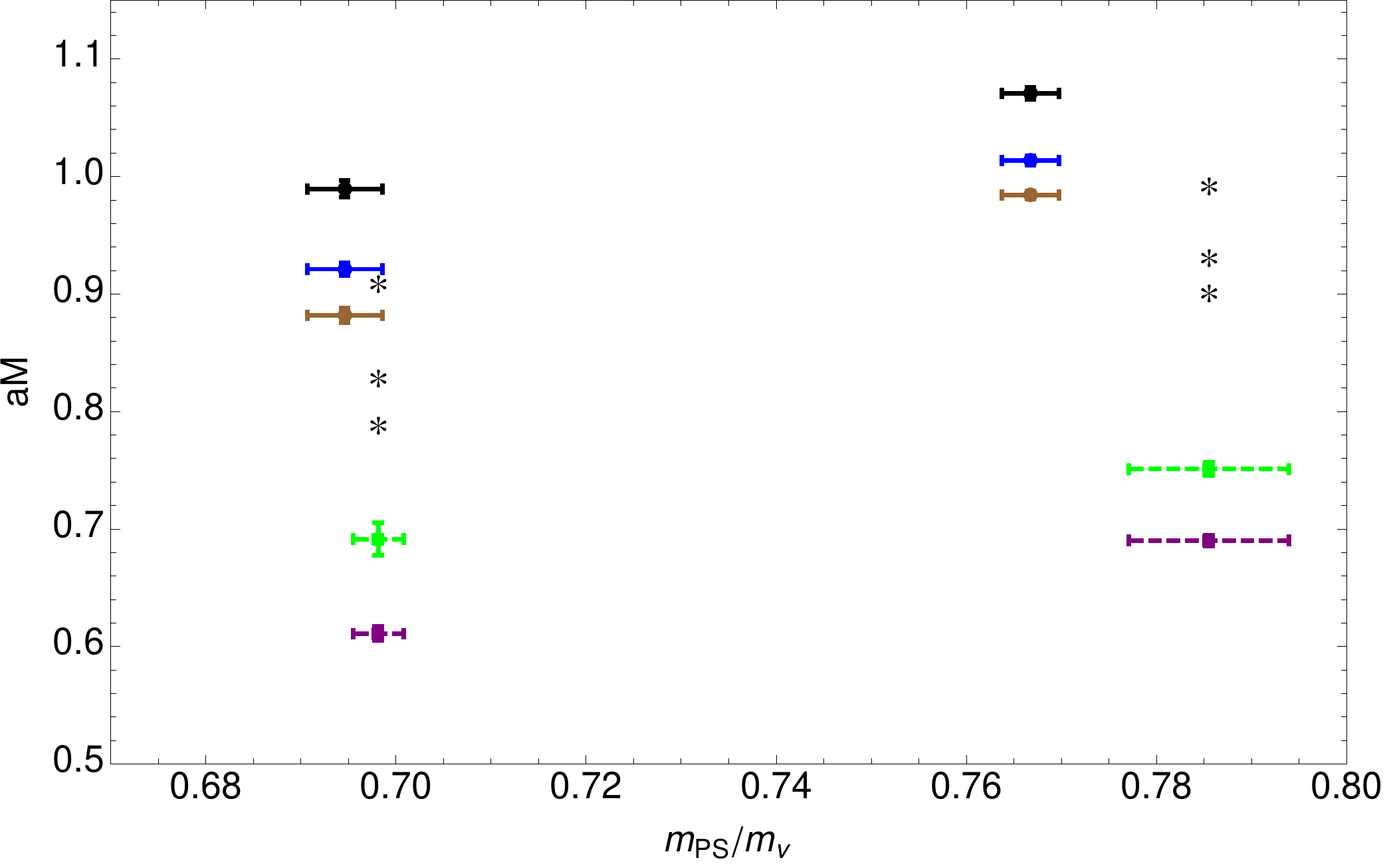}
      \includegraphics[width=0.46\textwidth]{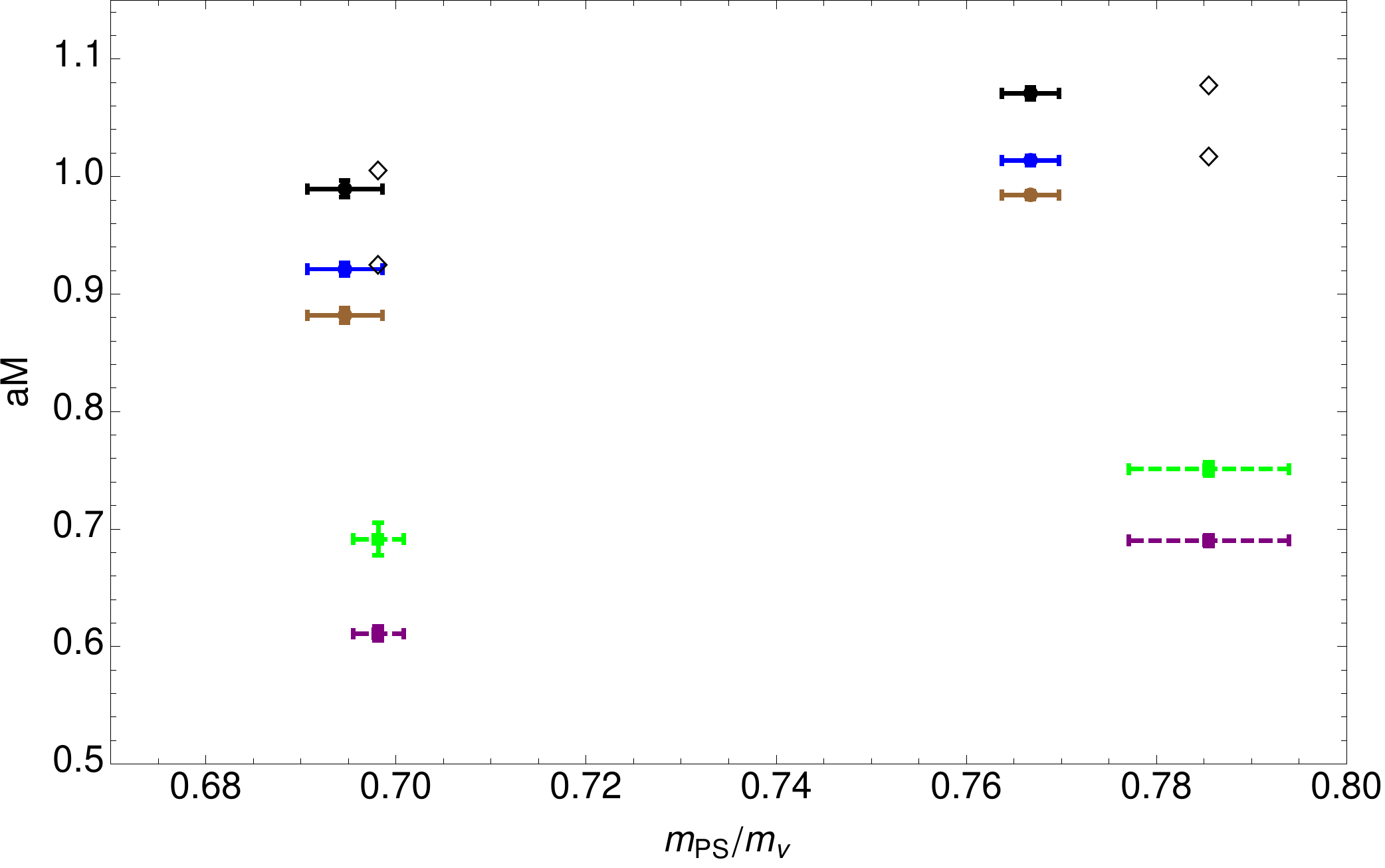}
   \caption{Comparison of SU(3) and SU(4) baryons to the large $N_c$ rotor spectra predictions.  In the top plot, the asterisks represent the two-parameter rotor spectrum predictions for SU(4) baryons, Eq.~\eqref{eq:rotor}, which uses the SU(3) baryon spectrum as input.  Similarly, in the bottom plot, the diamonds represent the three-parameter rotor spectrum predictions for SU(4) baryons, Eq.~\eqref{eq:rotor_c}, which uses the 3-color baryon spectrum and 4-color spin-0 baryon mass as input.}
\label{fig:3c4c_rotor}
\end{figure}

In Fig.~\ref{fig:3c4c}, 3-color (dashed squares) and 4-color (solid circles) results are compared for $\beta$ values in Ref.~\cite{Bali:2013kia}.  Also, fermion masses between these two theories were chosen in Ref.~\cite{Bali:2013kia} to match the pseudoscalar meson mass.  As a result, the vector mass, which is not expected to have any appreciable scaling at different $N_c$ matches quite well between the two theories.  Also, as expected, the baryon masses for the 4-color theory are all significantly larger than the 3-color theory.  Fig.~\ref{fig:3c4c_rotor} shows the comparisons of the baryon spectrum to the large $N_c$ rotor spectrum prediction.  What is worth noting here is that the two-parameter rotor spectrum predictions  (top figure, black asterisks) from Eq.~\eqref{eq:rotor} for the 4-color baryons using the 3-color baryon input do not align with the lattice 4-color results, while the three-parameter rotor-spectrum (bottom figure, black diamonds) in Eq.~\eqref{eq:rotor_c} is consistent with the lattice values for the spin-1 and spin-2 baryons.  Inherently, fermionic and bosonic baryons have different wave functions, which can shift the scale $m_0$.  However, when this $\mathcal{O}(1)$ term is included, the spectra appear to match well within the expected $\mathcal{O}(1/N_c^2)$ correction.  Also, it is worth noting that the numerical values of $m_0^{(0)}$ and $C$ are consistent within 15\%, indicating that this $C$ term really is a subleading correction to $m_0$, as expected from a systematic large $N_c$ expansion of $m_0$.

\section{Bounds from the effective Higgs coupling}
\label{sec:Higgs_Results}

\begin{table}[t]
   \centering
   {\footnotesize
   \begin{tabular}{|c|c|c|} 
   \hline
   $\kappa$ & $\frac{m_{PS}}{m_V}$ & $\frac{m_f}{m_B}\frac{\partial m_B}{\partial m_f}$ \\
   \hline
   0.1554 & 0.767(3)  & 0.338(17)   \\
    \hline
    0.15625 & 0.695(4)  & 0.262(13)   \\
     \hline
      0.1572 & 0.549(5)  & 0.153(8)   \\
     \hline
   \end{tabular} }
   \caption{Normalized sigma parameter results for $\beta=11.028$ on $32^3 \times 64$ lattices.}
   \label{tab:b11_data_sigma}
\end{table}

\begin{figure}[!t] 
   \centering
   \includegraphics[width=0.46\textwidth]{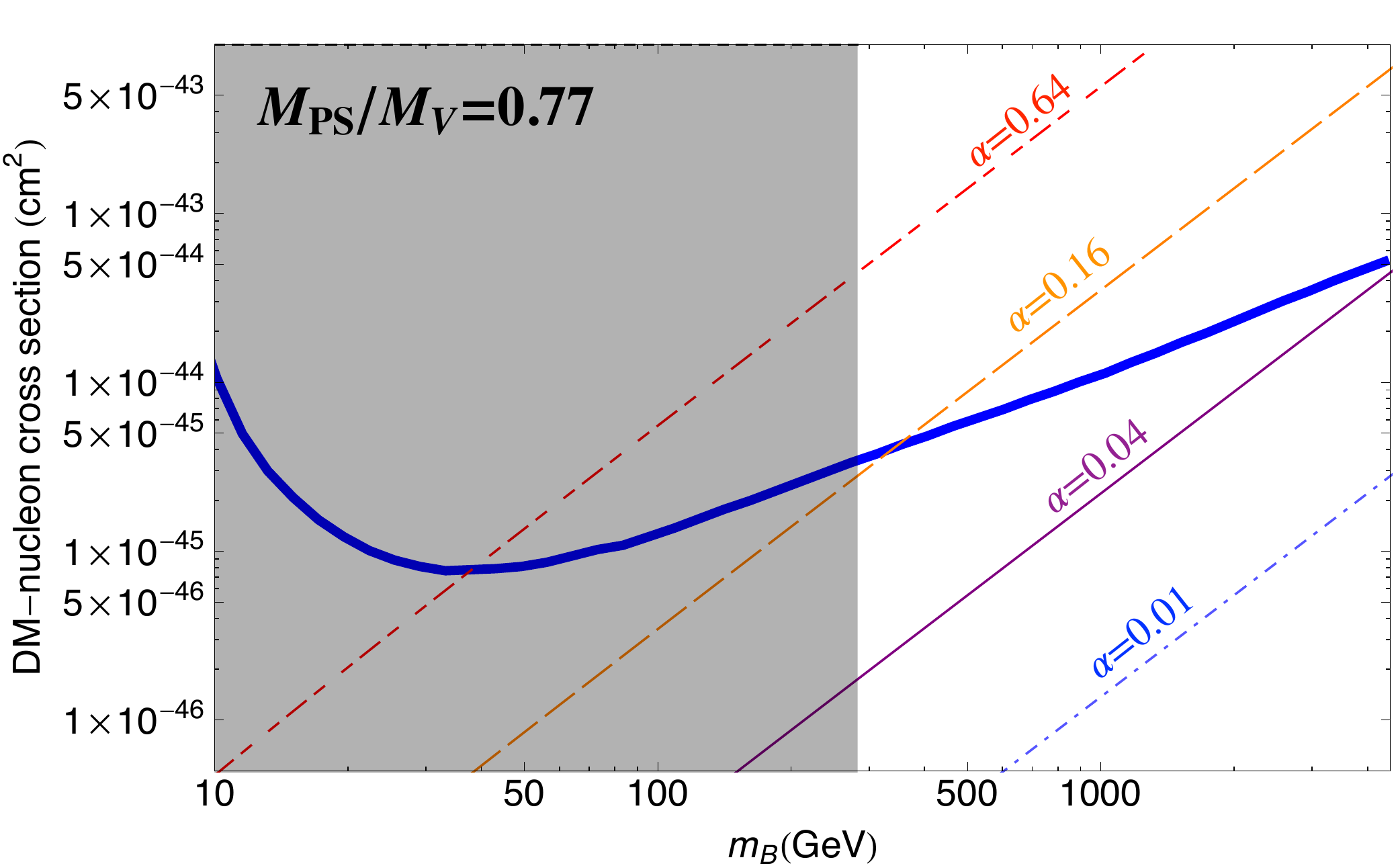}  \\
   \includegraphics[width=0.46\textwidth]{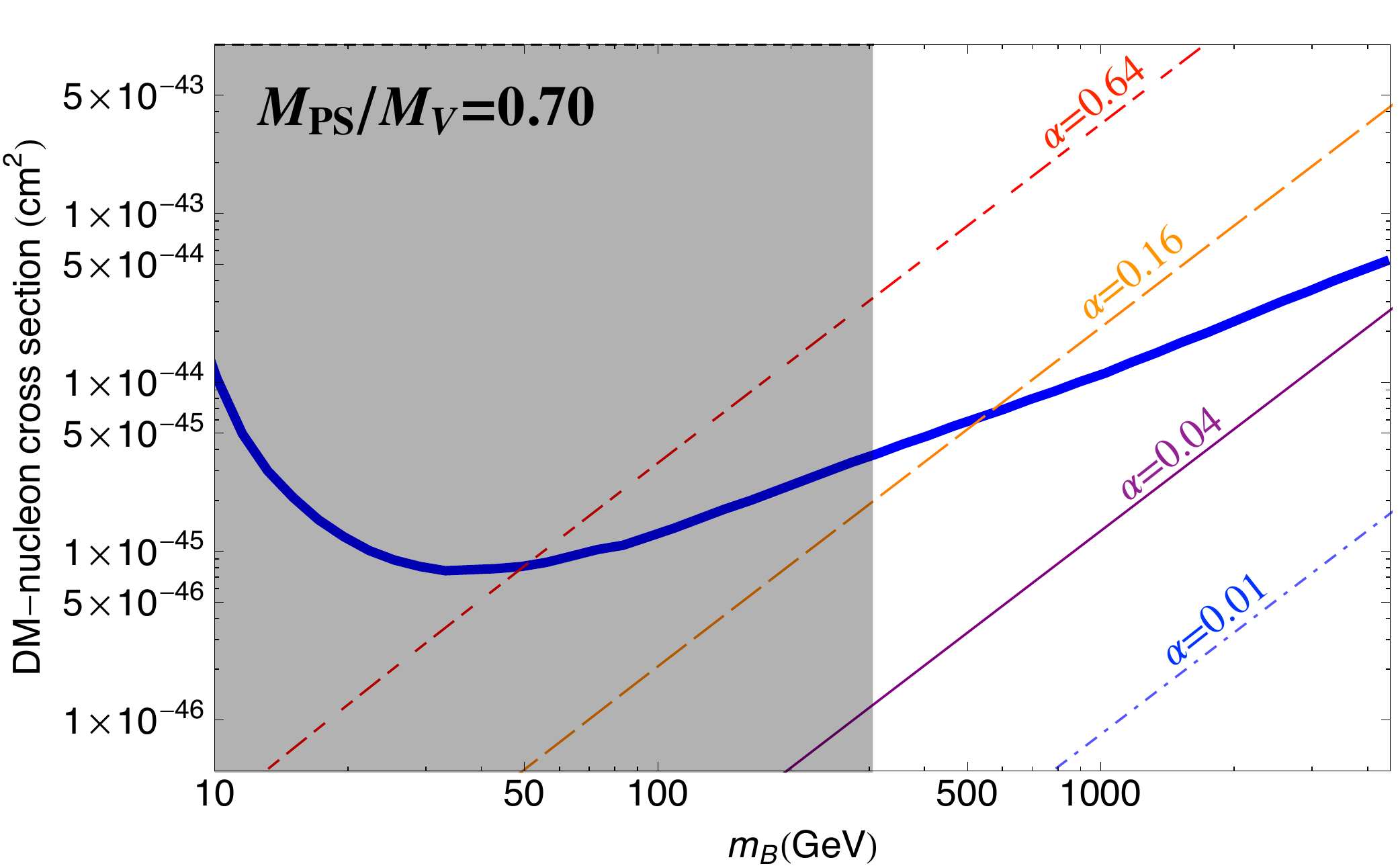} \\
   \includegraphics[width=0.46\textwidth]{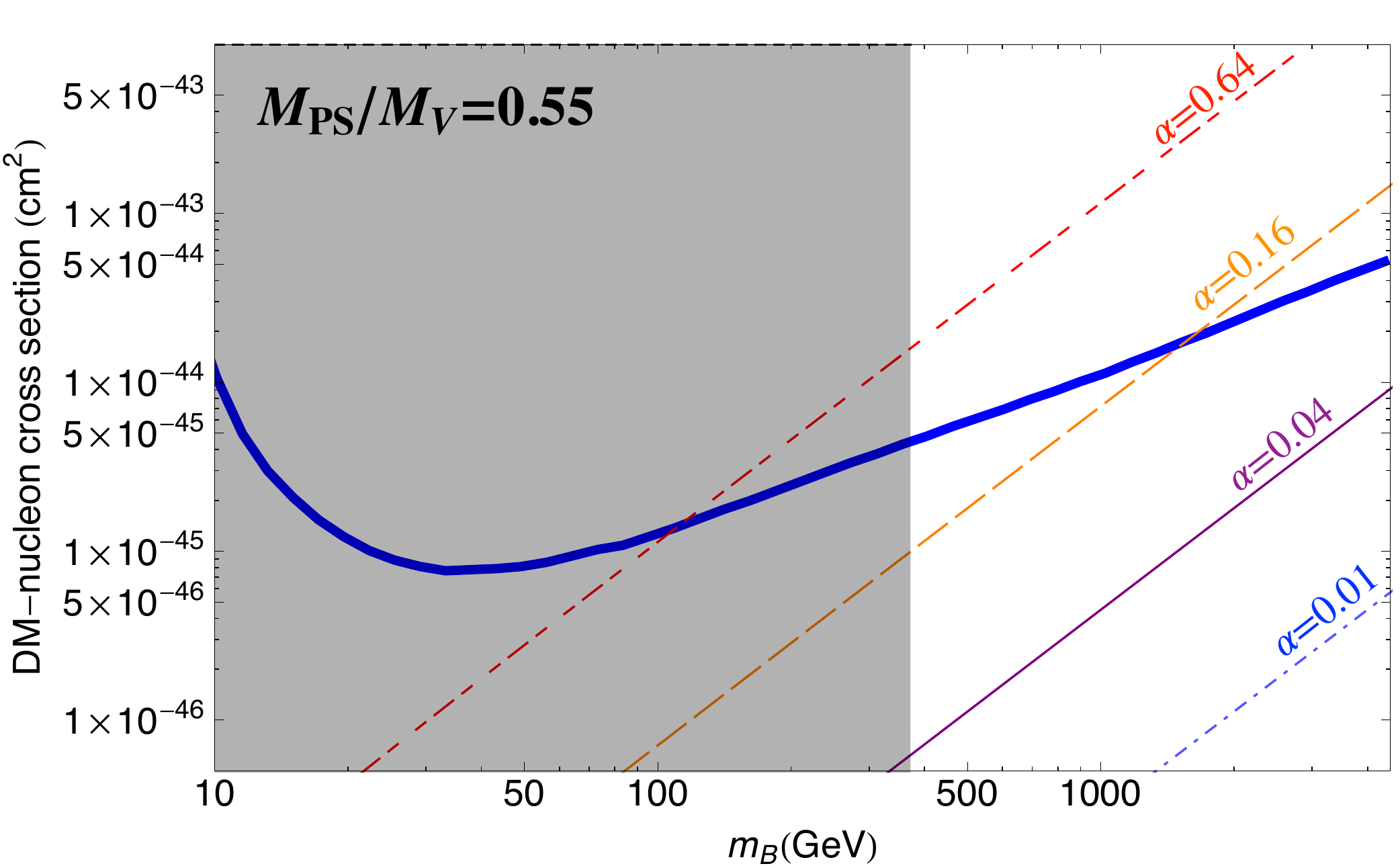} 
   \caption{The spin-independent dark matter direct detection scattering 
cross section per nucleon through Higgs exchange is shown.  
The solid blue curve is the upper bound set by LUX \cite{Akerib:2013tjd}.  
The three plots correspond to three different fermion masses: 
$m_{PS}/m_{V} \simeq 0.77$ (top), 
$m_{PS}/m_{V} \simeq 0.70$ (middle), and 
$m_{PS}/m_{V} \simeq 0.55$ (bottom). 
Each thin line represents the spin-independent scattering cross section predicted
for a particular effective Higgs coupling, given by $\alpha$ 
[defined in Eq.~\eqref{eq:alpha}].  The dark shaded region has 
pseudoscalar mesons with masses below $100$~GeV, which we anticipate 
are excluded by LEP II bounds.  Notice that the corresponding bound on 
the baryon mass (slightly) increases as $m_{PS}/m_V$ is lowered, 
from top to bottom.}
\label{fig:Higgs_Exclusion} 
\end{figure}

The dark baryon is comprised of fermions that acquire some of their mass
from the interactions with the Higgs field.  This means there is a 
model-dependent Higgs interaction with the dark baryon that we can
constrain using the non-perturbative information extracted from
the lattice.  The crucial input from our lattice simulations is the zero-momentum
scalar form factor, $\sigma_f$.  
While the entire momenta dependent form factor can be extracted on the lattice directly through 
an expensive three-point calculation with disconnected diagrams, 
$\sigma_f$ is more straightforward to extract via the Feynman-Hellmann theorem,
\beq
\sigma_f = m_f \langle B| \bar{f} f |B  \rangle 
         = m_f \frac{\partial m_B}{\partial m_f},
\label{eq:sig}
\eeq
where $am_B$ and $am_f$ are dimensionless numbers for the baryon mass and fermion mass extracted from the lattice calculation, and $a$ is the dimensionful lattice spacing, whose inverse represents the UV cutoff of the theory. 
In lattice simulations, $m_f$ is the standard (renormalized) fermion mass 
in the mass-diagonalized basis, which is defined in the Wilson fermion 
action used here in terms of the lattice input $\kappa$,
\beq
am_f = \frac{1}{2}\bigg(\frac{1}{\kappa}-\frac{1}{\kappa_c}\bigg),
\eeq
where $\kappa_c$ is the critical value where the fermion mass vanishes.  
Unlike the Higgs coupling to the nucleons in QCD, the effective coupling 
of the Higgs to the dark baryons is parameterized by
\begin{equation}
\alpha \equiv \frac{v}{m_f} \frac{\partial \, m_f(h)}{\partial \, h}\bigg|_{h=v} 
   \, . 
\label{eq:alpha}
\end{equation}
For a given $\alpha$ and $m_{PS}/m_V$, 
we can calculate the spin-independent scattering
cross section off nucleons and compare directly to bounds from 
dark matter direct detection experiments.
In principle, there is an $\alpha$ for each fermion $f$. 
However, since we assumed degenerate fermions in the quenched approximation, 
the lightest baryon is made of identical mass fermions with the identical
Higgs coupling, and so no flavor label is necessary. 

To illustrate how the constraints on $\alpha$ impact the model-dependent
fermion-Higgs couplings, we can parameterize the Higgs 
field-dependent mass as, 
\begin{equation}
m_f(h) = m + \frac{y h}{\sqrt{2}} \, ,
\label{eq:expandmass}
\end{equation}
that, for example, would arise from a model with both vector-like 
masses as well as electroweak symmetry breaking masses for the 
constituent fermions \cite{shortpaper}.  The expression for $\alpha$,
Eq.~(\ref{eq:alpha}), is then 
\begin{equation}
\alpha = \frac{y v}{\sqrt{2} m + y v} \le 1 \, .
\end{equation}
In the limit that $y \rightarrow 0$, the fermion masses become purely 
vector-like, corresponding to $\alpha = 0$; no bound can be placed on 
these models from Higgs exchange.  On the other hand, if $m \rightarrow 0$ 
at fixed $y$, then the fermion masses are purely from electroweak symmetry breaking, 
and the effective Higgs coupling is maximized, $\alpha = 1$.  For a fixed $m_B$, 
the resulting bounds from direct-detection experiments are therefore 
quite strong.  

The fact that the baryon mass is roughly linear in the fermion mass is 
universally observed in lattice QCD data  \cite{WalkerLoud:2008bp}.  
This trend helps us calculate $\sigma_f$  without many fermion mass values.  
In other words, $\partial m_B/\partial m_f$ is roughly constant.  
We have extracted $\partial m_B/\partial m_f$ from our lattice data, 
\begin{equation}
0.153 \; \lsim \;  
      \sum_f f^{(B)}_{f} \equiv \frac{m_f}{m_B}\frac{\partial m_B}{\partial m_f}
      \; \lsim \; 0.338  ,
\label{eq:Non-pert_value}
\end{equation}
for 
\begin{equation}\label{eq:meson_ratio_range}
0.55 \; \lsim \; \frac{m_{PS}}{m_V} \; \lsim \; 0.77 \, .
\end{equation}
We emphasize that the sum over valance fermions is implicitly done in 
our lattice extraction of $\sigma_f/m_B$, and thus 
$\sum_f f^{(B)}_{f} \le 1$.  The values for the coarsest lattice 
spacing/largest volume calculations 
(which have the least lattice artifacts as detailed in 
subsequent sections) are given in Table \ref{tab:b11_data_sigma}.

\begin{figure}[!t] 
\centering
\includegraphics[width=0.46\textwidth]{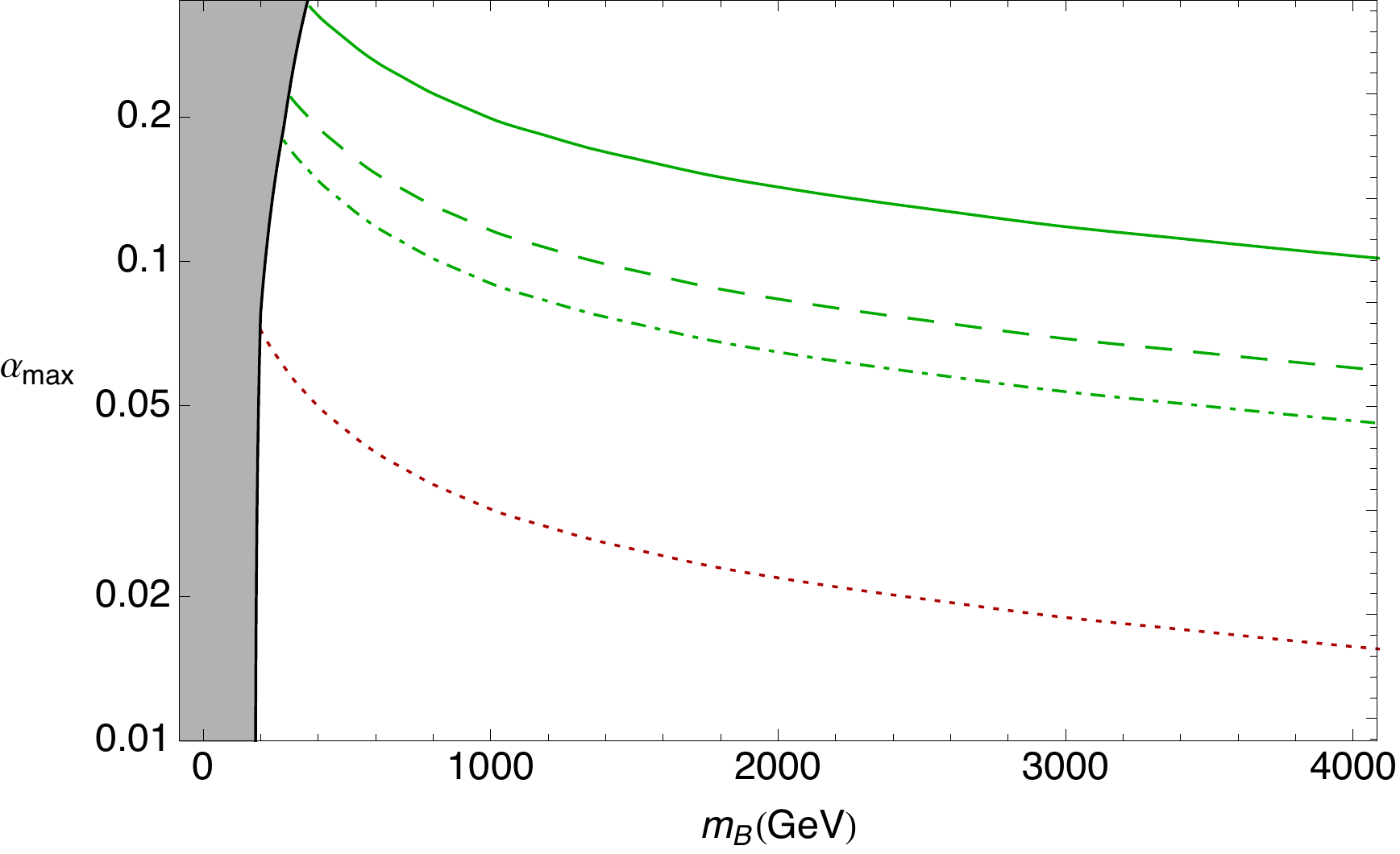}
\caption{The maximum allowed effective Higgs coupling, $\alpha$
[defined in Eq.~\eqref{eq:alpha}] to the dark baryon is shown.
The green (upper three) contours correspond to 
$m_{PS}/m_{V} \simeq 0.55$ (top, solid), 
$m_{PS}/m_{V} \simeq 0.70$ (upper middle, dashed), and 
$m_{PS}/m_{V} \simeq 0.77$ (lower middle, dot-dashed), 
obtained by finding the largest $\alpha$ allowed by the LUX bounds 
as a function of the dark matter mass for the three values 
of $m_{PS}/m_{V}$ simulated on the lattice in this paper.
The red (lower dotted) line is the maximum $\alpha$ in heavy fermion limit,  
$m_{PS}/m_{V} = 1$.  The dark shaded region has pseudoscalar mesons 
with masses below $100$~GeV, which we anticipate are excluded by LEP II bounds.}
\label{fig:alpha_max} 
\end{figure}

Using the lattice results for the baryon masses and $\sigma_f$, we employ
Eq.~\eqref{eq:cross_section} and the definitions in Eq.~\eqref{eq:f_tilde} 
to calculate the spin-independent Higgs-mediated cross section off nucleons 
for several values of $\alpha$ in the standard $(m_B,\sigma_0)$ plane.  
These results are shown in Fig.~\ref{fig:Higgs_Exclusion}. 
Additionally, the strongest limit to date from LUX \cite{Akerib:2013tjd} 
is also displayed.   In addition to the LUX exclusions, we also show the 
lower bound on $m_B$ resulting from requiring the lightest 
pseudoscalar meson satisfy the anticipated LEP II bound of $m_{PS} > 100$~GeV\@. 
The corresponding bound on $m_B$ is obtained using our lattice result 
for the ratio $m_B/m_{PS}$, given in Table~\ref{tab:b11_data}. 
These excluded regions are depicted by the dark, shaded region 
in Fig.~\ref{fig:Higgs_Exclusion}.

These results, when combined with the LUX exclusion curves, allow for an 
extraction of the maximum allowed $\alpha$ values as a function of 
dark matter mass and are presented in Fig.~\ref{fig:alpha_max}. 
The shape of the maximum $\alpha$ curve can be understood in the 
large $m_B$ limit:  The cross section $\sigma_0$, Eq.~(\ref{eq:pernucleon}), 
scales parametrically as $\alpha^2 m_B^2$ while the LUX bound on $\sigma_0$ 
is weakening proportional to $m_B$.  This implies that the maximum $\alpha$
decreases roughly proportional to $1/\sqrt{m_B}$.  
The plots are robust for any composite SU(4) model with the particular fermion mass range corresponding to Eq.~\eqref{eq:meson_ratio_range}.  Any such model with $\alpha$ above the green curves would be excluded for this class of theories.  Also, depicted on this figure is the result one would expect from the heavy fermion limit, where the fermion mass is much larger than the confinement scale. 

\subsection{Qualitative expectations at low fermion mass}
One limit not depicted in Fig.~\ref{fig:Higgs_Exclusion} or  Fig.~\ref{fig:alpha_max} is  $m_{PS}/m_{V} \rightarrow 0$. In this case, the sigma term should approach a constant (which can only be extracted with very low mass lattice simulations not explored in the work), but the allowed values will be greatly restricted by the bounds on the lightest mesons, since $m_B/m_{PS}\rightarrow \infty$ in this limit.  What this would imply for Fig.~\ref{fig:Higgs_Exclusion} is that each of the thin lines would move to smaller cross-sections as the fermion mass decreases (thus less excluded) at a perpetually slower rate, while the shaded excluded region would push to the right with a rapidly increasing rate.  For Fig.~\ref{fig:alpha_max}, the top green curve would move upward at a slower rate (thus allowing a higher allowed values of $\alpha$), while the shaded excluded section would push more and more to the right.   To answer this question quantitatively, additional fully-dynamical lattice calculations would need to be performed at smaller mass.  However, we do expect an ``ultimate'' bound  for each dark matter mass from the combination of the sigma parameter and LEP bound.  In future lattice calculations, we hope to address this region.

\section{Simulation Details}
\label{sec:Sim_Detail}

The 4-color calculations were performed on quenched lattices (10,000 trajectories each; configurations separated every 50 trajectories with thermalization cuts of 500 heatbath trajectories) at three different lattice spacings ($\beta=11.028, 11.5, 12.0$) at four different volumes ($16^3 \times 32$, $32^3 \times 64$, $48^3 \times 96$, $64^3 \times 128$).  Autocorrelations were found to be smaller than statistical errors when measurements were taken every 50 trajectories.  The 3-color quenched calculations were performed on $\beta=6.0175$, $32^3 \times 64$ lattices to compare three and four colors using the scale matching in Ref.~\cite{Bali:2008an}.  All heat bath gauge generation (Wilson gauge action) and inversions were performed using \textit{Chroma} \cite{Edwards:2004sx}.  For 4-colors, three fermion mass values were explored for $\beta=11.028$, five mass values for $\beta=11.5$, and six mass values were explored for $\beta=12.0$.   All the data and number of measurements are presented in Table~\ref{tab:meas}.

\begin{table}[t]
   \centering
   {\footnotesize
   \begin{tabular}{|c|c|c|c|c|}
   \hline
   $N_c$& $\beta$ & $\kappa$ & $N_s^3 \times N_t$ & \# Meas. \\
   \hline
   4 &11.028 &  0.1554 & $16^3 \times 32$ & 4878 \\
    & &   & $32^3 \times 64$ & 1126 \\
    \hline
     & &  0.15625 & $16^3 \times 32$ & 4765 \\
     & &   & $32^3 \times 64$ & 1146 \\
     & &   & $48^3 \times 96$ & 1091 \\
     \hline
     & &  0.1572 & $32^3 \times 64$ & 1075 \\
     \hline
     & 11.5 & 0.1515 &  $16^3 \times 32$ & 2975 \\
     & &  &  $32^3 \times 64$ & 1057 \\
     \hline
     & & 0.1520 &  $16^3 \times 32$ & 2872 \\
     & &  &  $32^3 \times 64$ & 1052 \\
     \hline
     & & 0.1523 &  $16^3 \times 32$ & 2976 \\
     & &  &  $32^3 \times 64$ & 914 \\
     & &  &  $48^3 \times 96$ & 637 \\
     & &  &  $64^3 \times 128$ & 489 \\
     \hline
     & & 0.1524 &  $16^3 \times 32$ & 2970 \\
     & &  &  $32^3 \times 64$ & 863 \\
     \hline
     & & 0.1527 &  $32^3 \times 64$ & 1011 \\
     \hline
     & 12.0 & 0.1475 &  $32^3 \times 64$ & 1125 \\
     \hline
      & & 0.1480 &  $32^3 \times 64$ & 1189 \\
      \hline
      & & 0.1486 &  $32^3 \times 64$ & 1055 \\
      \hline
      & & 0.1491 &  $16^3 \times 32$ & 411 \\
      & & 0.1491 &  $32^3 \times 64$ & 1050 \\
      & & 0.1491 &  $48^3 \times 96$ & 1150 \\
      && 0.1491 &  $64^3 \times 128$ & 928 \\
      \hline
      & & 0.1495&  $32^3 \times 64$ & 1043 \\
      \hline
       && 0.1496&  $32^3 \times 64$ & 1009 \\
       \hline
       3 & 6.0175 & 0.1537 & $32^3 \times 64$ & 1000 \\
       & & 0.1547 & $32^3 \times 64$ & 1000\\
       \hline
   \end{tabular} }
   \caption{Ensembles and number of measurements.}
   \label{tab:meas}
\end{table}

\begin{table}[t]
   \centering
   {\footnotesize
   \begin{tabular}{|c|c|c|c|c|c|} 
   \hline
   $\kappa$ & $am_{PS}$ & $am_V$ & $aM_{S0}$ & $aM_{S1}$ & $aM_{S2}$  \\
   \hline
   0.1554 & 0.3477(6)  & 0.4549(18) & 0.9828(33) & 1.0119(39) & 1.0668(45)  \\
    \hline
    0.15625 & 0.2886(7)  & 0.4170(20) & 0.8831(55) & 0.9183(55) & 0.9883(79)  \\
     \hline
      0.1572 & 0.2066(8)  & 0.3783(26) & 0.7687(92) & 0.8129(74) & 0.898(19)  \\
     \hline
   \end{tabular} }
   \caption{Spectrum results for $\beta=11.028$ on $32^3 \times 64$ lattices.}
   \label{tab:b11_data}
\end{table}

\section{Calculation and Fitting}
\label{sec:Calc_Fit}

\begin{figure}[!t] 
   \centering
   \includegraphics[width=0.46\textwidth]{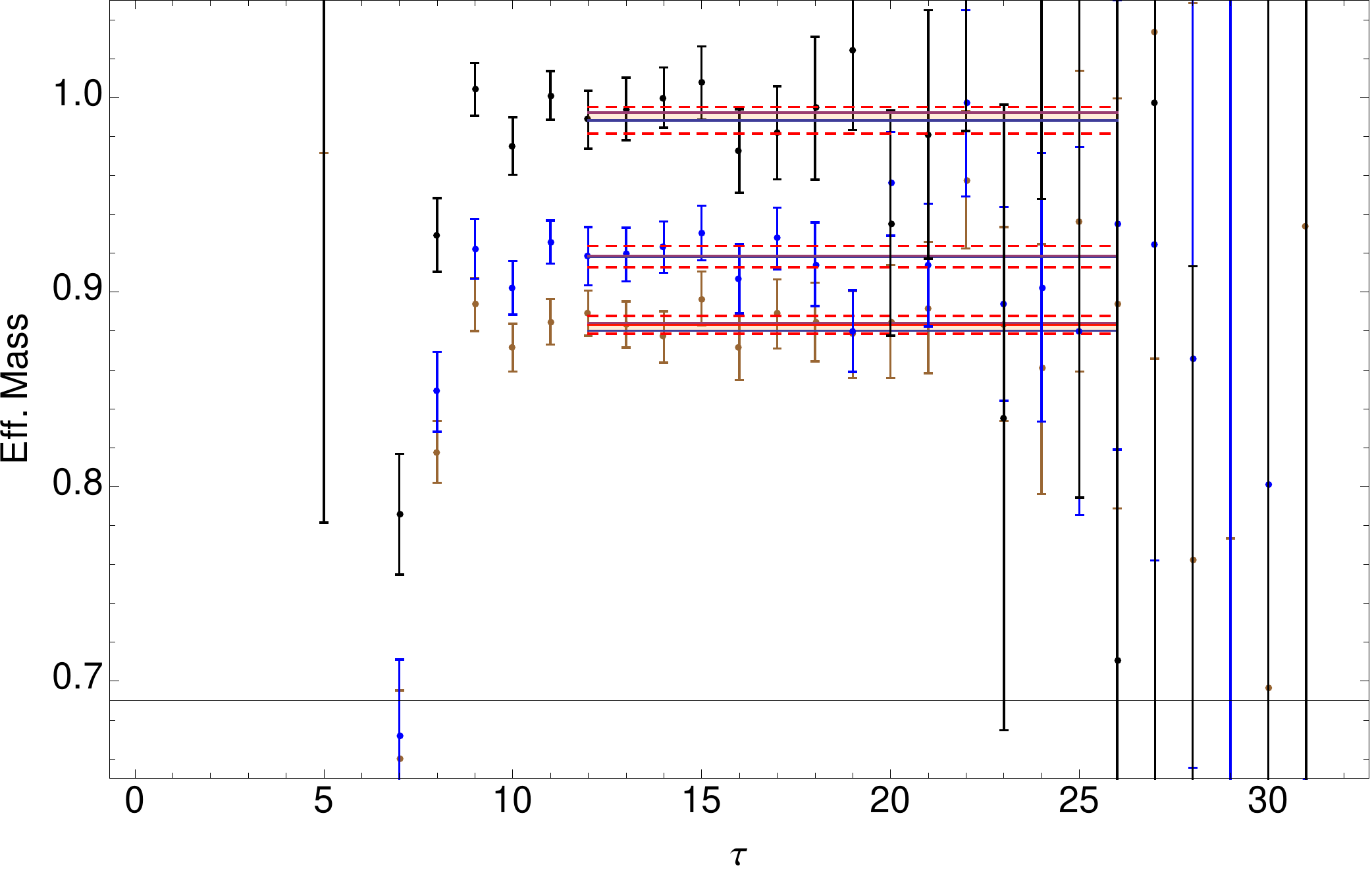}
   \caption{Example of folded 4-color baryon effective mass for $32^3 \times 64$, $\beta=11.028$, $\kappa=0.15625$ lattices.  Plotted on the figure are the effective mass for the spin-0 (bottom), spin-1 (middle), and spin-2 (top) baryons.}
\label{fig:eff_mass}
\end{figure}

The masses of the baryons are extracted from the long Euclidean time behavior of the baryon two point function projected onto zero momentum
\beqs
\label{eq:baryon_2_point}
  C_{BB}(\tau)&=&\sum_\mathbf{x} 
    \langle \mathcal{O}_B(\mathbf{x},\tau)  \bar{\mathcal{O}}_B(\mathbf{0},0)\rangle \nonumber\\
    &\rightarrow& A e^{-m_B \tau} + B e^{-m_{B}^\prime \tau},
\eeqs
where $M_{B}^\prime$ is the baryon mass of the first excited state with the same quantum numbers as the ground state.  In principle, one could remove this excited state by going to very long Euclidean time.    In practice, the long Euclidean time limit is marred by the exponential degradation of baryon signal to lattice noise (known as the Signal-to-Noise problem \cite{Endres:2011jm}).  As a result, only a small region of the correlator as a function of $\tau$ can be used to extract the desired signal.  However, this region can be greatly improved if one were to use a method to ``subtract off'' the first excited state's effects.

To remove the excited state effects, we calculate two sets of correlation functions for each observable.  Each measurement uses Gaussian smearing (shell) for the source and each sink has either no smearing (point) or Gaussian smearing (shell).  The  long Euclidean time behavior  for the ground and first excited states is given by
\beqs
C_{BB}^{SP}(\tau) &\rightarrow& A^{SP} e^{-m_B \tau} + B^{SP} e^{-M_{B}^\prime \tau} \nonumber\\
C_{BB}^{SS}(\tau) &\rightarrow& A^{SS} e^{-m_B \tau} + B^{SS} e^{-M_{B}^\prime \tau}.
\eeqs
By subtracting these correlators with appropriate coefficients, one can cancel off the excited state exponential and is left with a systematically improved plateau for the ground state at the cost of larger statistical errors.  
\begin{figure}[!t] 
   \centering
   \includegraphics[width=0.46\textwidth]{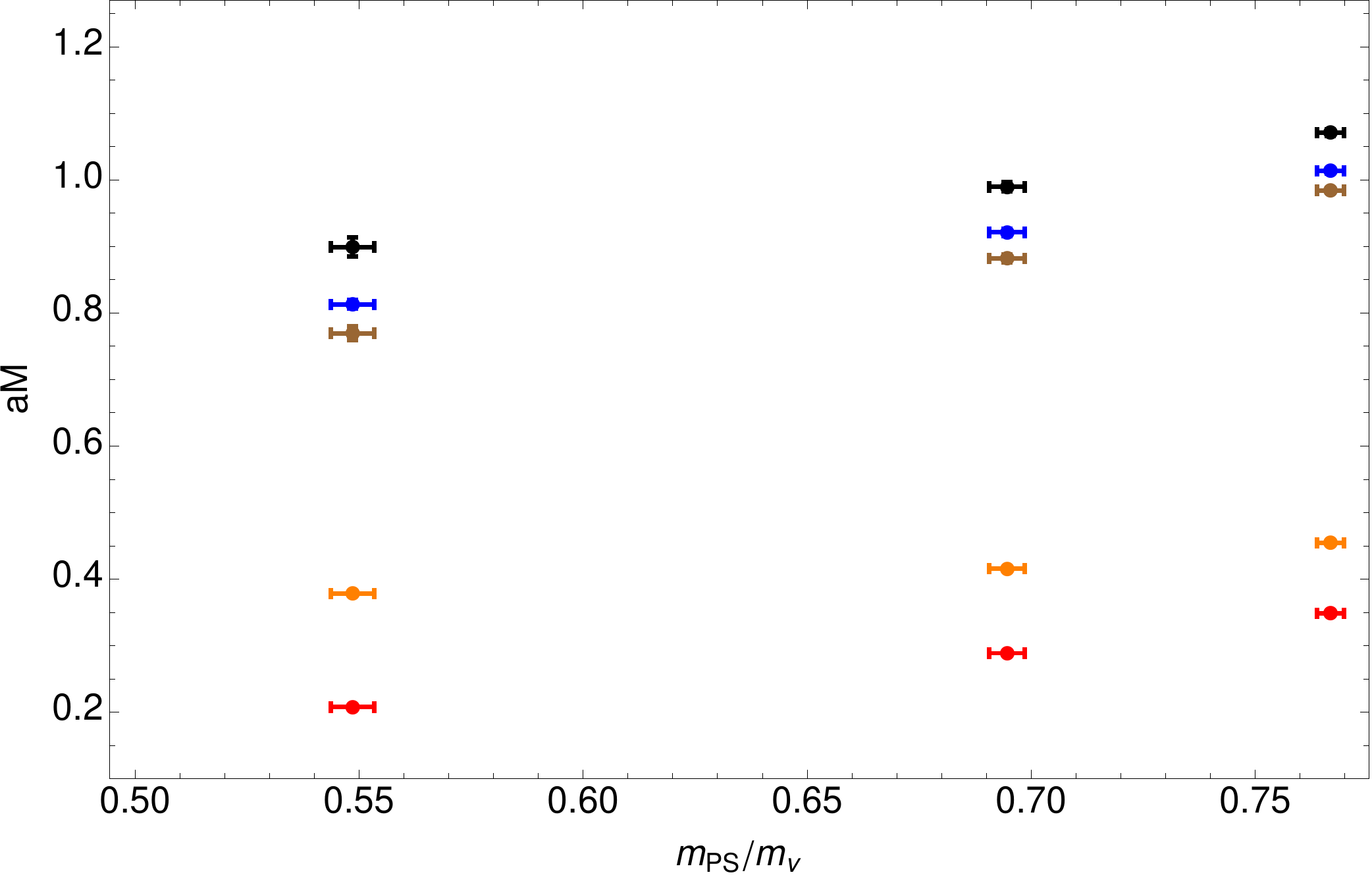}\\
   \includegraphics[width=0.46\textwidth]{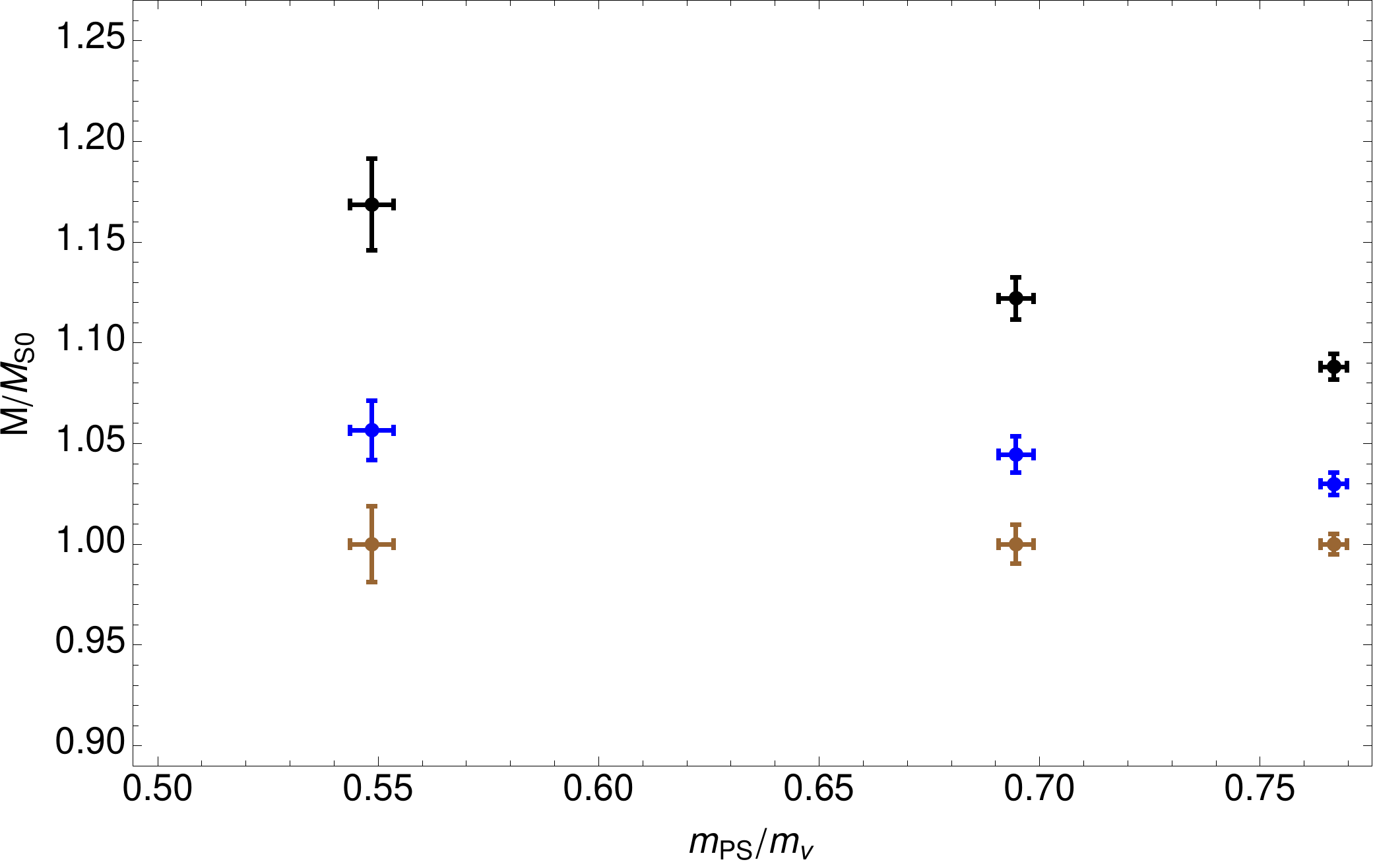}
   \caption{Lattice spectrum results for the coarse lattice spacing ($\beta=11.028$) on $32^3 \times 64$ lattices for three input quark masses.  \textit{(top)} Masses in lattice units of the pseudoscalar meson (red),  vector meson (orange), spin-0 baryon (brown), spin-1 baryon (blue), and spin-2 baryon (black) vs. the meson mass ratio (pseudoscalar over vector).  \textit{(bottom)} Masses in units of the spin-0 baryon mass for the spin-0 baryon mass (brown), spin-1 baryon mass (blue), and spin-2 baryon mass (black) vs.  the meson mass ratio.  Vertical error bars of spin-0 baryon mass represent the error on the scale setting for the dark matter mass.}
\label{fig:beta_11_results}
\end{figure}

One unique feature of baryons with even number of colors is that they are bosons.  As a result, bosonic effective masses, like meson effective masses, are symmetric about the lattice midpoint.  Hence, these correlators can be ``folded'' about the midpoint and averaged, a common procedure for mesons in lattice QCD.   Examples of these effective mass plots are in Fig.~\ref{fig:eff_mass}.

\section{Baryon Spectrum Lattice Results}
\label{sec:Baryon_Spectrum}

The primary lattice results in this work are the 4-color baryon spectrum, the baryon sigma term, and a  determination of volume and lattice spacing effects.  In particular, for each lattice spacing, we determine which lattice volume is required to keep the finite volume systematic smaller than the statistical errors, and then proceed to use the large volume results to quantify a systematic lattice spacing effect using a line of constant physics (LCP) defined by the meson mass ratio, $m_{PS}/m_V$.  Moreover, the slope of the baryon mass as a function of fermion mass is related to the strength of the Higgs coupling. This Higgs exchange prediction along with a direct comparison of 3-color and 4-color baryon spectrum results was discussed in previous sections.

For each ensemble, five spectrum quantities are calculated: pseudoscalar meson mass ($m_{PS}$), vector meson mass ($m_V$), spin-0 baryon mass ($M_{S0}$), spin-1 baryon mass ($M_{S1}$), and spin-2 baryon mass ($M_{S2}$).  The standard expectation for degenerate quark masses with two or more flavors is that the meson hierarchy will be as in QCD, $m_{PS}<m_V$, and the lower spin baryon states will be lighter than the heavier ones, $M_{S0} < M_{S1} < M_{S2}$.  This expectation holds for all the data presented in this work, however, large error bars in some ensembles make this observation less clear.  In this section, we present the $\beta=11.028, 11.5, 12.0$ results for the $32^3 \times 64$ lattice simulations.  Lattice artifacts will be quantified in subsequent sections.

\begin{table}[t]
   \centering
   {\footnotesize
   \begin{tabular}{|c|c|c|c|c|c|} 
   \hline
   $\kappa$ & $am_{PS}$ & $am_V$ & $aM_{S0}$ & $aM_{S1}$ & $aM_{S2}$  \\
   \hline
   0.1515 & 0.256(2)  & 0.328(3) & 0.700(13) & 0.724(8) & 0.754(9)  \\
    \hline
    0.1520 & 0.216(2)  & 0.302(6) & 0.632(11) & 0.663(10) & 0.691(20)  \\
     \hline
      0.1523 & 0.192(2)  & 0.280(5) & 0.590(11) & 0.622(12) & 0.672(12)  \\
     \hline
       0.1524 & 0.182(2)  & 0.283(4) & 0.570(15) & 0.610(10) & 0.679(10)  \\
     \hline
       0.1527 & 0.152(2)  & 0.262(7) & 0.554(12) & 0.589(10) & 0.649(11)  \\
     \hline
   \end{tabular} }
   \caption{Spectrum results for $\beta=11.5$ on $32^3 \times 64$ lattices.}
   \label{tab:b11p5_data}
\end{table}

\begin{table}[t]
   \centering
   {\footnotesize
   \begin{tabular}{|c|c|c|c|c|c|} 
   \hline
   $\kappa$ & $am_{PS}$ & $am_V$ & $aM_{S0}$ & $aM_{S1}$ & $aM_{S2}$  \\
   \hline
   0.1475 & 0.280(1)  & 0.310(3) & 0.660(6) & 0.672(5) & 0.692(6)  \\
    \hline
    0.1480 & 0.247(2)  & 0.288(3) & 0.607(7) & 0.623(7) & 0.648(7)  \\
     \hline
      0.1486 & 0.204(2)  & 0.248(6) & 0.538(7) & 0.543(8) & 0.569(11)  \\
     \hline
       0.1491 & 0.159(4)  & 0.223(5) & 0.481(10) & 0.498(10) & 0.528(11)  \\
     \hline
       0.1495 & 0.114(5)  & 0.195(9) & 0.421(15) & 0.443(12) & 0.495(12)  \\
     \hline
       0.1496 & 0.109(5)  & 0.192(9) & 0.413(18) & 0.434(12) & 0.495(12)  \\
     \hline
   \end{tabular} }
   \caption{Spectrum results for $\beta=12.0$ on $32^3 \times 64$ lattices.}
   \label{tab:b12_data}
\end{table}

\begin{figure}[!t] 
   \centering
   \includegraphics[width=0.46\textwidth]{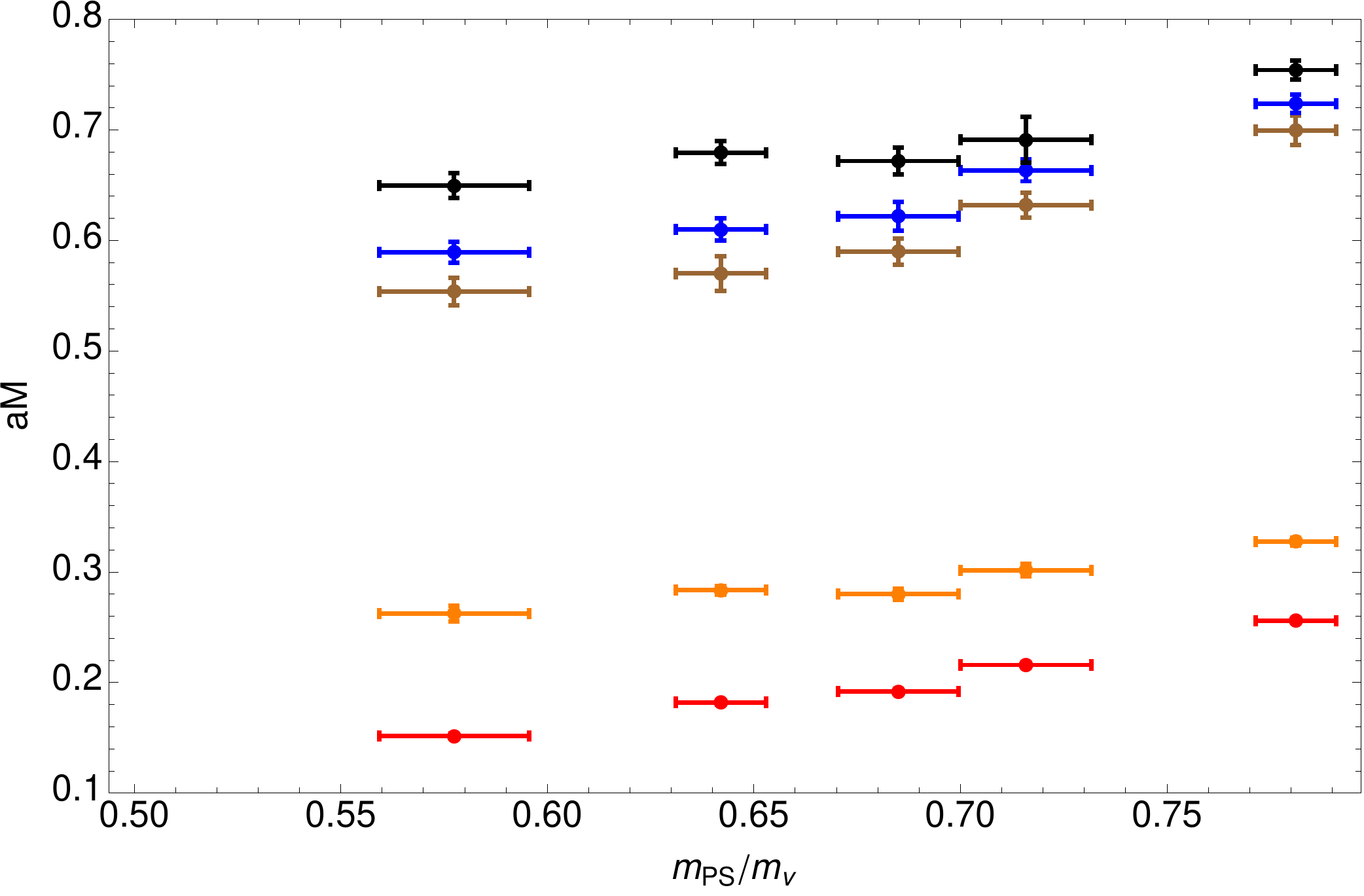}\\
   \includegraphics[width=0.46\textwidth]{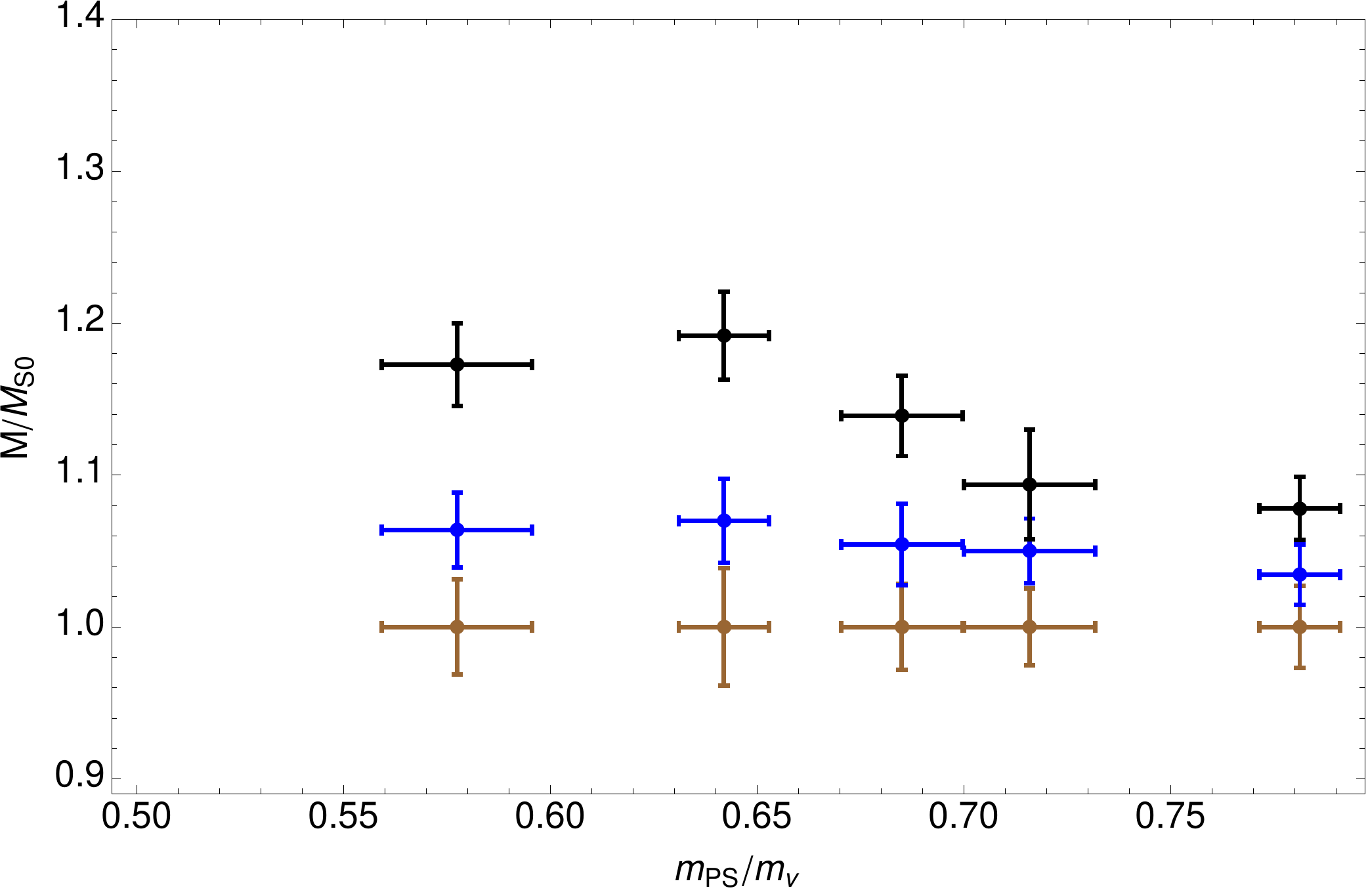}
   \caption{Lattice spectrum results for the intermediate lattice spacing ($\beta=11.5$) on $32^3 \times 64$ lattices for five input quark masses.  \textit{(top)} Masses in lattice units of the pseudoscalar meson (red),  vector meson (orange), spin-0 baryon (brown), spin-1 baryon (blue), and spin-2 baryon (black) vs. the meson mass ratio (pseudoscalar over vector).  \textit{(bottom)} Masses in units of the spin-0 baryon mass for the spin-0 baryon mass (brown), spin-1 baryon mass (blue), and spin-2 baryon mass (black) vs.  the meson mass ratio.  Vertical error bars of spin-0 baryon mass represent the error on the scale setting for the dark matter mass.}
\label{fig:beta_11p5_results}
\end{figure}

The cleanest, most statistically controlled results  are the results on the coarsest lattice spacing, $\beta=11.028$, with the largest $32^3$ physical volume.  The numerical results are presented in Table \ref{tab:b11_data}.  In Fig.~\ref{fig:beta_11_results}, these results are presented in two formats; the spectrum measurements in lattice units vs. $m_{PS}/m_V$ and the baryon mass ratio to the (lightest) spin-0 baryon mass vs.  $m_{PS}/m_V$.  Presenting the results as a function the meson mass ratio gives an optimal sense of the relative magnitude of the fermion mass.  In the heavy quark limit, this ratio approaches 1 and in the chiral limit, this ratio approaches 0 (for reference, this value in QCD is $m_{PS}/m_V\approx 0.18$).   On the second plot in Fig.~\ref{fig:beta_11_results}, the baryon masses are given in units of the $M_{S0}$ mass, which sets the scale of our dark matter mass in exclusion plots, Fig.~\ref{fig:Higgs_Exclusion}.  The ratio $M_{S0}/M_{S0}$ is trivially 1, but the associated errors here correspond to the error on the scale setting.  For these coarse lattice spacing results, the scale setting error is no more than 1.7\%.  It is clear (from this plot in particular) that the relative separation is growing as the pseudoscalar meson mass is decreased.  This is to be expected, as all three baryon states should have equal mass in the heavy fermion mass limit (four times the fermion mass), and are thus expected to separate as fermion mass is decreased.  What is not as predictable \textit{a priori} is the relative separation of the states.  In particular, the spin-1/spin-0 separation is much smaller than the spin-2/spin-1 separation (i.e. the spin-2 state separation grows faster with decreasing quark mass).  These relative separations follow from the large $N_c$ rotor spectrum for baryons as was previously discussed in Sec.~\ref{sec:3_4_comp}.  While volume effects on these lattices are under control, finite lattice spacing effects will need to be quantified.

\begin{figure}[!t] 
   \centering
   \includegraphics[width=0.46\textwidth]{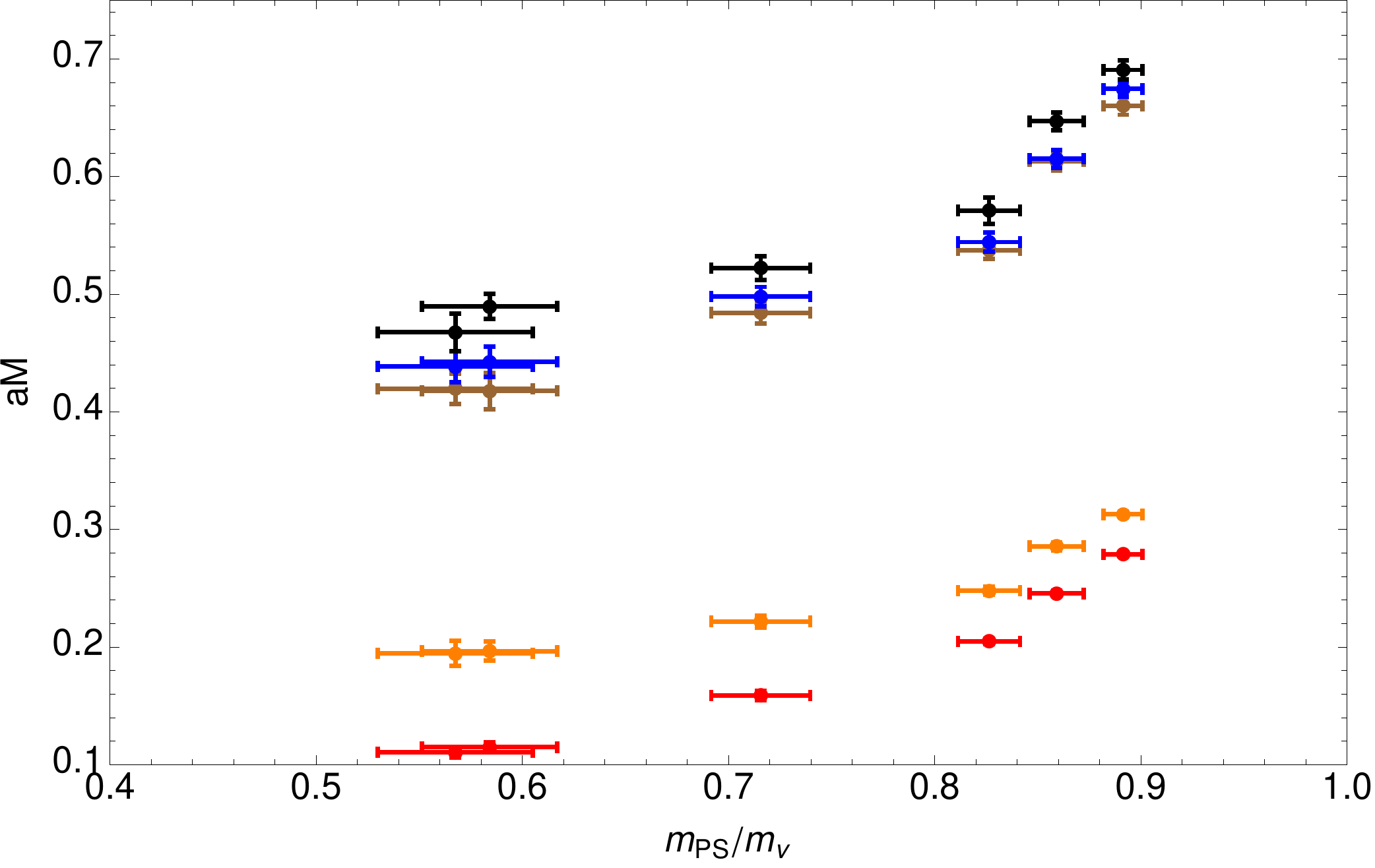}\\
   \includegraphics[width=0.46\textwidth]{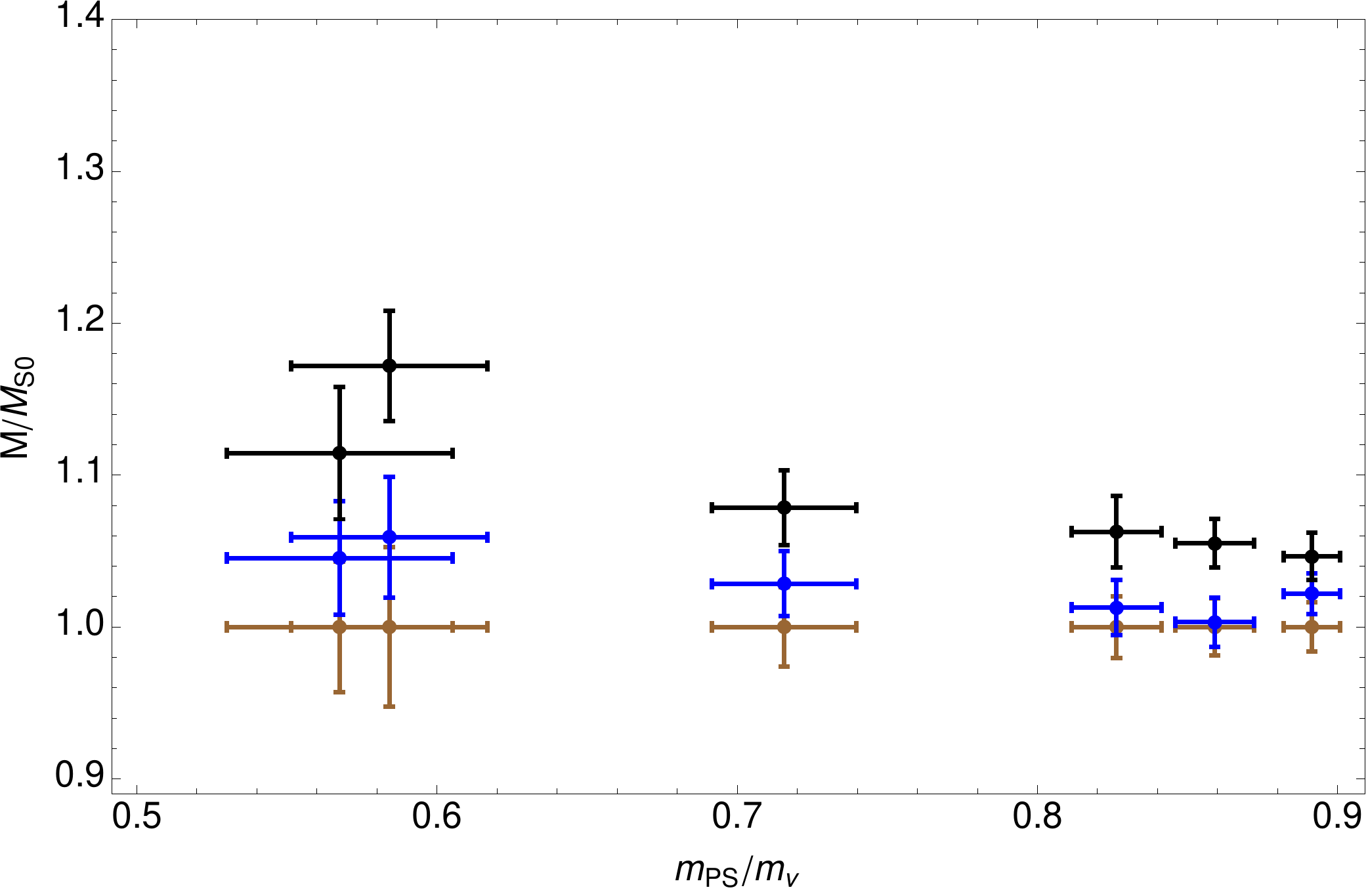}
   \caption{Lattice spectrum results for the fine lattice spacing ($\beta=12.0$) on $32^3 \times 64$ lattices for six input quark masses.  \textit{(top)} Masses in lattice units of the pseudoscalar meson (red),  vector meson (orange), spin-0 baryon (brown), spin-1 baryon (blue), and spin-2 baryon (black) vs. the meson mass ratio (pseudoscalar over vector).  \textit{(bottom)} Masses in units of the spin-0 baryon mass for the spin-0 baryon mass (brown), spin-1 baryon mass (blue), and spin-2 baryon mass (black) vs.  the meson mass ratio.  Vertical error bars of spin-0 baryon mass represent the error on the scale setting for the dark matter mass.}
\label{fig:beta_12_results}
\end{figure}

The results for the intermediate lattice spacing ($\beta=11.5$) are presented in Table~\ref{tab:b11p5_data} and shown in Fig.~\ref{fig:beta_11p5_results}. This lattice spacing (and corresponding volume) is roughly 2/3 the size of the coarse lattice spacing.  As a result, for these $32^3$ lattices, larger volume effects are expected.  Still, the baryon mass ratios to the spin-0 masses are consistent for a given value of $m_{PS}/m_V$.  The size of the volume effects (which will be discussed more in a later section) are within the current statistical errors, but could prove to be a several percent effect with more statistics.

For the finest lattice spacing, the numerical masses in lattice units are presented in Table~\ref{tab:b12_data} and the corresponding plots are in Fig.~\ref{fig:beta_12_results}.  More fermion masses have been explored here with comparable measurements, but due to the smaller physical volume (by roughly a factor of $2^4$) as compared to the $\beta=11.028$, the resulting errors are larger.  For that reason, our results are not as conclusive on these lattices.  Nevertheless, the usual trends of the state separation are still observed and the spin-1 state stays close to the spin-0 state (even more than the $\beta=11.028$ results).  However, as will be discussed, the volume effects are expected to be non-trivial for these measurements.

\section{Calculation of baryon mass derivative}
\label{sec:Baryon_Deriv}

\begin{figure}[!t] 
   \centering
   \includegraphics[width=0.46\textwidth]{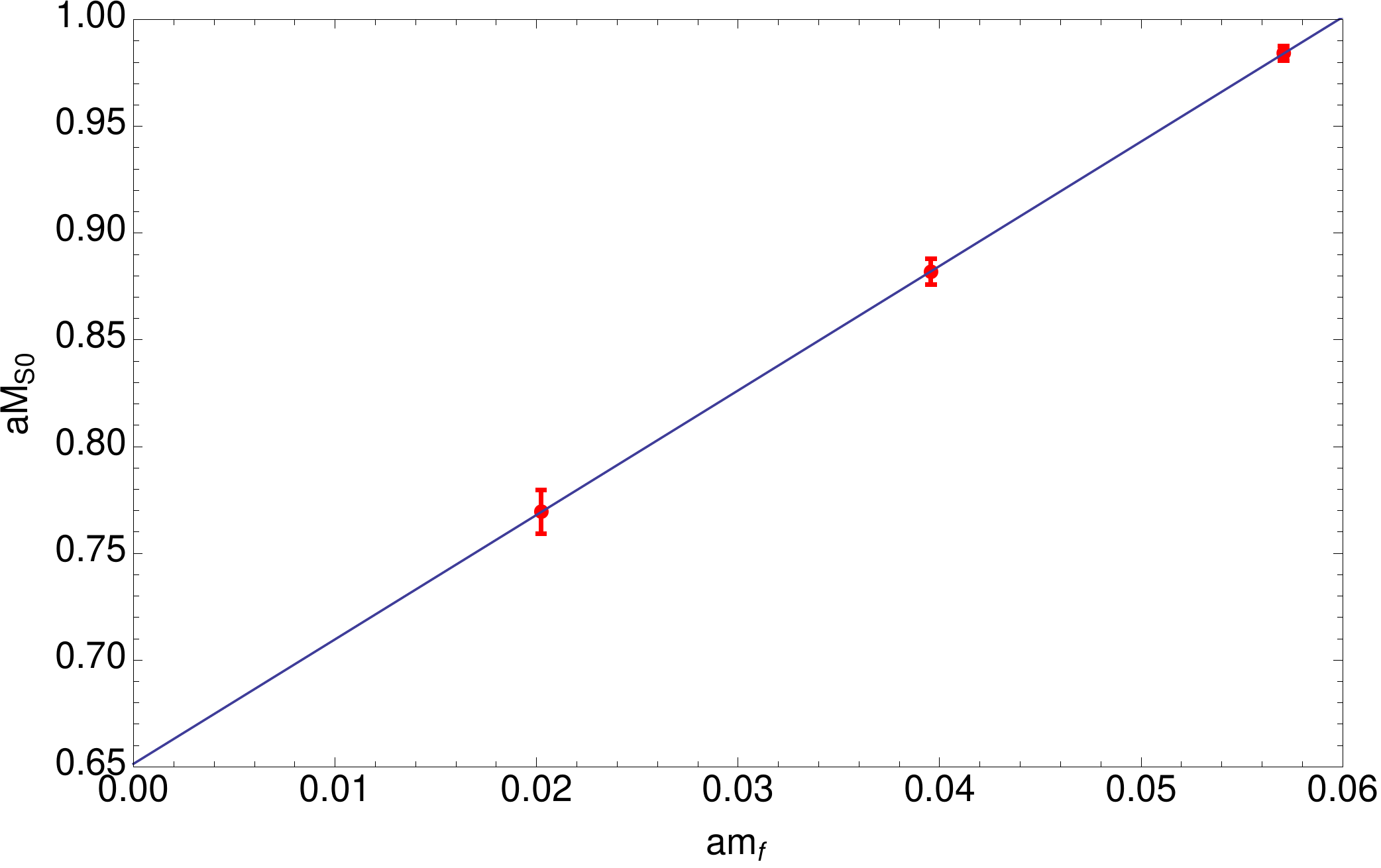}\\ 
   \includegraphics[width=0.46\textwidth]{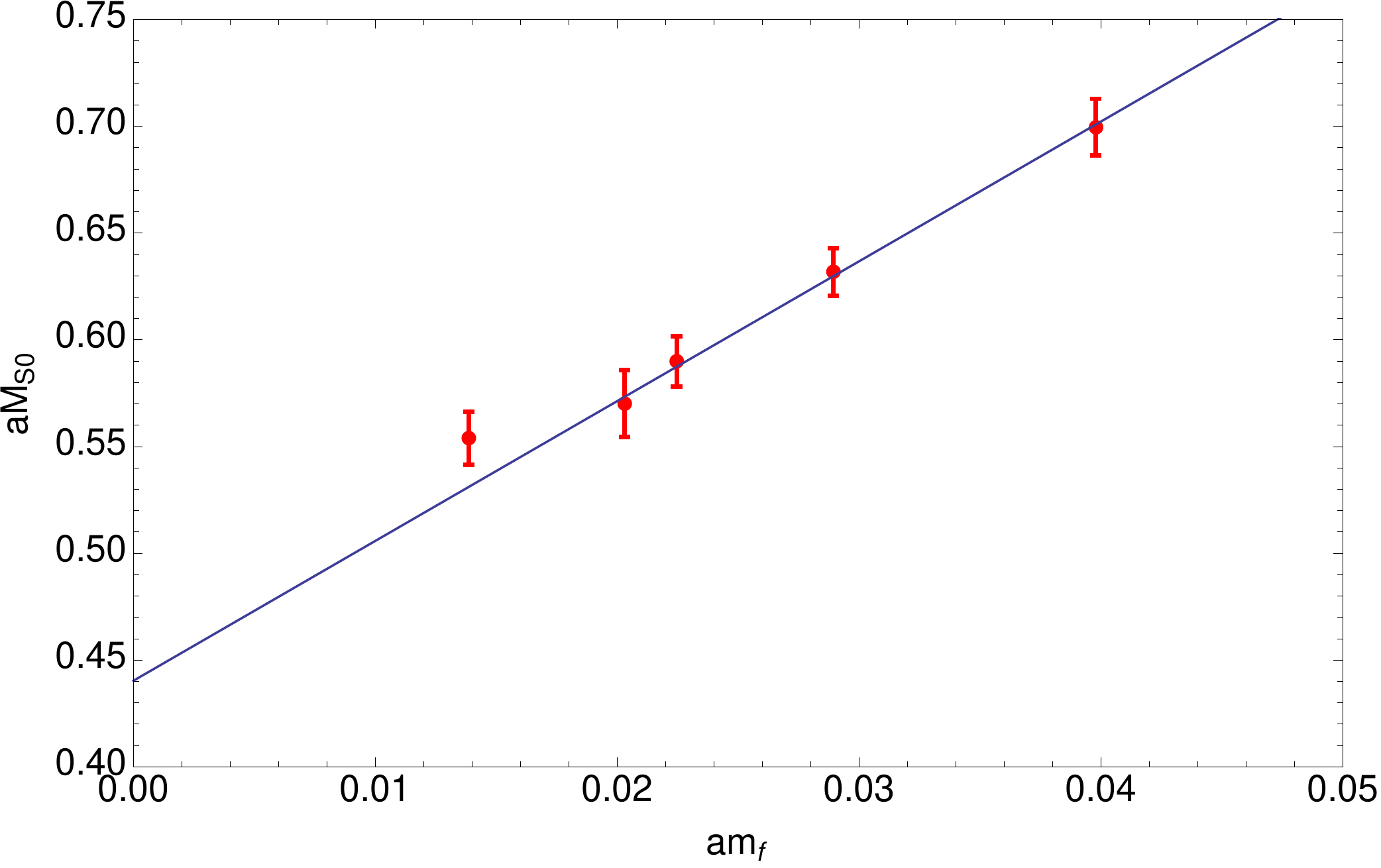}\\
   \includegraphics[width=0.46\textwidth]{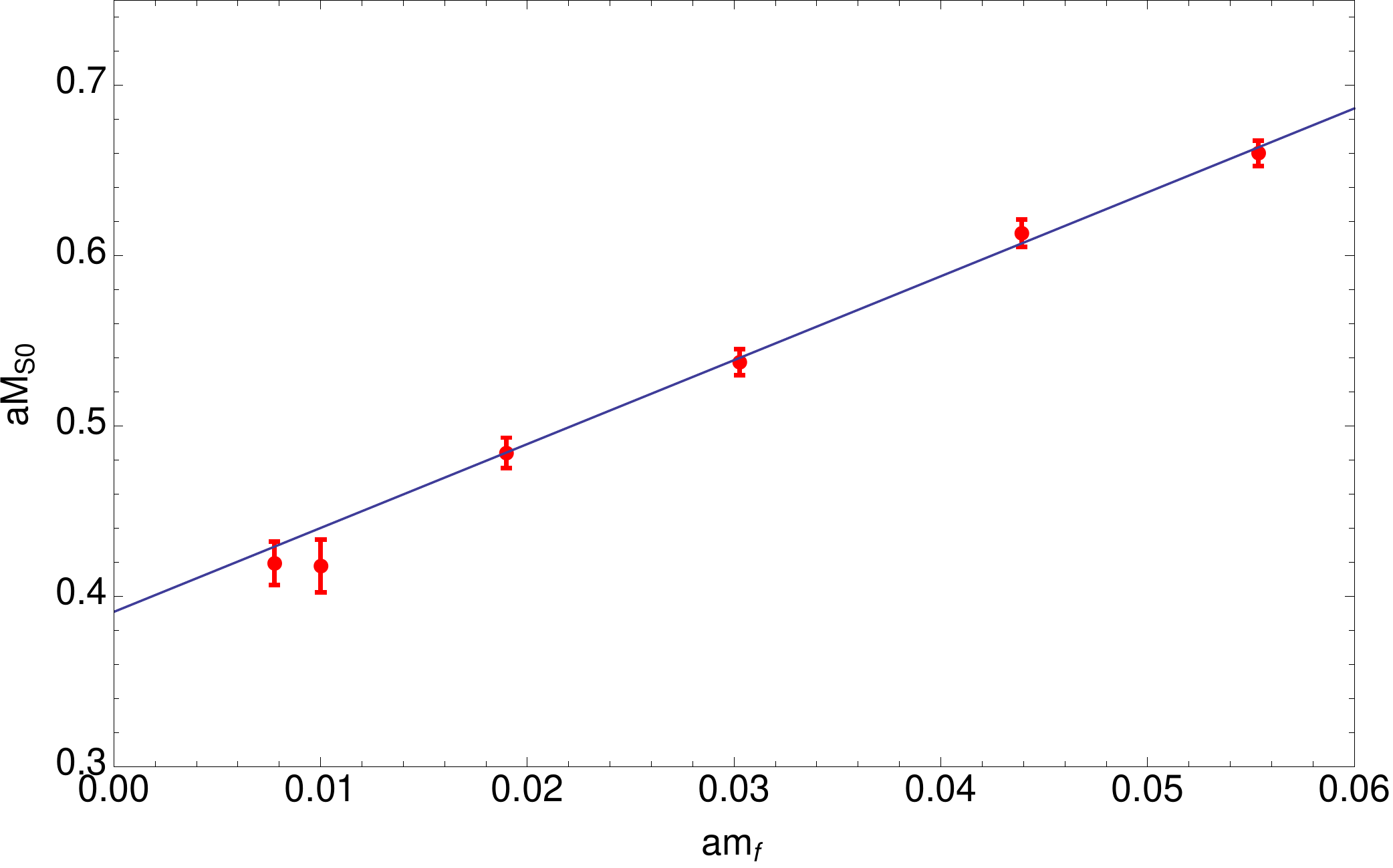}
   \caption{Calculation of $\partial m_B/\partial m_f$ from baryon spectrum.  Plots of $am_{S0}$ vs. $am_f$ are displayed for the coarsest lattice spacing (top), intermediate lattice spacing (middle), and finest lattice spacing (bottom).  As seen in lattice QCD calculations, the fermion mass dependence of the baryon mass is primarily linear.}
\label{fig:Data_Fits}
\end{figure}

\begin{table}[t]
   \centering
   {\footnotesize
   \begin{tabular}{|c|c|c|} 
   \hline
   $\kappa$ & $\frac{m_{PS}}{m_V}$ & $\frac{m_f}{m_B}\frac{\partial m_B}{\partial m_f}$ \\
   \hline
   0.1515 & 0.781(10)  & 0.372(52)   \\
    \hline
    0.1520 & 0.716(16)  & 0.300(42)   \\
     \hline
      0.1523 & 0.685(15)  & 0.249(35)   \\
     \hline
      0.1524 & 0.641(11)  & 0.244(33)   \\
     \hline
      0.1527 & 0.577(18)  & 0.164(23)   \\
     \hline
   \end{tabular} }
   \caption{Normalized sigma parameter results for $\beta=11.5$ on $32^3 \times 64$ lattices.}
   \label{tab:b11p5_data_sigma}
\end{table}

\begin{table}[t]
   \centering
   {\footnotesize
   \begin{tabular}{|c|c|c|} 
   \hline
   $\kappa$ & $\frac{m_{PS}}{m_V}$ & $\frac{m_f}{m_B}\frac{\partial m_B}{\partial m_f}$ \\
   \hline
   0.1475 & 0.891(9)  & 0.413(25)   \\
    \hline
    0.1480 & 0.859(13)  & 0.353(22)   \\
     \hline
      0.1486 & 0.826(15)  & 0.277(17)   \\
     \hline
      0.1491 & 0.716(24)  & 0.193(12)   \\
     \hline
      0.1495 & 0.584(33)  & 0.118(8)   \\
     \hline
       0.1496 & 0.568(38)  & 0.091(6)   \\
     \hline
   \end{tabular} }
   \caption{Normalized sigma parameter results for $\beta=12.0$ on $32^3 \times 64$ lattices.}
   \label{tab:b12_data_sigma}
\end{table}

From the baryon mass spectrum as function of the fermion mass, the baryon mass derivative needed for the sigma term can be extracted.  The visual depictions of these linear fits on the $32^3$ data are shown in Fig~\ref{fig:Data_Fits} and the results are shown in Table~\ref{tab:b11_data_sigma}, Table~\ref{tab:b11p5_data_sigma}, and Table~\ref{tab:b12_data_sigma}.  Clearly, the more mass ensembles one has for a given lattice spacing will allow a more complete extraction of the derivative.  However, due to the linear nature of the data, the derivative can be estimated as linear.  For each beta value, the derivative is given by
\beqs
\frac{\partial m_B}{\partial m_f}&=&5.83(30) \quad \text{For}\ \beta=11.026 \nonumber\\
\frac{\partial m_B}{\partial m_f}&=& 6.55(90) \quad \text{For}\ \beta=11.5\nonumber\\
\frac{\partial m_B}{\partial m_f}&=& 4.92(30)\quad \text{For}\ \beta=12.0.
\eeqs
It is worth mentioning that at this stage, there is an overall normalization of $m_f$ that is left undetermined.  However, ultimately we are going to multiply this derivative by $m_f/m_B$, canceling this normalization.  One curiosity is that the $\beta=12.0$ result is below those of the coarser lattice spacings.  We will argue in subsequent sections that the $\beta=12.0$ results are significantly more sensitive to lattice artifacts (in particular, volume effects) than the other two lattice spacings.  

Comparisons between the coarse and intermediate lattice spacing can be made for $m_{PS}/m_V \approx 0.69$ and $m_{PS}/m_V \approx 0.77$ from Table~\ref{tab:b11_data_sigma} and Table~\ref{tab:b11p5_data_sigma}.  As expected, the results are constant within errors.  This helps strengthen the conclusion that lattice artifact systematics for these masses for these lattice spacings on the $32^3 \times 64$ lattices are smaller than the statistical errors.

\section{Estimation of lattice artifacts}
\label{sec:Lattice_Artifacts}

As in any calculation in lattice field theory, there are several sets of unphysical lattice artifacts that need to be quantified.  Since chiral extrapolations to low masses are not strictly necessary for the applications to composite dark matter theory, the two primary unphysical contributions are the discretization effects in terms of our lattice spacing, $a$, and finite volume ``wrap-around'' effects, where the lattice extent is given by number of sites times the lattice spacing.  One systematic error that will remain uncontrolled in this work is the use of quenched lattices, which corresponds to unphysically dropping dynamical sea fermion loops.   This approximation works better as one goes to larger fermion masses and larger number of colors, which is the regime we are currently in.  For the QCD calculation of the light quark sigma term, the quenched results are entirely consistent with state-of-the-art dynamical simulations (see Fig.~3 in Ref.~\cite{Young:2013nn}).  The statistical errors are less than 10\% (the largest possible systematic error), and, again, we would expect our systematic errors to be smaller than this due to a larger number of colors and heavier fermions.  With that being said, we hope to produce several unquenched SU(4) ensembles in the future, to more directly quantify this effect.

\subsection{\textit{Volume systematic}}

One approach often employed when only two volumes are explored is to assign a systematic error between them.  When one has three or more large volumes, it is useful to define ``nearly infinite volume'' points where the finite volume systematic is below that of the statistical error.  Defining such volumes is also advantageous for doing comparisons of lattice spacing systematics, so as to appropriately decouple these two lattice artifacts.

\begin{figure}[!t] 
   \centering
   \includegraphics[width=0.46\textwidth]{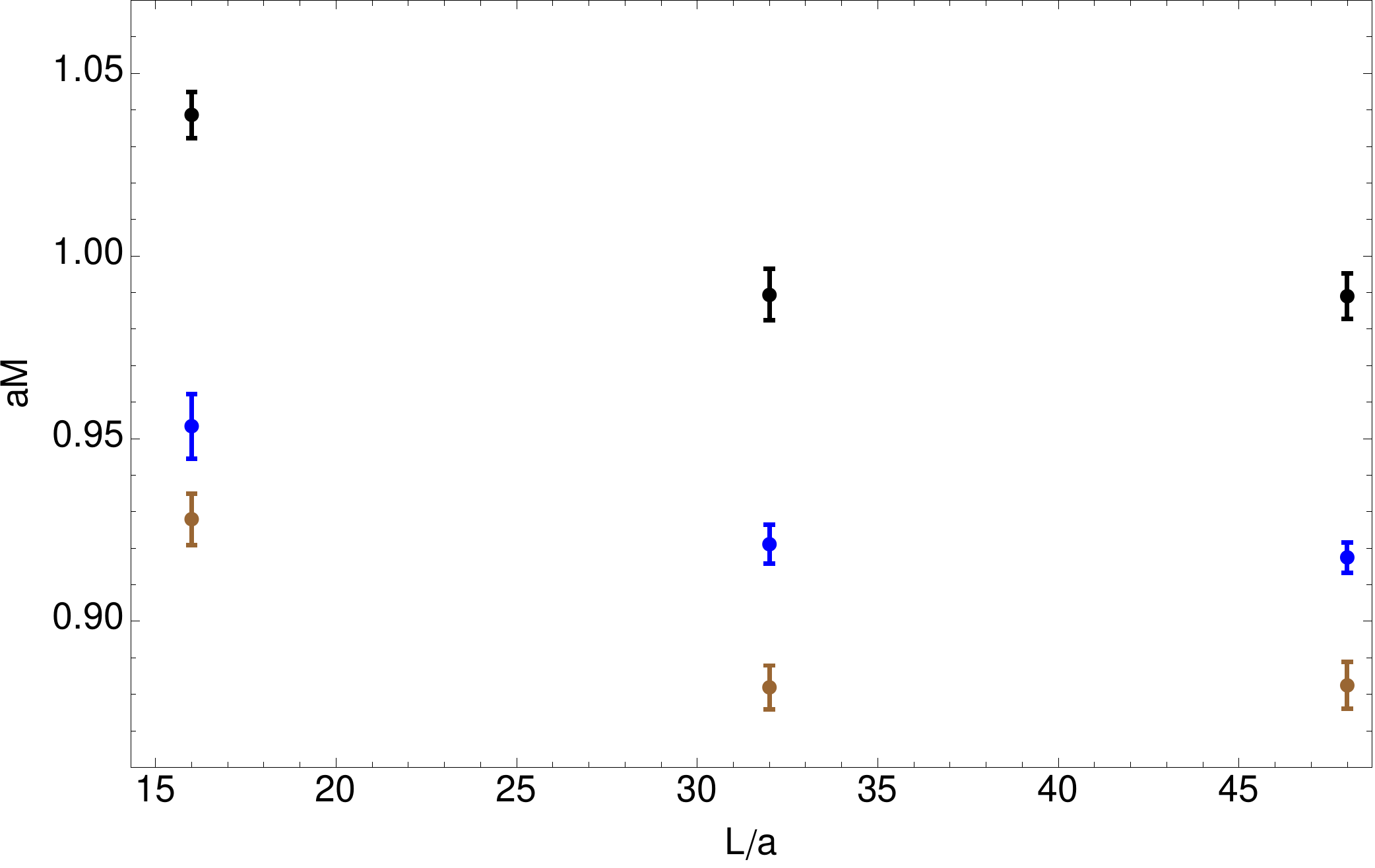}
   \caption{Volume scaling of the spin-0 (brown), spin-1 (blue), and spin-2 (black) baryon masses in lattice units for the coarse lattice spacing ($\beta=11.028$) and middle quark mass ($m_{PS}/m_V \sim 0.7$) for lattice sizes of $16^3 \times 32$, $32^3 \times 64$, and $48^3 \times 96$.  Volume effects between $32^3$ and $48^3$ lattices are  smaller than the statistical error.}
\label{fig:beta_11_volume}
\end{figure}

\begin{figure}[!t] 
   \centering
   \includegraphics[width=0.46\textwidth]{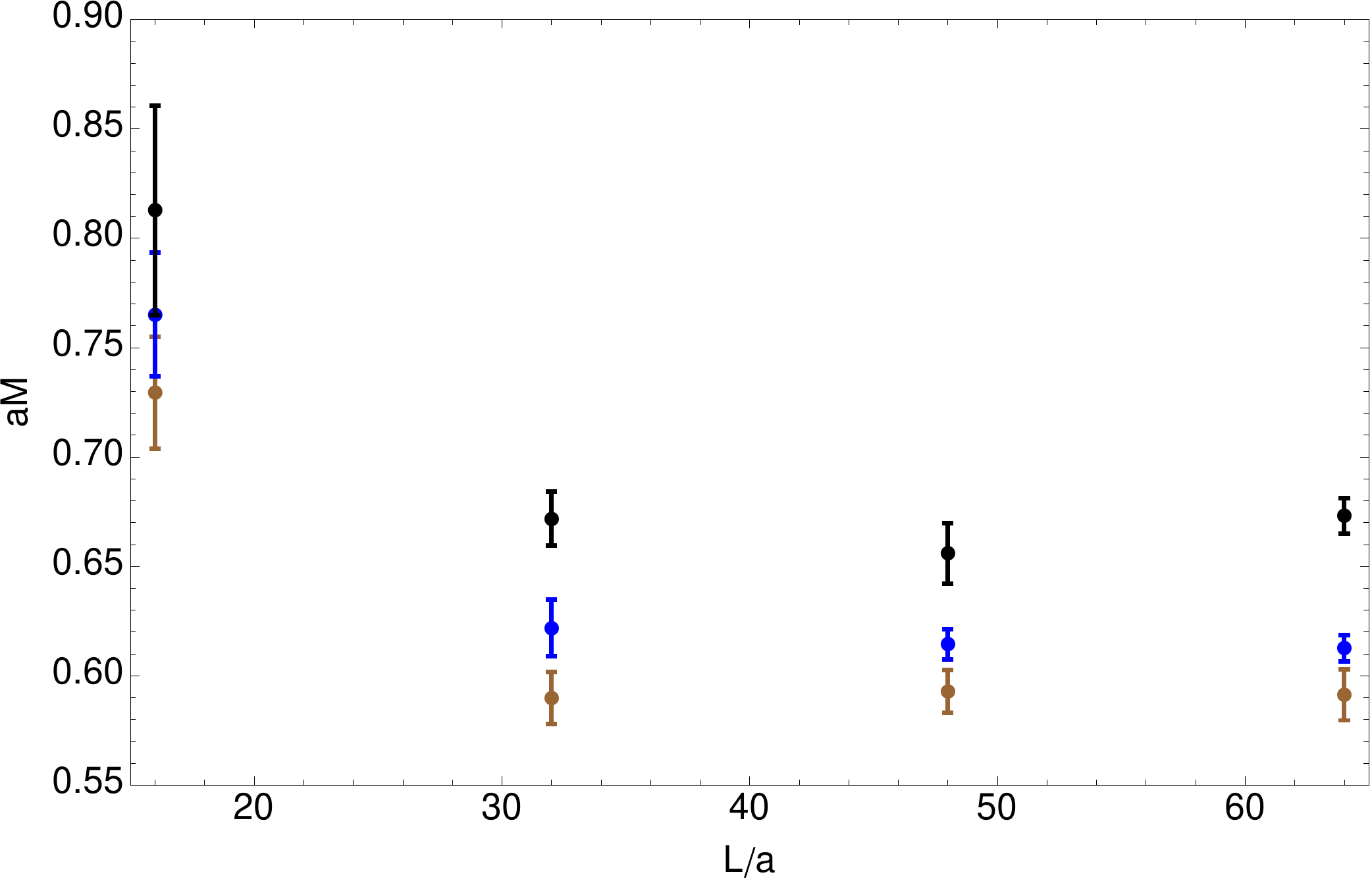}\\
   \includegraphics[width=0.46\textwidth]{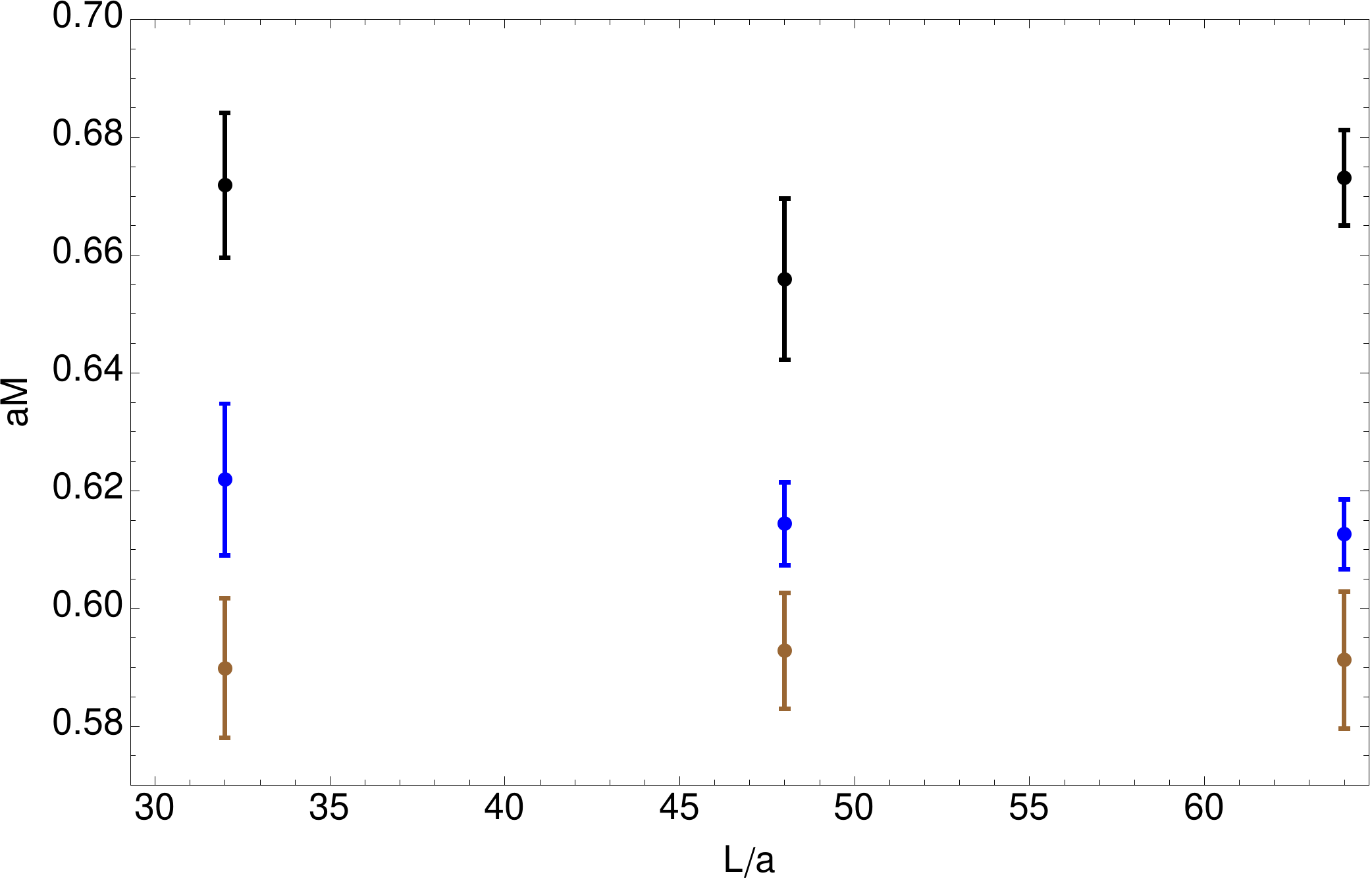}
   \caption{Volume scaling of the spin-0 (brown), spin-1 (blue), and spin-2 (black) baryon masses in lattice units for the intermediate lattice spacing ($\beta=11.5$) and middle quark mass ($m_{PS}/m_V \sim 0.7$) for lattice sizes of $16^3 \times 32$, $32^3 \times 64$, $48^3 \times 96$, and $64^3 \times 128$ (bottom figure zoomed in on the latter three).  Volume effects between $32^3$ and $48^3$ lattices are smaller than the statistical error.}
\label{fig:beta_11p5_volume}
\end{figure}

\begin{figure}[!t] 
   \centering
   \includegraphics[width=0.46\textwidth]{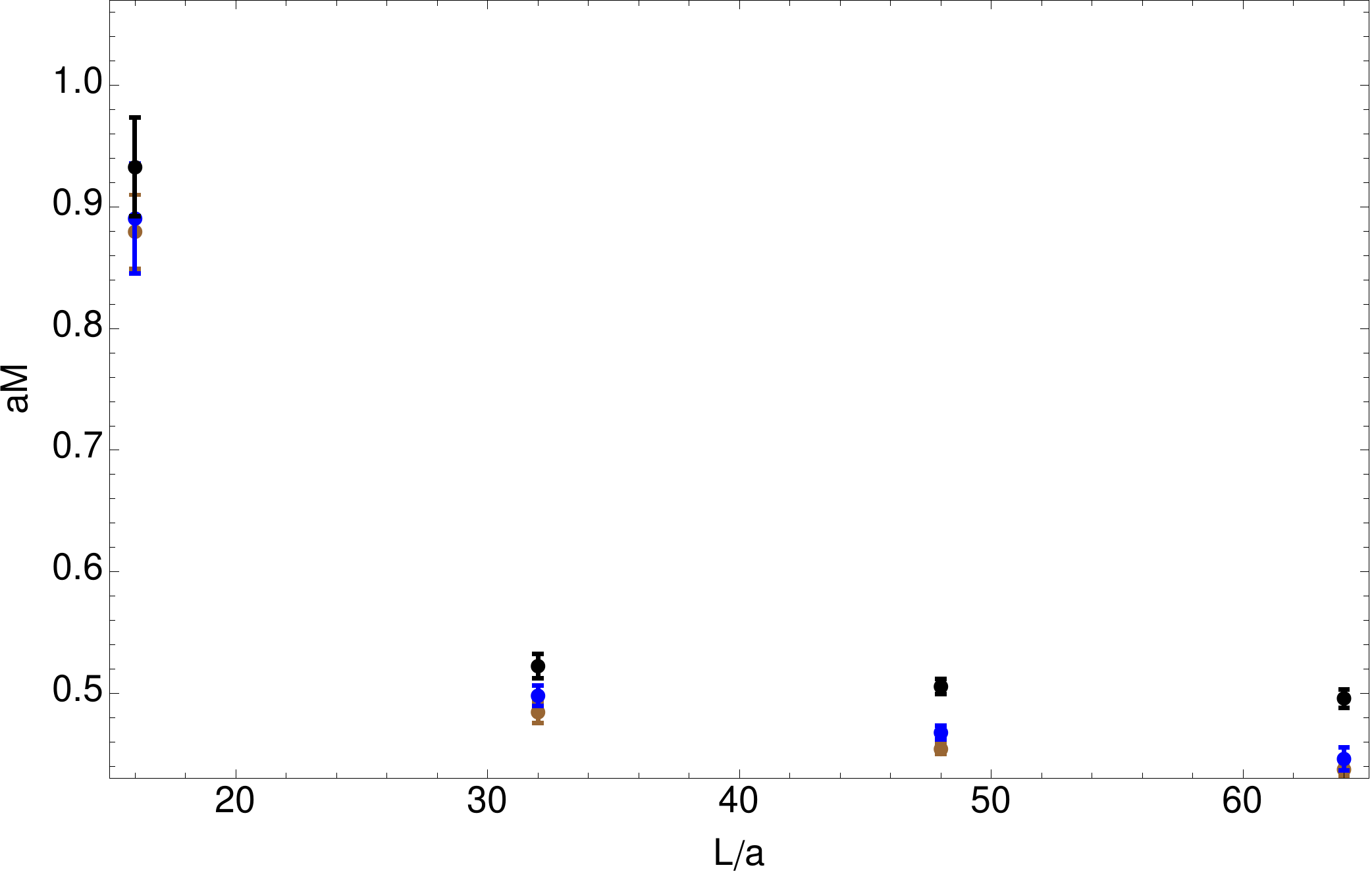}\\
   \includegraphics[width=0.46\textwidth]{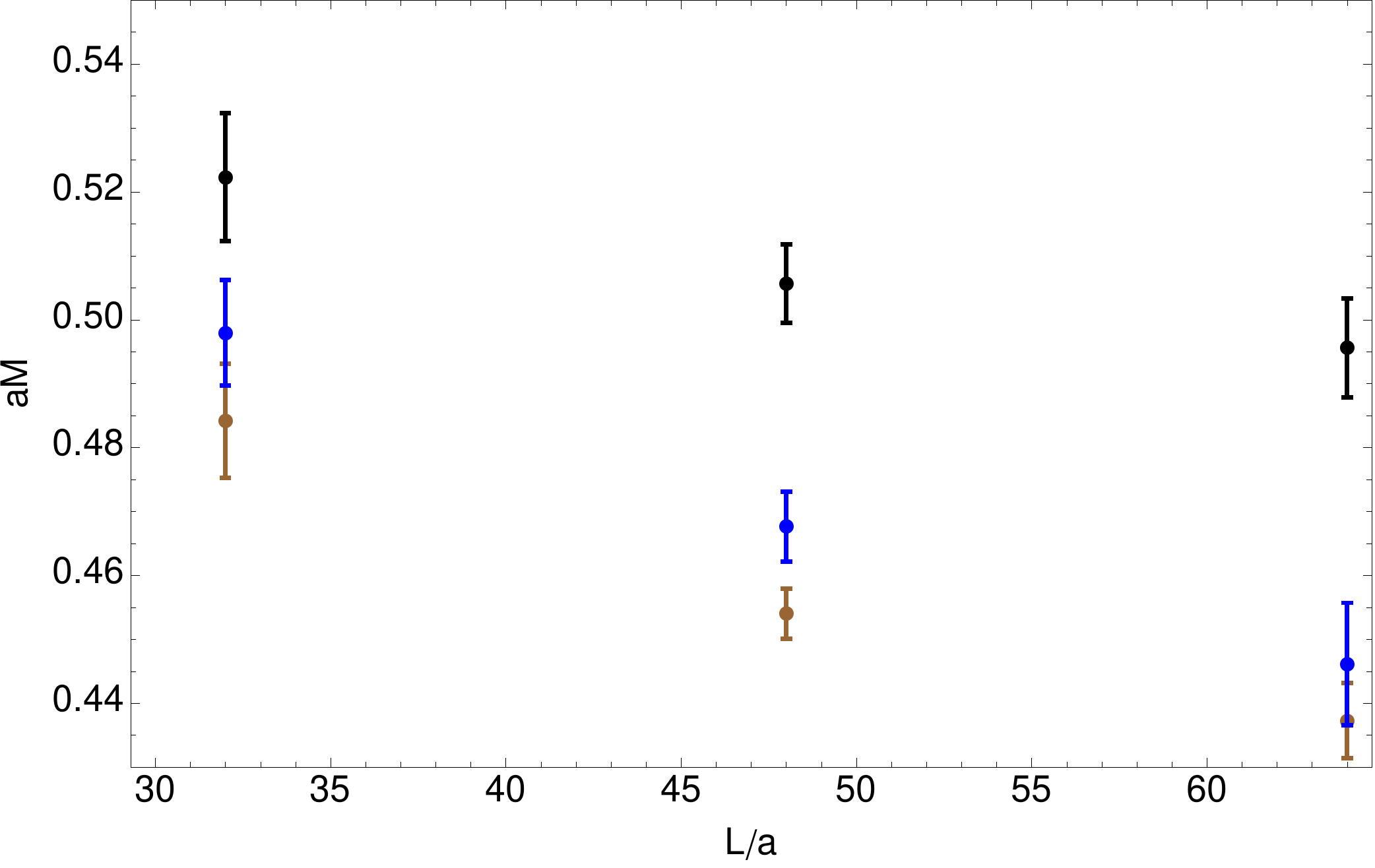}
   \caption{Volume scaling of the spin-0 (brown), spin-1 (blue), and spin-2 (black) baryon masses in lattice units for the fine lattice spacing ($\beta=12.0$) and middle quark mass ($m_{PS}/m_V \sim 0.7$) for lattice sizes of $16^3 \times 32$, $32^3 \times 64$, $48^3 \times 96$, and $64^3 \times 128$ (bottom figure zoomed in on the latter three).  Volume effects between $32^3$ and $48^3$ lattices are  roughly $7\%$ for the spin-0 baryon mass and larger than 3\% from $48^3$ to $64^3$.}
\label{fig:beta_12_volume}
\end{figure}

The volume effects are expected to be best behaved for the coarsest lattice spacing on a given number of lattice sites.   In Fig.~\ref{fig:beta_11_volume}, the middle fermion mass ($\kappa=0.15625$) is explored for three volumes corresponding to $L/a=16, 32, 48$.   Comparing $L/a=16$ to $L/a=32$, finite volume effects cause the baryon masses to be roughly 7\% larger for the smaller volume as compared to the larger volume, well larger than the statistical error bars.   Comparing $L/a=32$ to $L/a=48$, the volume are within the statistical error.  This implies that the $L/a=32$ results are suitably at the ``nearly infinite volume'' point.

The intermediate ($\beta=11.5$) lattice spacing at its middle fermion mass ($\kappa=0.1523$) is presented in Fig.~\ref{fig:beta_11p5_volume} for lattices corresponding to $L/a=16, 32, 48,64$.  Since the total spatial extent is roughly two-thirds of the corresponding spatial extent at coarse lattice spacing, the volume effects are expected to be more significant here.  This is clearly the case for $L/a=16$, where the extracted baryon masses are significantly heavier than their larger volume counterparts.  For the statistics considered here, the $L/a=32$ results are constant within uncertainties with the $L/a=48$ and $L/a=64$ results.  A slight systematic drop could arise between $L/a=32$ and $L/a=48$  data sets with a moderate improvement in statistics.  However, such effects cannot be inferred with the current data set and $L/a=32$ should be a sufficiently large enough volume at this level of statistics.

For the finest lattice spacing ($\beta=12.0$), the baryon spectrum is shown for $\kappa=0.1491$ in Fig.~\ref{fig:beta_12_volume} for four volumes, $L/a=16, 32, 48, 64$.  Comparing $L/a=16$ to $L/a=32$, there are clearly enormous volume effects on the order of 100\%.  For this reason, the $L/a=16$ data at this lattice spacing is essentially unusable.  The more informative comparison is between $L/a=32$ to $L/a=48$, where the volume effects are much more manageable, but still on the order of 7\% and larger than the statistical uncertainty.  For this reason, $L/a=32$  cannot be considered ``nearly infinite volume'' and $L/a=48$ or larger is required.  To tell if $L/a=48$  is sufficiently close to infinite volume, a larger $L/a=64$ volume is required. While the volume effects between $L/a=48$ and $L/a=64$ are smaller, there is still a clear systematic decrease due to finite volume of roughly 4\%.  In other words, through $L/a=64$, all quantifiable volume effects are non-negligible.  

\begin{figure}[!t] 
   \centering
   \includegraphics[width=0.46\textwidth]{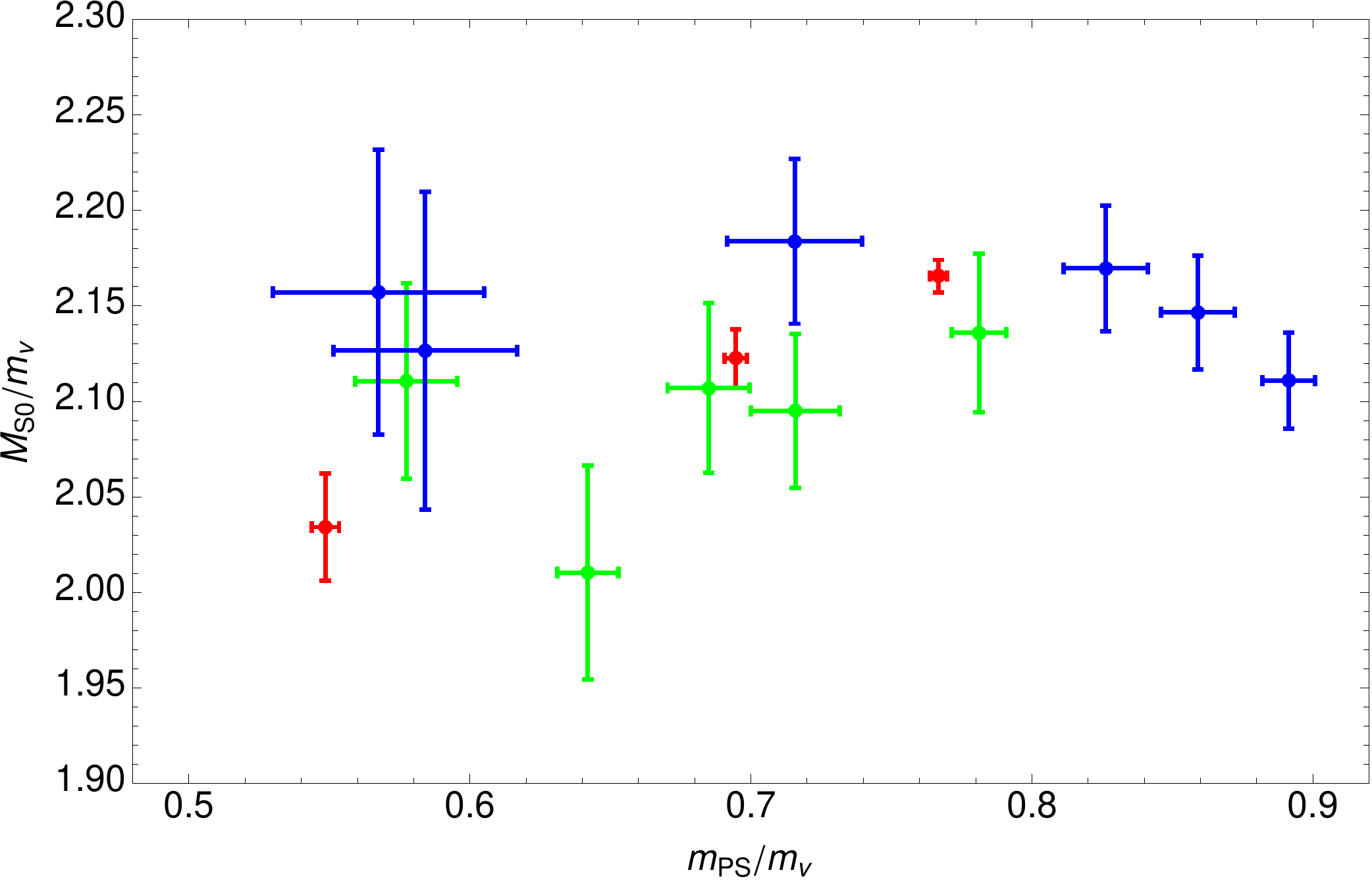}
   \caption{Edinburgh-style plot for coarse ($\beta=11.028$) (red), intermediate ($\beta=11.5$) (green) and fine ($\beta=12.0$) (blue) lattice spacing for the $L/a=32$ lattices.  Results in the large volume limit should decrease in $M_{S0}/m_V$ as $m_{PS}/m_V$ decreases.}
\label{fig:Edin}
\end{figure}

Also, it is useful to examine the data on an Edinburgh-style plot in Fig.~\ref{fig:Edin}, where quantities of different lattice spacings can be compared directly.  This plot displays the mass ratio $M_{S0}/m_V$ vs. $m_{PS}/m_V$ for the coarse ($\beta=11.028$), intermediate ($\beta=11.5$) and fine ($\beta=12.0$) lattice spacings for $32^3 \times 64$ lattices.  In the absence of lattice volume effects, one would expect these ratios to decrease as the fermion mass decreases.  This behavior is clearly visible in the coarse lattice results and the heavier four points on the intermediate lattice spacing.  However, for the fine lattice spacing, the ratio $M_{S0}/m_V$ is roughly independent of fermion mass.  This is often an indication that volume effects are significant.  This figure, once again, supports the hypothesis that $32^3$ lattices are large enough volumes for the coarse  and intermediate lattice spacing, but not large enough for the fine lattice spacing.

\subsection{\textit{Lattice spacing systematic}}
Before discussing lattice spacing effects, one must first determine some physics that remains constant between two different lattice spacings, often referred to as a line of constant physics (LCP) with minimal volume effects, and then proceed to compare other quantities directly.   The quantity that we choose as our LCP is the meson mass ratio $m_{PS}/m_V$.  In most dark matter models of interest, the vector and the pseudoscalar mesons are not of direct interest, making them the ideal physical quantity to match in order to see the lattice spacing effects in the more interesting baryon sector.
\begin{figure}[!t] 
   \centering
   \includegraphics[width=0.46\textwidth]{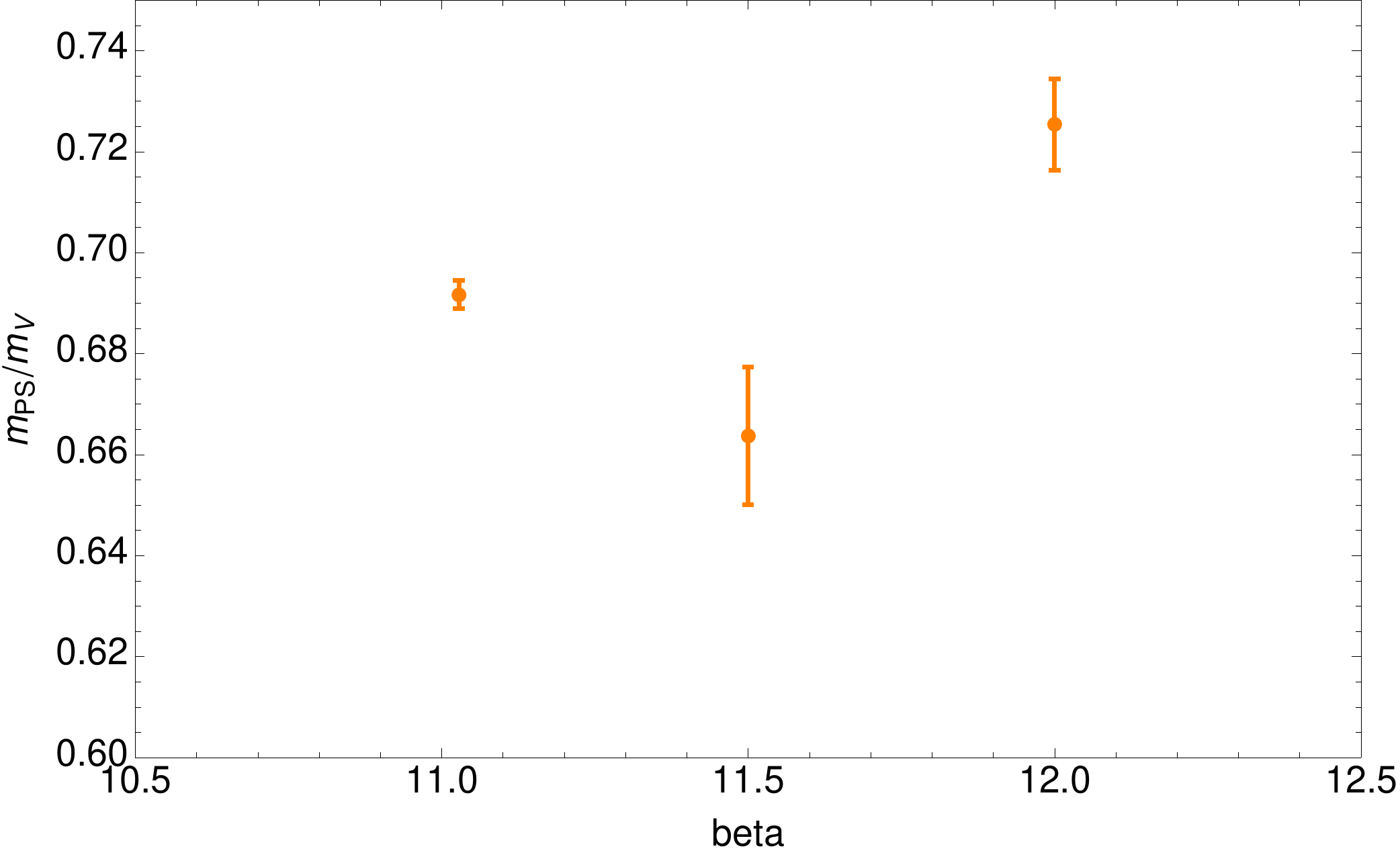}
   \caption{Pseudoscalar to vector meson mass ratio vs. $\beta$ for $L/a = 48$.  For these the three chosen kappa values, the ratio agrees within 3\%.  Assuming infinite volume, these points can be used as line-of-constant physics up to 5\% systematic error. }
\label{fig:LCP}
\end{figure}
In Fig.~\ref{fig:LCP}, the meson ratio for three fermion masses is checked to see how good of a candidate they are for being an LCP.  The error in this quantity sets the lower limit as to what can be quoted as a lattice spacing systematic.  In this figure, the three quantities compared on the $48^3$ volumes, show roughly a 5\% difference.   Thus, any statement on lattice spacing effects cannot reliably be determined below the 5\% level.  Since it is an inherently computationally expensive procedure to vary the valance fermion masses in lattice calculations, the $5\%$ level is the best that could be achieved with our current data sets.

With an LCP determined within 5\%, it is useful to roughly understand the relative size of the lattice spacings.  
\begin{figure}[!t] 
   \centering
   \includegraphics[width=0.46\textwidth]{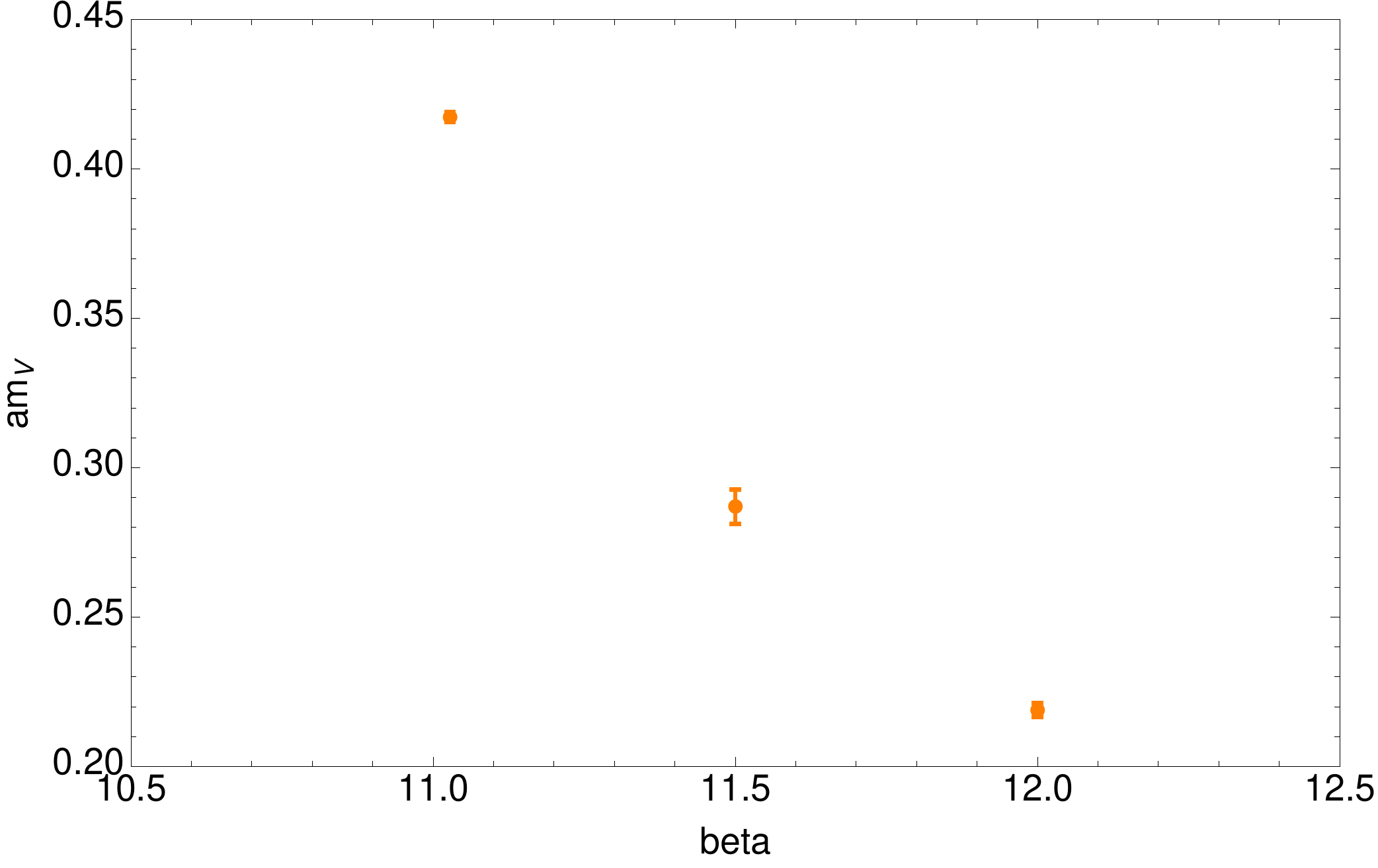}\\
   \includegraphics[width=0.46\textwidth]{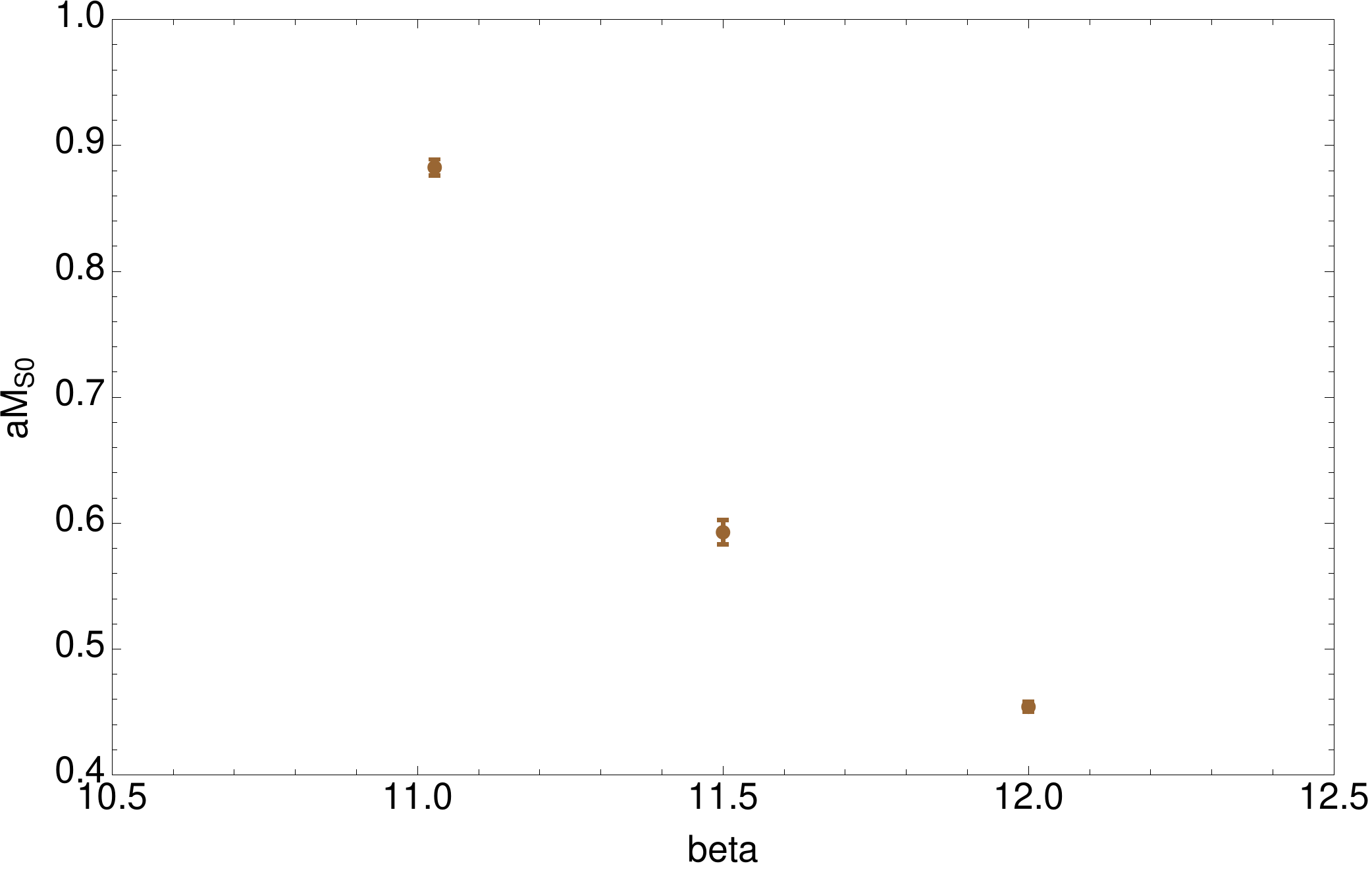}
   \caption{Vector meson and spin-0 baryon masses in lattice units vs. $\beta$ for $L/a = 48$.  This is indicative of lattice spacing differing by a factor of 2 between $\beta=11.028$ and $\beta=12.0$, with $\beta=11.5$ having a lattice spacing roughly 30\% larger than $\beta=12.0$. }
\label{fig:Rel_scale}
\end{figure}
To examine this, we compare two common scale setting quantities, $m_V$ and $M_{S0}$, in lattice units in Fig.~\ref{fig:Rel_scale}.  It is immediately clear from these comparisons that the lattice spacing difference is close to a factor of two. 

\begin{figure}[!t] 
   \centering
   \includegraphics[width=0.46\textwidth]{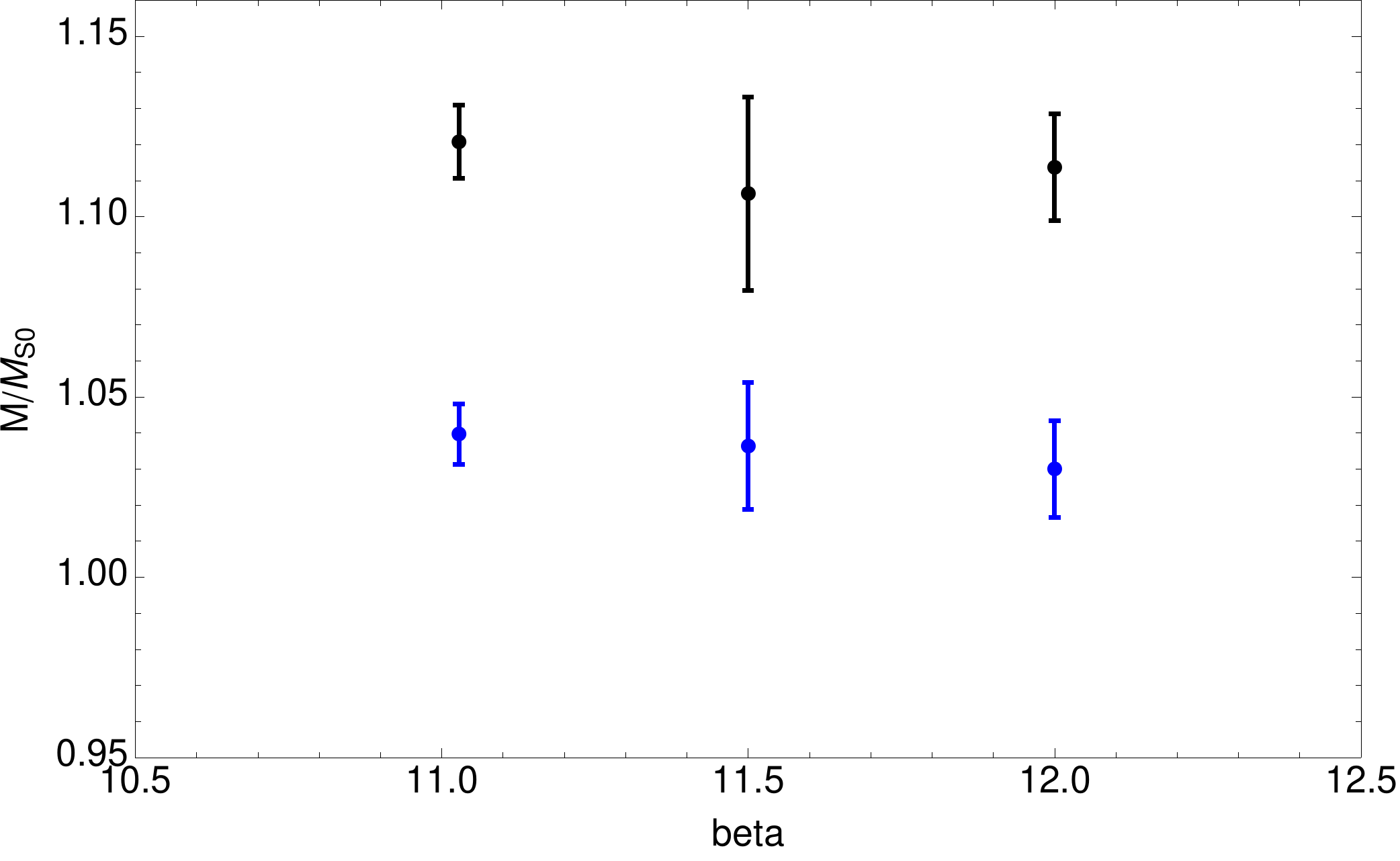}
   \caption{Spin-1 (blue) and spin-2 (black) to spin-0 baryon mass ratio vs $\beta$.  Lattice systematics appear to be small compared to the statistical errors.}
\label{fig:M_MSO_v_beta}
\end{figure}

To put a quantitative error on the lattice spacing systematic, a comparison of ratios of quantities is the most efficient method.  Since the spin-0 baryon will ultimately set the scale, it is natural to be in the denominator for the comparison.  In Fig.~\ref{fig:M_MSO_v_beta}, the baryon ratios of the spin-1 and spin-2 states to the spin-0 state are shown as a function of $\beta$.  In this figure, it is apparent that all of the baryon mass ratios agree within statistical errors.   Since the LCP matching is roughly 5\%, we can say that the lattice spacing systematic for the baryons should be well within 5\%.

\section{Discussion}

We have presented the spectrum and effective Higgs interaction for a 
composite dark matter theory based on the bosonic baryons of an SU(4) 
confining gauge theory.  The constituent fermions acquire both
vector-like and electroweak symmetry breaking masses, leading to 
a suppressed effective Higgs interaction characterized by the
parameter $0 \le \alpha \le 1$, defined in Eq.~(\ref{eq:alpha}).
Our primary results extracted from our lattice simulations are 
the SU(4) hadron spectrum and the $\sigma_f$ terms for the 
lightest baryon. 
This enables us to calculate the spin-independent scattering 
cross section in Figs.~\ref{fig:Higgs_Exclusion} and compare
to the latest results from LUX\@.  Our results are expressed as
bounds on the effective Higgs interaction strength $\alpha$, 
shown in Fig.~\ref{fig:alpha_max}, where we find $\alpha \lsim 0.34$ 
throughout the range of fermion masses that were simulated on the lattice
in this paper.  The least constrained-in-$\alpha$ theories have 
pseudoscalar mesons with masses at the anticipated LEP II bound 
and scalar baryon dark matter with a mass of several hundred GeV\@.  
The bounds on the effective Higgs interaction scale approximately as 
\begin{equation}
\alpha \lsim \left( \frac{370 \; {\rm GeV}}{m_{B}} \right)^{1/2} \times 
\left\{ \begin{array}{ll} 0.34 & \quad m_{PS}/m_{V} = 0.55 \\
                          0.05 & \quad m_{PS}/m_{V} = 1 \, ,
        \end{array} \right. 
\end{equation} 
where the bound on the baryon mass is $m_B > 370$~GeV
(for $m_{PS}/m_{V} = 0.55$) and $m_B > 200$~GeV (for $m_{PS}/m_{V} = 1$) 
given the anticipated LEP II bound on the pseudoscalar meson masses 
of $m_{PS} > 100$~GeV\@.
We conclude that composite dark matter theories with fermion masses 
purely from electroweak symmetry breaking 
($\alpha = 1$) appear to be strongly disfavored.  

We compared the hadron spectrum
between SU(3) and SU(4), shown in Fig.~\ref{fig:3c4c} and Fig.~\ref{fig:3c4c_rotor}. 
The latter figure shows that the two-parameter large $N_c$ rotor spectrum prediction 
in Eq.~\eqref{eq:rotor} is insufficient to define the 4-color baryon spectrum, 
but the three-parameter version in Eq.~\eqref{eq:rotor_c}, which appropriately 
accounts for $N_c^{-1}$ corrections, does match the higher spin SU(4) 
baryon masses well.  It would be interesting to probe baryons in SU(6) 
to see if this relation continues with all of its coefficients 
fixed by the current data set.

There are several future directions to pursue.  We have assumed 
custodial SU(2) symmetry of the fermion spectra, causing the 
charge radius of the lightest baryon to vanish.  It would be
interesting to determine the bounds on custodial SU(2) breaking
from the spin-independent scattering cross section that would be
induced through the charge radius.  Even if custodial SU(2) symmetry
is exact and the effective Higgs coupling vanishes 
($\alpha = 0$), spin-independent dark matter 
scattering can occur through the scalar baryon polarizability.  
As a prerequisite to attacking this very computationally and theoretically 
difficult problem, one must first understand the lattice systematics 
at a high precision, all of which are expected to be challenging 
for extracted polarizabilities \cite{Tiburzi:2008pa,Detmold:2010ts}.  
To that end, we performed an extensive study of volume and 
lattice spacing effects on three lattice spacings ($\beta=11.028,11.5, 12.0$) 
and four volumes ($L/a=16,32,48,64$).   In particular, we first investigated 
the minimum number of sites for volume effects to be negligible in the 
baryon spectrum. 
For the coarsest lattice spacing and intermediate lattice spacing, 
$L/a=32$ was found to be sufficient, but for the finest lattice spacing 
even $L/a=48$ was found to have too small a volume.  With this in mind, 
the polarizability calculation should not have volumes below these sizes. 
The other systematic that had to be quantified is the lattice spacing systematic. 
The results show at most 5\% lattice spacing effects on the 
coarsest lattice.  For that reason, $\beta=11.028$ and $\beta=11.5$ 
will likely prove to be the best ensembles for polarizabilities as 
both lattice spacing and volume effects are effectively 
controlled within errors. 

It should also be emphasized that the polarizability calculations can 
benefit from larger volume, as the quantized background fields can be made 
finer and be better used to extract the quadratic contribution of the 
energy proportional to the polarizability.  However, even more statistics 
will be required at each background field value (including zero field) 
to resolve these differences.  Initial estimates state that the 
baryon uncertainty will need to be at least a factor of two smaller 
than the current values.  For that reason, at least a factor of four 
increase of statistics will likely be required for each ensemble to 
reliably perform that calculation.  Also, the validity 
of the quenched approximation is still in question.  We plan to 
perform at least one unquenched ensemble to estimate the size of these 
effects as well.

\section{Acknowledgments}

MIB would like to thank Tom DeGrand, Rich Lebed, and Martin Savage for 
very enlightening discussions that greatly aided the direction and 
presentation of this work.

We thank the Lawrence Livermore National Laboratory (LLNL) Multiprogrammatic 
and Institutional Computing program for Grand Challenge allocations and 
time on the LLNL BlueGene/Q (rzuseq and vulcan) supercomputer. 
We thank LLNL for funding from LDRD~13-ERD-023 
``Illuminating the Dark Universe with PetaFlops Supercomputing''.  
Computing support for this work comes from the LLNL Institutional 
Computing Grand Challenge program.

GDK thanks the Ambrose Monell Foundation for support while at the 
Institute for Advanced Study.  MIB and ETN thank the Institute for Advanced Study 
and the University of Oregon for hospitality during the course of this work.  

This work has been supported by the U.~S.~Department of Energy under 
Grant Nos.\ 
DE-FC02-12ER41877 and DE-FG02-85ER40231 (D.S.), 
DE-SC0010025 (R.C.B., C.R., E.W.), 
DE-FG02-92ER-40704 (T.A.),
DE-FG02-96ER40969 (G.D.K.), 
DE-FG02-00ER41132 (M.I.B.),
and Contracts 
DE-AC52-07NA27344 (LLNL), 
DE-AC02-06CH11357 (Argonne Leadership Computing Facility), 
and by the National Science Foundation under Grant Nos.\ 
NSF PHY11-00905 (G.F., G.V.), 
PHY09-18108 (G.D.K.), 
OCI-0749300 (O.W.).
S.N.S was supported by the Office of Nuclear Physics in the 
US Department of Energy's Office of Science under Contract DE-AC02-05CH11231.

\bibliography{LSD-SU4}

\begin{thebibliography}{85}%
\makeatletter
\providecommand \@ifxundefined [1]{%
 \@ifx{#1\undefined}
}%
\providecommand \@ifnum [1]{%
 \ifnum #1\expandafter \@firstoftwo
 \else \expandafter \@secondoftwo
 \fi
}%
\providecommand \@ifx [1]{%
 \ifx #1\expandafter \@firstoftwo
 \else \expandafter \@secondoftwo
 \fi
}%
\providecommand \natexlab [1]{#1}%
\providecommand \enquote  [1]{``#1''}%
\providecommand \bibnamefont  [1]{#1}%
\providecommand \bibfnamefont [1]{#1}%
\providecommand \citenamefont [1]{#1}%
\providecommand \href@noop [0]{\@secondoftwo}%
\providecommand \href [0]{\begingroup \@sanitize@url \@href}%
\providecommand \@href[1]{\@@startlink{#1}\@@href}%
\providecommand \@@href[1]{\endgroup#1\@@endlink}%
\providecommand \@sanitize@url [0]{\catcode `\\12\catcode `\$12\catcode
  `\&12\catcode `\#12\catcode `\^12\catcode `\_12\catcode `\%12\relax}%
\providecommand \@@startlink[1]{}%
\providecommand \@@endlink[0]{}%
\providecommand \url  [0]{\begingroup\@sanitize@url \@url }%
\providecommand \@url [1]{\endgroup\@href {#1}{\urlprefix }}%
\providecommand \urlprefix  [0]{URL }%
\providecommand \Eprint [0]{\href }%
\providecommand \doibase [0]{http://dx.doi.org/}%
\providecommand \selectlanguage [0]{\@gobble}%
\providecommand \bibinfo  [0]{\@secondoftwo}%
\providecommand \bibfield  [0]{\@secondoftwo}%
\providecommand \translation [1]{[#1]}%
\providecommand \BibitemOpen [0]{}%
\providecommand \bibitemStop [0]{}%
\providecommand \bibitemNoStop [0]{.\EOS\space}%
\providecommand \EOS [0]{\spacefactor3000\relax}%
\providecommand \BibitemShut  [1]{\csname bibitem#1\endcsname}%
\let\auto@bib@innerbib\@empty
\bibitem [{\citenamefont {Nussinov}(1985)}]{Nussinov:1985xr}%
  \BibitemOpen
  \bibfield  {author} {\bibinfo {author} {\bibfnamefont {S.}~\bibnamefont
  {Nussinov}},\ }\href {\doibase 10.1016/0370-2693(85)90689-6} {\bibfield
  {journal} {\bibinfo  {journal} {Phys.Lett.}\ }\textbf {\bibinfo {volume}
  {B165}},\ \bibinfo {pages} {55} (\bibinfo {year} {1985})}\BibitemShut
  {NoStop}%
\bibitem [{\citenamefont {Chivukula}\ and\ \citenamefont
  {Walker}(1990)}]{Chivukula:1989qb}%
  \BibitemOpen
  \bibfield  {author} {\bibinfo {author} {\bibfnamefont {R.~S.}\ \bibnamefont
  {Chivukula}}\ and\ \bibinfo {author} {\bibfnamefont {T.~P.}\ \bibnamefont
  {Walker}},\ }\href {\doibase 10.1016/0550-3213(90)90151-3} {\bibfield
  {journal} {\bibinfo  {journal} {Nucl.Phys.}\ }\textbf {\bibinfo {volume}
  {B329}},\ \bibinfo {pages} {445} (\bibinfo {year} {1990})}\BibitemShut
  {NoStop}%
\bibitem [{\citenamefont {Barr}\ \emph {et~al.}(1990)\citenamefont {Barr},
  \citenamefont {Chivukula},\ and\ \citenamefont {Farhi}}]{Barr:1990ca}%
  \BibitemOpen
  \bibfield  {author} {\bibinfo {author} {\bibfnamefont {S.~M.}\ \bibnamefont
  {Barr}}, \bibinfo {author} {\bibfnamefont {R.~S.}\ \bibnamefont {Chivukula}},
  \ and\ \bibinfo {author} {\bibfnamefont {E.}~\bibnamefont {Farhi}},\ }\href
  {\doibase 10.1016/0370-2693(90)91661-T} {\bibfield  {journal} {\bibinfo
  {journal} {Phys.Lett.}\ }\textbf {\bibinfo {volume} {B241}},\ \bibinfo
  {pages} {387} (\bibinfo {year} {1990})}\BibitemShut {NoStop}%
\bibitem [{\citenamefont {Kaplan}(1992)}]{Kaplan:1991ah}%
  \BibitemOpen
  \bibfield  {author} {\bibinfo {author} {\bibfnamefont {D.~B.}\ \bibnamefont
  {Kaplan}},\ }\href {\doibase 10.1103/PhysRevLett.68.741} {\bibfield
  {journal} {\bibinfo  {journal} {Phys.Rev.Lett.}\ }\textbf {\bibinfo {volume}
  {68}},\ \bibinfo {pages} {741} (\bibinfo {year} {1992})}\BibitemShut
  {NoStop}%
\bibitem [{\citenamefont {Aad}\ \emph {et~al.}(2012)\citenamefont {Aad} \emph
  {et~al.}}]{Aad:2012tfa}%
  \BibitemOpen
  \bibfield  {author} {\bibinfo {author} {\bibfnamefont {G.}~\bibnamefont
  {Aad}} \emph {et~al.} (\bibinfo {collaboration} {ATLAS Collaboration}),\
  }\href {\doibase 10.1016/j.physletb.2012.08.020} {\bibfield  {journal}
  {\bibinfo  {journal} {Phys.Lett.}\ }\textbf {\bibinfo {volume} {B716}},\
  \bibinfo {pages} {1} (\bibinfo {year} {2012})},\ \Eprint
  {http://arxiv.org/abs/1207.7214} {arXiv:1207.7214 [hep-ex]} \BibitemShut
  {NoStop}%
\bibitem [{\citenamefont {Chatrchyan}\ \emph {et~al.}(2012)\citenamefont
  {Chatrchyan} \emph {et~al.}}]{Chatrchyan:2012ufa}%
  \BibitemOpen
  \bibfield  {author} {\bibinfo {author} {\bibfnamefont {S.}~\bibnamefont
  {Chatrchyan}} \emph {et~al.} (\bibinfo {collaboration} {CMS Collaboration}),\
  }\href {\doibase 10.1016/j.physletb.2012.08.021} {\bibfield  {journal}
  {\bibinfo  {journal} {Phys.Lett.}\ }\textbf {\bibinfo {volume} {B716}},\
  \bibinfo {pages} {30} (\bibinfo {year} {2012})},\ \Eprint
  {http://arxiv.org/abs/1207.7235} {arXiv:1207.7235 [hep-ex]} \BibitemShut
  {NoStop}%
\bibitem [{\citenamefont {Aprile}\ \emph {et~al.}(2012)\citenamefont {Aprile}
  \emph {et~al.}}]{Aprile:2012nq}%
  \BibitemOpen
  \bibfield  {author} {\bibinfo {author} {\bibfnamefont {E.}~\bibnamefont
  {Aprile}} \emph {et~al.} (\bibinfo {collaboration} {XENON100
  Collaboration}),\ }\href {\doibase 10.1103/PhysRevLett.109.181301} {\bibfield
   {journal} {\bibinfo  {journal} {Phys.Rev.Lett.}\ }\textbf {\bibinfo {volume}
  {109}},\ \bibinfo {pages} {181301} (\bibinfo {year} {2012})},\ \Eprint
  {http://arxiv.org/abs/1207.5988} {arXiv:1207.5988 [astro-ph.CO]} \BibitemShut
  {NoStop}%
\bibitem [{\citenamefont {Agnese}\ \emph {et~al.}(2013)\citenamefont {Agnese}
  \emph {et~al.}}]{Agnese:2013rvf}%
  \BibitemOpen
  \bibfield  {author} {\bibinfo {author} {\bibfnamefont {R.}~\bibnamefont
  {Agnese}} \emph {et~al.} (\bibinfo {collaboration} {CDMS Collaboration}),\
  }\href@noop {} {\bibfield  {journal} {\bibinfo  {journal} {Phys.Rev.Lett.}\ }
  (\bibinfo {year} {2013})},\ \Eprint {http://arxiv.org/abs/1304.4279}
  {arXiv:1304.4279 [hep-ex]} \BibitemShut {NoStop}%
\bibitem [{\citenamefont {Akerib}\ \emph {et~al.}(2013)\citenamefont {Akerib}
  \emph {et~al.}}]{Akerib:2013tjd}%
  \BibitemOpen
  \bibfield  {author} {\bibinfo {author} {\bibfnamefont {D.}~\bibnamefont
  {Akerib}} \emph {et~al.} (\bibinfo {collaboration} {LUX Collaboration}),\
  }\href@noop {} {\  (\bibinfo {year} {2013})},\ \Eprint
  {http://arxiv.org/abs/1310.8214} {arXiv:1310.8214 [astro-ph.CO]} \BibitemShut
  {NoStop}%
\bibitem [{\citenamefont {Chivukula}\ \emph {et~al.}(1993)\citenamefont
  {Chivukula}, \citenamefont {Cohen}, \citenamefont {Luke},\ and\ \citenamefont
  {Savage}}]{Chivukula:1992pn}%
  \BibitemOpen
  \bibfield  {author} {\bibinfo {author} {\bibfnamefont {R.~S.}\ \bibnamefont
  {Chivukula}}, \bibinfo {author} {\bibfnamefont {A.~G.}\ \bibnamefont
  {Cohen}}, \bibinfo {author} {\bibfnamefont {M.~E.}\ \bibnamefont {Luke}}, \
  and\ \bibinfo {author} {\bibfnamefont {M.~J.}\ \bibnamefont {Savage}},\
  }\href {\doibase 10.1016/0370-2693(93)91836-C} {\bibfield  {journal}
  {\bibinfo  {journal} {Phys.Lett.}\ }\textbf {\bibinfo {volume} {B298}},\
  \bibinfo {pages} {380} (\bibinfo {year} {1993})},\ \Eprint
  {http://arxiv.org/abs/hep-ph/9210274} {arXiv:hep-ph/9210274 [hep-ph]}
  \BibitemShut {NoStop}%
\bibitem [{\citenamefont {Bagnasco}\ \emph {et~al.}(1994)\citenamefont
  {Bagnasco}, \citenamefont {Dine},\ and\ \citenamefont
  {Thomas}}]{Bagnasco:1993st}%
  \BibitemOpen
  \bibfield  {author} {\bibinfo {author} {\bibfnamefont {J.}~\bibnamefont
  {Bagnasco}}, \bibinfo {author} {\bibfnamefont {M.}~\bibnamefont {Dine}}, \
  and\ \bibinfo {author} {\bibfnamefont {S.~D.}\ \bibnamefont {Thomas}},\
  }\href {\doibase 10.1016/0370-2693(94)90830-3} {\bibfield  {journal}
  {\bibinfo  {journal} {Phys.Lett.}\ }\textbf {\bibinfo {volume} {B320}},\
  \bibinfo {pages} {99} (\bibinfo {year} {1994})},\ \Eprint
  {http://arxiv.org/abs/hep-ph/9310290} {arXiv:hep-ph/9310290 [hep-ph]}
  \BibitemShut {NoStop}%
\bibitem [{\citenamefont {Pospelov}\ and\ \citenamefont {ter
  Veldhuis}(2000)}]{Pospelov:2000bq}%
  \BibitemOpen
  \bibfield  {author} {\bibinfo {author} {\bibfnamefont {M.}~\bibnamefont
  {Pospelov}}\ and\ \bibinfo {author} {\bibfnamefont {T.}~\bibnamefont {ter
  Veldhuis}},\ }\href {\doibase 10.1016/S0370-2693(00)00358-0} {\bibfield
  {journal} {\bibinfo  {journal} {Phys.Lett.}\ }\textbf {\bibinfo {volume}
  {B480}},\ \bibinfo {pages} {181} (\bibinfo {year} {2000})},\ \Eprint
  {http://arxiv.org/abs/hep-ph/0003010} {arXiv:hep-ph/0003010 [hep-ph]}
  \BibitemShut {NoStop}%
\bibitem [{\citenamefont {Sigurdson}\ \emph {et~al.}(2004)\citenamefont
  {Sigurdson}, \citenamefont {Doran}, \citenamefont {Kurylov}, \citenamefont
  {Caldwell},\ and\ \citenamefont {Kamionkowski}}]{Sigurdson:2004zp}%
  \BibitemOpen
  \bibfield  {author} {\bibinfo {author} {\bibfnamefont {K.}~\bibnamefont
  {Sigurdson}}, \bibinfo {author} {\bibfnamefont {M.}~\bibnamefont {Doran}},
  \bibinfo {author} {\bibfnamefont {A.}~\bibnamefont {Kurylov}}, \bibinfo
  {author} {\bibfnamefont {R.~R.}\ \bibnamefont {Caldwell}}, \ and\ \bibinfo
  {author} {\bibfnamefont {M.}~\bibnamefont {Kamionkowski}},\ }\href {\doibase
  10.1103/PhysRevD.70.083501, 10.1103/PhysRevD.73.089903} {\bibfield  {journal}
  {\bibinfo  {journal} {Phys.Rev.}\ }\textbf {\bibinfo {volume} {D70}},\
  \bibinfo {pages} {083501} (\bibinfo {year} {2004})},\ \Eprint
  {http://arxiv.org/abs/astro-ph/0406355} {arXiv:astro-ph/0406355 [astro-ph]}
  \BibitemShut {NoStop}%
\bibitem [{\citenamefont {Gudnason}\ \emph {et~al.}(2006)\citenamefont
  {Gudnason}, \citenamefont {Kouvaris},\ and\ \citenamefont
  {Sannino}}]{Gudnason:2006ug}%
  \BibitemOpen
  \bibfield  {author} {\bibinfo {author} {\bibfnamefont {S.~B.}\ \bibnamefont
  {Gudnason}}, \bibinfo {author} {\bibfnamefont {C.}~\bibnamefont {Kouvaris}},
  \ and\ \bibinfo {author} {\bibfnamefont {F.}~\bibnamefont {Sannino}},\ }\href
  {\doibase 10.1103/PhysRevD.73.115003} {\bibfield  {journal} {\bibinfo
  {journal} {Phys.Rev.}\ }\textbf {\bibinfo {volume} {D73}},\ \bibinfo {pages}
  {115003} (\bibinfo {year} {2006})},\ \Eprint
  {http://arxiv.org/abs/hep-ph/0603014} {arXiv:hep-ph/0603014 [hep-ph]}
  \BibitemShut {NoStop}%
\bibitem [{\citenamefont {Alves}\ \emph {et~al.}(2010)\citenamefont {Alves},
  \citenamefont {Behbahani}, \citenamefont {Schuster},\ and\ \citenamefont
  {Wacker}}]{Alves:2009nf}%
  \BibitemOpen
  \bibfield  {author} {\bibinfo {author} {\bibfnamefont {D.~S.}\ \bibnamefont
  {Alves}}, \bibinfo {author} {\bibfnamefont {S.~R.}\ \bibnamefont
  {Behbahani}}, \bibinfo {author} {\bibfnamefont {P.}~\bibnamefont {Schuster}},
  \ and\ \bibinfo {author} {\bibfnamefont {J.~G.}\ \bibnamefont {Wacker}},\
  }\href {\doibase 10.1016/j.physletb.2010.08.006} {\bibfield  {journal}
  {\bibinfo  {journal} {Phys.Lett.}\ }\textbf {\bibinfo {volume} {B692}},\
  \bibinfo {pages} {323} (\bibinfo {year} {2010})},\ \Eprint
  {http://arxiv.org/abs/0903.3945} {arXiv:0903.3945 [hep-ph]} \BibitemShut
  {NoStop}%
\bibitem [{\citenamefont {Kribs}\ \emph {et~al.}(2010)\citenamefont {Kribs},
  \citenamefont {Roy}, \citenamefont {Terning},\ and\ \citenamefont
  {Zurek}}]{Kribs:2009fy}%
  \BibitemOpen
  \bibfield  {author} {\bibinfo {author} {\bibfnamefont {G.~D.}\ \bibnamefont
  {Kribs}}, \bibinfo {author} {\bibfnamefont {T.~S.}\ \bibnamefont {Roy}},
  \bibinfo {author} {\bibfnamefont {J.}~\bibnamefont {Terning}}, \ and\
  \bibinfo {author} {\bibfnamefont {K.~M.}\ \bibnamefont {Zurek}},\ }\href
  {\doibase 10.1103/PhysRevD.81.095001} {\bibfield  {journal} {\bibinfo
  {journal} {Phys. Rev.}\ }\textbf {\bibinfo {volume} {D81}},\ \bibinfo {pages}
  {095001} (\bibinfo {year} {2010})},\ \Eprint {http://arxiv.org/abs/0909.2034}
  {arXiv:0909.2034 [hep-ph]} \BibitemShut {NoStop}%
\bibitem [{\citenamefont {Barbieri}\ \emph {et~al.}(2010)\citenamefont
  {Barbieri}, \citenamefont {Rychkov},\ and\ \citenamefont
  {Torre}}]{Barbieri:2010mn}%
  \BibitemOpen
  \bibfield  {author} {\bibinfo {author} {\bibfnamefont {R.}~\bibnamefont
  {Barbieri}}, \bibinfo {author} {\bibfnamefont {S.}~\bibnamefont {Rychkov}}, \
  and\ \bibinfo {author} {\bibfnamefont {R.}~\bibnamefont {Torre}},\ }\href
  {\doibase 10.1016/j.physletb.2010.04.010} {\bibfield  {journal} {\bibinfo
  {journal} {Phys.Lett.}\ }\textbf {\bibinfo {volume} {B688}},\ \bibinfo
  {pages} {212} (\bibinfo {year} {2010})},\ \Eprint
  {http://arxiv.org/abs/1001.3149} {arXiv:1001.3149 [hep-ph]} \BibitemShut
  {NoStop}%
\bibitem [{\citenamefont {Banks}\ \emph {et~al.}(2010)\citenamefont {Banks},
  \citenamefont {Fortin},\ and\ \citenamefont {Thomas}}]{Banks:2010eh}%
  \BibitemOpen
  \bibfield  {author} {\bibinfo {author} {\bibfnamefont {T.}~\bibnamefont
  {Banks}}, \bibinfo {author} {\bibfnamefont {J.-F.}\ \bibnamefont {Fortin}}, \
  and\ \bibinfo {author} {\bibfnamefont {S.}~\bibnamefont {Thomas}},\
  }\href@noop {} {\  (\bibinfo {year} {2010})},\ \Eprint
  {http://arxiv.org/abs/1007.5515} {arXiv:1007.5515 [hep-ph]} \BibitemShut
  {NoStop}%
\bibitem [{\citenamefont {Chang}\ \emph {et~al.}(2010)\citenamefont {Chang},
  \citenamefont {Weiner},\ and\ \citenamefont {Yavin}}]{Chang:2010en}%
  \BibitemOpen
  \bibfield  {author} {\bibinfo {author} {\bibfnamefont {S.}~\bibnamefont
  {Chang}}, \bibinfo {author} {\bibfnamefont {N.}~\bibnamefont {Weiner}}, \
  and\ \bibinfo {author} {\bibfnamefont {I.}~\bibnamefont {Yavin}},\ }\href
  {\doibase 10.1103/PhysRevD.82.125011} {\bibfield  {journal} {\bibinfo
  {journal} {Phys.Rev.}\ }\textbf {\bibinfo {volume} {D82}},\ \bibinfo {pages}
  {125011} (\bibinfo {year} {2010})},\ \Eprint {http://arxiv.org/abs/1007.4200}
  {arXiv:1007.4200 [hep-ph]} \BibitemShut {NoStop}%
\bibitem [{\citenamefont {Barger}\ \emph {et~al.}(2011)\citenamefont {Barger},
  \citenamefont {Keung},\ and\ \citenamefont {Marfatia}}]{Barger:2010gv}%
  \BibitemOpen
  \bibfield  {author} {\bibinfo {author} {\bibfnamefont {V.}~\bibnamefont
  {Barger}}, \bibinfo {author} {\bibfnamefont {W.-Y.}\ \bibnamefont {Keung}}, \
  and\ \bibinfo {author} {\bibfnamefont {D.}~\bibnamefont {Marfatia}},\ }\href
  {\doibase 10.1016/j.physletb.2010.12.008} {\bibfield  {journal} {\bibinfo
  {journal} {Phys.Lett.}\ }\textbf {\bibinfo {volume} {B696}},\ \bibinfo
  {pages} {74} (\bibinfo {year} {2011})},\ \Eprint
  {http://arxiv.org/abs/1007.4345} {arXiv:1007.4345 [hep-ph]} \BibitemShut
  {NoStop}%
\bibitem [{\citenamefont {Weiner}\ and\ \citenamefont
  {Yavin}(2012)}]{Weiner:2012cb}%
  \BibitemOpen
  \bibfield  {author} {\bibinfo {author} {\bibfnamefont {N.}~\bibnamefont
  {Weiner}}\ and\ \bibinfo {author} {\bibfnamefont {I.}~\bibnamefont {Yavin}},\
  }\href {\doibase 10.1103/PhysRevD.86.075021} {\bibfield  {journal} {\bibinfo
  {journal} {Phys.Rev.}\ }\textbf {\bibinfo {volume} {D86}},\ \bibinfo {pages}
  {075021} (\bibinfo {year} {2012})},\ \Eprint {http://arxiv.org/abs/1206.2910}
  {arXiv:1206.2910 [hep-ph]} \BibitemShut {NoStop}%
\bibitem [{\citenamefont {Fortin}\ and\ \citenamefont
  {Tait}(2012)}]{Fortin:2011hv}%
  \BibitemOpen
  \bibfield  {author} {\bibinfo {author} {\bibfnamefont {J.-F.}\ \bibnamefont
  {Fortin}}\ and\ \bibinfo {author} {\bibfnamefont {T.~M.}\ \bibnamefont
  {Tait}},\ }\href {\doibase 10.1103/PhysRevD.85.063506} {\bibfield  {journal}
  {\bibinfo  {journal} {Phys.Rev.}\ }\textbf {\bibinfo {volume} {D85}},\
  \bibinfo {pages} {063506} (\bibinfo {year} {2012})},\ \Eprint
  {http://arxiv.org/abs/1103.3289} {arXiv:1103.3289 [hep-ph]} \BibitemShut
  {NoStop}%
\bibitem [{\citenamefont {Appelquist}\ \emph {et~al.}(2013)\citenamefont
  {Appelquist}, \citenamefont {Brower}, \citenamefont {Buchoff}, \citenamefont
  {Cheng}, \citenamefont {Cohen} \emph {et~al.}}]{Appelquist:2013ms}%
  \BibitemOpen
  \bibfield  {author} {\bibinfo {author} {\bibfnamefont {T.}~\bibnamefont
  {Appelquist}}, \bibinfo {author} {\bibfnamefont {R.}~\bibnamefont {Brower}},
  \bibinfo {author} {\bibfnamefont {M.}~\bibnamefont {Buchoff}}, \bibinfo
  {author} {\bibfnamefont {M.}~\bibnamefont {Cheng}}, \bibinfo {author}
  {\bibfnamefont {S.}~\bibnamefont {Cohen}},  \emph {et~al.},\ }\href {\doibase
  10.1103/PhysRevD.88.014502} {\bibfield  {journal} {\bibinfo  {journal}
  {Phys.Rev.}\ }\textbf {\bibinfo {volume} {D88}},\ \bibinfo {pages} {014502}
  (\bibinfo {year} {2013})},\ \Eprint {http://arxiv.org/abs/1301.1693}
  {arXiv:1301.1693 [hep-ph]} \BibitemShut {NoStop}%
\bibitem [{\citenamefont {Lewis}\ \emph {et~al.}(2012)\citenamefont {Lewis},
  \citenamefont {Pica},\ and\ \citenamefont {Sannino}}]{Lewis:2011zb}%
  \BibitemOpen
  \bibfield  {author} {\bibinfo {author} {\bibfnamefont {R.}~\bibnamefont
  {Lewis}}, \bibinfo {author} {\bibfnamefont {C.}~\bibnamefont {Pica}}, \ and\
  \bibinfo {author} {\bibfnamefont {F.}~\bibnamefont {Sannino}},\ }\href
  {\doibase 10.1103/PhysRevD.85.014504} {\bibfield  {journal} {\bibinfo
  {journal} {Phys.Rev.}\ }\textbf {\bibinfo {volume} {D85}},\ \bibinfo {pages}
  {014504} (\bibinfo {year} {2012})},\ \Eprint {http://arxiv.org/abs/1109.3513}
  {arXiv:1109.3513 [hep-ph]} \BibitemShut {NoStop}%
\bibitem [{\citenamefont {Buckley}\ and\ \citenamefont
  {Neil}(2013)}]{Buckley:2012ky}%
  \BibitemOpen
  \bibfield  {author} {\bibinfo {author} {\bibfnamefont {M.~R.}\ \bibnamefont
  {Buckley}}\ and\ \bibinfo {author} {\bibfnamefont {E.~T.}\ \bibnamefont
  {Neil}},\ }\href {\doibase 10.1103/PhysRevD.87.043510} {\bibfield  {journal}
  {\bibinfo  {journal} {Phys.Rev.}\ }\textbf {\bibinfo {volume} {D87}},\
  \bibinfo {pages} {043510} (\bibinfo {year} {2013})},\ \Eprint
  {http://arxiv.org/abs/1209.6054} {arXiv:1209.6054 [hep-ph]} \BibitemShut
  {NoStop}%
\bibitem [{\citenamefont {Kitano}\ and\ \citenamefont
  {Low}(2005)}]{Kitano:2004sv}%
  \BibitemOpen
  \bibfield  {author} {\bibinfo {author} {\bibfnamefont {R.}~\bibnamefont
  {Kitano}}\ and\ \bibinfo {author} {\bibfnamefont {I.}~\bibnamefont {Low}},\
  }\href {\doibase 10.1103/PhysRevD.71.023510} {\bibfield  {journal} {\bibinfo
  {journal} {Phys.Rev.}\ }\textbf {\bibinfo {volume} {D71}},\ \bibinfo {pages}
  {023510} (\bibinfo {year} {2005})},\ \Eprint
  {http://arxiv.org/abs/hep-ph/0411133} {arXiv:hep-ph/0411133 [hep-ph]}
  \BibitemShut {NoStop}%
\bibitem [{\citenamefont {Farrar}\ and\ \citenamefont
  {Zaharijas}(2006)}]{Farrar:2005zd}%
  \BibitemOpen
  \bibfield  {author} {\bibinfo {author} {\bibfnamefont {G.~R.}\ \bibnamefont
  {Farrar}}\ and\ \bibinfo {author} {\bibfnamefont {G.}~\bibnamefont
  {Zaharijas}},\ }\href {\doibase 10.1103/PhysRevLett.96.041302} {\bibfield
  {journal} {\bibinfo  {journal} {Phys.Rev.Lett.}\ }\textbf {\bibinfo {volume}
  {96}},\ \bibinfo {pages} {041302} (\bibinfo {year} {2006})},\ \Eprint
  {http://arxiv.org/abs/hep-ph/0510079} {arXiv:hep-ph/0510079 [hep-ph]}
  \BibitemShut {NoStop}%
\bibitem [{\citenamefont {Banks}\ \emph {et~al.}(2005)\citenamefont {Banks},
  \citenamefont {Mason},\ and\ \citenamefont {O'Neil}}]{Banks:2005hc}%
  \BibitemOpen
  \bibfield  {author} {\bibinfo {author} {\bibfnamefont {T.}~\bibnamefont
  {Banks}}, \bibinfo {author} {\bibfnamefont {J.}~\bibnamefont {Mason}}, \ and\
  \bibinfo {author} {\bibfnamefont {D.}~\bibnamefont {O'Neil}},\ }\href
  {\doibase 10.1103/PhysRevD.72.043530} {\bibfield  {journal} {\bibinfo
  {journal} {Phys.Rev.}\ }\textbf {\bibinfo {volume} {D72}},\ \bibinfo {pages}
  {043530} (\bibinfo {year} {2005})},\ \Eprint
  {http://arxiv.org/abs/hep-ph/0506015} {arXiv:hep-ph/0506015 [hep-ph]}
  \BibitemShut {NoStop}%
\bibitem [{\citenamefont {Kitano}\ \emph {et~al.}(2008)\citenamefont {Kitano},
  \citenamefont {Murayama},\ and\ \citenamefont {Ratz}}]{Kitano:2008tk}%
  \BibitemOpen
  \bibfield  {author} {\bibinfo {author} {\bibfnamefont {R.}~\bibnamefont
  {Kitano}}, \bibinfo {author} {\bibfnamefont {H.}~\bibnamefont {Murayama}}, \
  and\ \bibinfo {author} {\bibfnamefont {M.}~\bibnamefont {Ratz}},\ }\href
  {\doibase 10.1016/j.physletb.2008.09.049} {\bibfield  {journal} {\bibinfo
  {journal} {Phys.Lett.}\ }\textbf {\bibinfo {volume} {B669}},\ \bibinfo
  {pages} {145} (\bibinfo {year} {2008})},\ \Eprint
  {http://arxiv.org/abs/0807.4313} {arXiv:0807.4313 [hep-ph]} \BibitemShut
  {NoStop}%
\bibitem [{\citenamefont {Kaplan}\ \emph {et~al.}(2009)\citenamefont {Kaplan},
  \citenamefont {Luty},\ and\ \citenamefont {Zurek}}]{Kaplan:2009ag}%
  \BibitemOpen
  \bibfield  {author} {\bibinfo {author} {\bibfnamefont {D.~E.}\ \bibnamefont
  {Kaplan}}, \bibinfo {author} {\bibfnamefont {M.~A.}\ \bibnamefont {Luty}}, \
  and\ \bibinfo {author} {\bibfnamefont {K.~M.}\ \bibnamefont {Zurek}},\ }\href
  {\doibase 10.1103/PhysRevD.79.115016} {\bibfield  {journal} {\bibinfo
  {journal} {Phys.Rev.}\ }\textbf {\bibinfo {volume} {D79}},\ \bibinfo {pages}
  {115016} (\bibinfo {year} {2009})},\ \Eprint {http://arxiv.org/abs/0901.4117}
  {arXiv:0901.4117 [hep-ph]} \BibitemShut {NoStop}%
\bibitem [{\citenamefont {An}\ \emph {et~al.}(2010)\citenamefont {An},
  \citenamefont {Chen}, \citenamefont {Mohapatra},\ and\ \citenamefont
  {Zhang}}]{An:2009vq}%
  \BibitemOpen
  \bibfield  {author} {\bibinfo {author} {\bibfnamefont {H.}~\bibnamefont
  {An}}, \bibinfo {author} {\bibfnamefont {S.-L.}\ \bibnamefont {Chen}},
  \bibinfo {author} {\bibfnamefont {R.~N.}\ \bibnamefont {Mohapatra}}, \ and\
  \bibinfo {author} {\bibfnamefont {Y.}~\bibnamefont {Zhang}},\ }\href
  {\doibase 10.1007/JHEP03(2010)124} {\bibfield  {journal} {\bibinfo  {journal}
  {JHEP}\ }\textbf {\bibinfo {volume} {1003}},\ \bibinfo {pages} {124}
  (\bibinfo {year} {2010})},\ \Eprint {http://arxiv.org/abs/0911.4463}
  {arXiv:0911.4463 [hep-ph]} \BibitemShut {NoStop}%
\bibitem [{\citenamefont {Spier Moreira~Alves}\ \emph
  {et~al.}(2010)\citenamefont {Spier Moreira~Alves}, \citenamefont {Behbahani},
  \citenamefont {Schuster},\ and\ \citenamefont {Wacker}}]{Alves:2010dd}%
  \BibitemOpen
  \bibfield  {author} {\bibinfo {author} {\bibfnamefont {D.}~\bibnamefont
  {Spier Moreira~Alves}}, \bibinfo {author} {\bibfnamefont {S.~R.}\
  \bibnamefont {Behbahani}}, \bibinfo {author} {\bibfnamefont {P.}~\bibnamefont
  {Schuster}}, \ and\ \bibinfo {author} {\bibfnamefont {J.~G.}\ \bibnamefont
  {Wacker}},\ }\href {\doibase 10.1007/JHEP06(2010)113} {\bibfield  {journal}
  {\bibinfo  {journal} {JHEP}\ }\textbf {\bibinfo {volume} {1006}},\ \bibinfo
  {pages} {113} (\bibinfo {year} {2010})},\ \Eprint
  {http://arxiv.org/abs/1003.4729} {arXiv:1003.4729 [hep-ph]} \BibitemShut
  {NoStop}%
\bibitem [{\citenamefont {Dulaney}\ \emph {et~al.}(2011)\citenamefont
  {Dulaney}, \citenamefont {Fileviez~Perez},\ and\ \citenamefont
  {Wise}}]{Dulaney:2010dj}%
  \BibitemOpen
  \bibfield  {author} {\bibinfo {author} {\bibfnamefont {T.~R.}\ \bibnamefont
  {Dulaney}}, \bibinfo {author} {\bibfnamefont {P.}~\bibnamefont
  {Fileviez~Perez}}, \ and\ \bibinfo {author} {\bibfnamefont {M.~B.}\
  \bibnamefont {Wise}},\ }\href {\doibase 10.1103/PhysRevD.83.023520}
  {\bibfield  {journal} {\bibinfo  {journal} {Phys.Rev.}\ }\textbf {\bibinfo
  {volume} {D83}},\ \bibinfo {pages} {023520} (\bibinfo {year} {2011})},\
  \Eprint {http://arxiv.org/abs/1005.0617} {arXiv:1005.0617 [hep-ph]}
  \BibitemShut {NoStop}%
\bibitem [{\citenamefont {Cohen}\ \emph {et~al.}(2010)\citenamefont {Cohen},
  \citenamefont {Phalen}, \citenamefont {Pierce},\ and\ \citenamefont
  {Zurek}}]{Cohen:2010kn}%
  \BibitemOpen
  \bibfield  {author} {\bibinfo {author} {\bibfnamefont {T.}~\bibnamefont
  {Cohen}}, \bibinfo {author} {\bibfnamefont {D.~J.}\ \bibnamefont {Phalen}},
  \bibinfo {author} {\bibfnamefont {A.}~\bibnamefont {Pierce}}, \ and\ \bibinfo
  {author} {\bibfnamefont {K.~M.}\ \bibnamefont {Zurek}},\ }\href {\doibase
  10.1103/PhysRevD.82.056001} {\bibfield  {journal} {\bibinfo  {journal}
  {Phys.Rev.}\ }\textbf {\bibinfo {volume} {D82}},\ \bibinfo {pages} {056001}
  (\bibinfo {year} {2010})},\ \Eprint {http://arxiv.org/abs/1005.1655}
  {arXiv:1005.1655 [hep-ph]} \BibitemShut {NoStop}%
\bibitem [{\citenamefont {Shelton}\ and\ \citenamefont
  {Zurek}(2010)}]{Shelton:2010ta}%
  \BibitemOpen
  \bibfield  {author} {\bibinfo {author} {\bibfnamefont {J.}~\bibnamefont
  {Shelton}}\ and\ \bibinfo {author} {\bibfnamefont {K.~M.}\ \bibnamefont
  {Zurek}},\ }\href {\doibase 10.1103/PhysRevD.82.123512} {\bibfield  {journal}
  {\bibinfo  {journal} {Phys.Rev.}\ }\textbf {\bibinfo {volume} {D82}},\
  \bibinfo {pages} {123512} (\bibinfo {year} {2010})},\ \Eprint
  {http://arxiv.org/abs/1008.1997} {arXiv:1008.1997 [hep-ph]} \BibitemShut
  {NoStop}%
\bibitem [{\citenamefont {Buckley}\ and\ \citenamefont
  {Randall}(2011)}]{Buckley:2010ui}%
  \BibitemOpen
  \bibfield  {author} {\bibinfo {author} {\bibfnamefont {M.~R.}\ \bibnamefont
  {Buckley}}\ and\ \bibinfo {author} {\bibfnamefont {L.}~\bibnamefont
  {Randall}},\ }\href {\doibase 10.1007/JHEP09(2011)009} {\bibfield  {journal}
  {\bibinfo  {journal} {JHEP}\ }\textbf {\bibinfo {volume} {1109}},\ \bibinfo
  {pages} {009} (\bibinfo {year} {2011})},\ \Eprint
  {http://arxiv.org/abs/1009.0270} {arXiv:1009.0270 [hep-ph]} \BibitemShut
  {NoStop}%
\bibitem [{\citenamefont {Haba}\ and\ \citenamefont
  {Matsumoto}(2011)}]{Haba:2010bm}%
  \BibitemOpen
  \bibfield  {author} {\bibinfo {author} {\bibfnamefont {N.}~\bibnamefont
  {Haba}}\ and\ \bibinfo {author} {\bibfnamefont {S.}~\bibnamefont
  {Matsumoto}},\ }\href {\doibase 10.1143/PTP.125.1311} {\bibfield  {journal}
  {\bibinfo  {journal} {Prog.Theor.Phys.}\ }\textbf {\bibinfo {volume} {125}},\
  \bibinfo {pages} {1311} (\bibinfo {year} {2011})},\ \Eprint
  {http://arxiv.org/abs/1008.2487} {arXiv:1008.2487 [hep-ph]} \BibitemShut
  {NoStop}%
\bibitem [{\citenamefont {Blennow}\ \emph {et~al.}(2011)\citenamefont
  {Blennow}, \citenamefont {Dasgupta}, \citenamefont {Fernandez-Martinez},\
  and\ \citenamefont {Rius}}]{Blennow:2010qp}%
  \BibitemOpen
  \bibfield  {author} {\bibinfo {author} {\bibfnamefont {M.}~\bibnamefont
  {Blennow}}, \bibinfo {author} {\bibfnamefont {B.}~\bibnamefont {Dasgupta}},
  \bibinfo {author} {\bibfnamefont {E.}~\bibnamefont {Fernandez-Martinez}}, \
  and\ \bibinfo {author} {\bibfnamefont {N.}~\bibnamefont {Rius}},\ }\href
  {\doibase 10.1007/JHEP03(2011)014} {\bibfield  {journal} {\bibinfo  {journal}
  {JHEP}\ }\textbf {\bibinfo {volume} {1103}},\ \bibinfo {pages} {014}
  (\bibinfo {year} {2011})},\ \Eprint {http://arxiv.org/abs/1009.3159}
  {arXiv:1009.3159 [hep-ph]} \BibitemShut {NoStop}%
\bibitem [{\citenamefont {Hall}\ \emph {et~al.}(2010)\citenamefont {Hall},
  \citenamefont {March-Russell},\ and\ \citenamefont {West}}]{Hall:2010jx}%
  \BibitemOpen
  \bibfield  {author} {\bibinfo {author} {\bibfnamefont {L.~J.}\ \bibnamefont
  {Hall}}, \bibinfo {author} {\bibfnamefont {J.}~\bibnamefont {March-Russell}},
  \ and\ \bibinfo {author} {\bibfnamefont {S.~M.}\ \bibnamefont {West}},\
  }\href@noop {} {\  (\bibinfo {year} {2010})},\ \Eprint
  {http://arxiv.org/abs/1010.0245} {arXiv:1010.0245 [hep-ph]} \BibitemShut
  {NoStop}%
\bibitem [{\citenamefont {Falkowski}\ \emph {et~al.}(2011)\citenamefont
  {Falkowski}, \citenamefont {Ruderman},\ and\ \citenamefont
  {Volansky}}]{Falkowski:2011xh}%
  \BibitemOpen
  \bibfield  {author} {\bibinfo {author} {\bibfnamefont {A.}~\bibnamefont
  {Falkowski}}, \bibinfo {author} {\bibfnamefont {J.~T.}\ \bibnamefont
  {Ruderman}}, \ and\ \bibinfo {author} {\bibfnamefont {T.}~\bibnamefont
  {Volansky}},\ }\href {\doibase 10.1007/JHEP05(2011)106} {\bibfield  {journal}
  {\bibinfo  {journal} {JHEP}\ }\textbf {\bibinfo {volume} {1105}},\ \bibinfo
  {pages} {106} (\bibinfo {year} {2011})},\ \Eprint
  {http://arxiv.org/abs/1101.4936} {arXiv:1101.4936 [hep-ph]} \BibitemShut
  {NoStop}%
\bibitem [{\citenamefont {Graesser}\ \emph {et~al.}(2011)\citenamefont
  {Graesser}, \citenamefont {Shoemaker},\ and\ \citenamefont
  {Vecchi}}]{Graesser:2011wi}%
  \BibitemOpen
  \bibfield  {author} {\bibinfo {author} {\bibfnamefont {M.~L.}\ \bibnamefont
  {Graesser}}, \bibinfo {author} {\bibfnamefont {I.~M.}\ \bibnamefont
  {Shoemaker}}, \ and\ \bibinfo {author} {\bibfnamefont {L.}~\bibnamefont
  {Vecchi}},\ }\href {\doibase 10.1007/JHEP10(2011)110} {\bibfield  {journal}
  {\bibinfo  {journal} {JHEP}\ }\textbf {\bibinfo {volume} {1110}},\ \bibinfo
  {pages} {110} (\bibinfo {year} {2011})},\ \Eprint
  {http://arxiv.org/abs/1103.2771} {arXiv:1103.2771 [hep-ph]} \BibitemShut
  {NoStop}%
\bibitem [{\citenamefont {Kaplan}\ \emph {et~al.}(2011)\citenamefont {Kaplan},
  \citenamefont {Krnjaic}, \citenamefont {Rehermann},\ and\ \citenamefont
  {Wells}}]{Kaplan:2011yj}%
  \BibitemOpen
  \bibfield  {author} {\bibinfo {author} {\bibfnamefont {D.~E.}\ \bibnamefont
  {Kaplan}}, \bibinfo {author} {\bibfnamefont {G.~Z.}\ \bibnamefont {Krnjaic}},
  \bibinfo {author} {\bibfnamefont {K.~R.}\ \bibnamefont {Rehermann}}, \ and\
  \bibinfo {author} {\bibfnamefont {C.~M.}\ \bibnamefont {Wells}},\ }\href
  {\doibase 10.1088/1475-7516/2011/10/011} {\bibfield  {journal} {\bibinfo
  {journal} {JCAP}\ }\textbf {\bibinfo {volume} {1110}},\ \bibinfo {pages}
  {011} (\bibinfo {year} {2011})},\ \Eprint {http://arxiv.org/abs/1105.2073}
  {arXiv:1105.2073 [hep-ph]} \BibitemShut {NoStop}%
\bibitem [{\citenamefont {Cui}\ \emph {et~al.}(2011)\citenamefont {Cui},
  \citenamefont {Randall},\ and\ \citenamefont {Shuve}}]{Cui:2011qe}%
  \BibitemOpen
  \bibfield  {author} {\bibinfo {author} {\bibfnamefont {Y.}~\bibnamefont
  {Cui}}, \bibinfo {author} {\bibfnamefont {L.}~\bibnamefont {Randall}}, \ and\
  \bibinfo {author} {\bibfnamefont {B.}~\bibnamefont {Shuve}},\ }\href
  {\doibase 10.1007/JHEP08(2011)073} {\bibfield  {journal} {\bibinfo  {journal}
  {JHEP}\ }\textbf {\bibinfo {volume} {1108}},\ \bibinfo {pages} {073}
  (\bibinfo {year} {2011})},\ \Eprint {http://arxiv.org/abs/1106.4834}
  {arXiv:1106.4834 [hep-ph]} \BibitemShut {NoStop}%
\bibitem [{\citenamefont {Petraki}\ and\ \citenamefont
  {Volkas}(2013)}]{Petraki:2013wwa}%
  \BibitemOpen
  \bibfield  {author} {\bibinfo {author} {\bibfnamefont {K.}~\bibnamefont
  {Petraki}}\ and\ \bibinfo {author} {\bibfnamefont {R.~R.}\ \bibnamefont
  {Volkas}},\ }\href {\doibase 10.1142/S0217751X13300287} {\bibfield  {journal}
  {\bibinfo  {journal} {Int.J.Mod.Phys.}\ }\textbf {\bibinfo {volume} {A28}},\
  \bibinfo {pages} {1330028} (\bibinfo {year} {2013})},\ \Eprint
  {http://arxiv.org/abs/1305.4939} {arXiv:1305.4939 [hep-ph]} \BibitemShut
  {NoStop}%
\bibitem [{\citenamefont {Zurek}(2013)}]{Zurek:2013wia}%
  \BibitemOpen
  \bibfield  {author} {\bibinfo {author} {\bibfnamefont {K.~M.}\ \bibnamefont
  {Zurek}},\ }\href@noop {} {\  (\bibinfo {year} {2013})},\ \Eprint
  {http://arxiv.org/abs/1308.0338} {arXiv:1308.0338 [hep-ph]} \BibitemShut
  {NoStop}%
\bibitem [{\citenamefont {Kaplan}\ \emph {et~al.}(2010)\citenamefont {Kaplan},
  \citenamefont {Krnjaic}, \citenamefont {Rehermann},\ and\ \citenamefont
  {Wells}}]{Kaplan:2009de}%
  \BibitemOpen
  \bibfield  {author} {\bibinfo {author} {\bibfnamefont {D.~E.}\ \bibnamefont
  {Kaplan}}, \bibinfo {author} {\bibfnamefont {G.~Z.}\ \bibnamefont {Krnjaic}},
  \bibinfo {author} {\bibfnamefont {K.~R.}\ \bibnamefont {Rehermann}}, \ and\
  \bibinfo {author} {\bibfnamefont {C.~M.}\ \bibnamefont {Wells}},\ }\href
  {\doibase 10.1088/1475-7516/2010/05/021} {\bibfield  {journal} {\bibinfo
  {journal} {JCAP}\ }\textbf {\bibinfo {volume} {1005}},\ \bibinfo {pages}
  {021} (\bibinfo {year} {2010})},\ \Eprint {http://arxiv.org/abs/0909.0753}
  {arXiv:0909.0753 [hep-ph]} \BibitemShut {NoStop}%
\bibitem [{\citenamefont {Hietanen}\ \emph
  {et~al.}(2013{\natexlab{a}})\citenamefont {Hietanen}, \citenamefont {Pica},
  \citenamefont {Sannino},\ and\ \citenamefont
  {Sondergaard}}]{Hietanen:2012sz}%
  \BibitemOpen
  \bibfield  {author} {\bibinfo {author} {\bibfnamefont {A.}~\bibnamefont
  {Hietanen}}, \bibinfo {author} {\bibfnamefont {C.}~\bibnamefont {Pica}},
  \bibinfo {author} {\bibfnamefont {F.}~\bibnamefont {Sannino}}, \ and\
  \bibinfo {author} {\bibfnamefont {U.~I.}\ \bibnamefont {Sondergaard}},\
  }\href {\doibase 10.1103/PhysRevD.87.034508} {\bibfield  {journal} {\bibinfo
  {journal} {Phys.Rev.}\ }\textbf {\bibinfo {volume} {D87}},\ \bibinfo {pages}
  {034508} (\bibinfo {year} {2013}{\natexlab{a}})},\ \Eprint
  {http://arxiv.org/abs/1211.5021} {arXiv:1211.5021 [hep-lat]} \BibitemShut
  {NoStop}%
\bibitem [{\citenamefont {Hietanen}\ \emph
  {et~al.}(2013{\natexlab{b}})\citenamefont {Hietanen}, \citenamefont {Lewis},
  \citenamefont {Pica},\ and\ \citenamefont {Sannino}}]{Hietanen:2013fya}%
  \BibitemOpen
  \bibfield  {author} {\bibinfo {author} {\bibfnamefont {A.}~\bibnamefont
  {Hietanen}}, \bibinfo {author} {\bibfnamefont {R.}~\bibnamefont {Lewis}},
  \bibinfo {author} {\bibfnamefont {C.}~\bibnamefont {Pica}}, \ and\ \bibinfo
  {author} {\bibfnamefont {F.}~\bibnamefont {Sannino}},\ }\href@noop {} {\
  (\bibinfo {year} {2013}{\natexlab{b}})},\ \Eprint
  {http://arxiv.org/abs/1308.4130} {arXiv:1308.4130 [hep-ph]} \BibitemShut
  {NoStop}%
\bibitem [{\citenamefont {DeGrand}(2012)}]{DeGrand:2012hd}%
  \BibitemOpen
  \bibfield  {author} {\bibinfo {author} {\bibfnamefont {T.}~\bibnamefont
  {DeGrand}},\ }\href {\doibase 10.1103/PhysRevD.86.034508} {\bibfield
  {journal} {\bibinfo  {journal} {Phys.Rev.}\ }\textbf {\bibinfo {volume}
  {D86}},\ \bibinfo {pages} {034508} (\bibinfo {year} {2012})},\ \Eprint
  {http://arxiv.org/abs/1205.0235} {arXiv:1205.0235 [hep-lat]} \BibitemShut
  {NoStop}%
\bibitem [{\citenamefont {DeGrand}(2014)}]{DeGrand:2013nna}%
  \BibitemOpen
  \bibfield  {author} {\bibinfo {author} {\bibfnamefont {T.}~\bibnamefont
  {DeGrand}},\ }\href {\doibase 10.1103/PhysRevD.89.014506} {\bibfield
  {journal} {\bibinfo  {journal} {Phys.Rev.}\ }\textbf {\bibinfo {volume}
  {D89}},\ \bibinfo {pages} {014506} (\bibinfo {year} {2014})},\ \Eprint
  {http://arxiv.org/abs/1308.4114} {arXiv:1308.4114 [hep-lat]} \BibitemShut
  {NoStop}%
\bibitem [{\citenamefont {Heister}\ \emph {et~al.}(2002)\citenamefont {Heister}
  \emph {et~al.}}]{Heister:2001nk}%
  \BibitemOpen
  \bibfield  {author} {\bibinfo {author} {\bibfnamefont {A.}~\bibnamefont
  {Heister}} \emph {et~al.} (\bibinfo {collaboration} {ALEPH Collaboration}),\
  }\href {\doibase 10.1016/S0370-2693(01)01494-0} {\bibfield  {journal}
  {\bibinfo  {journal} {Phys.Lett.}\ }\textbf {\bibinfo {volume} {B526}},\
  \bibinfo {pages} {206} (\bibinfo {year} {2002})},\ \Eprint
  {http://arxiv.org/abs/hep-ex/0112011} {arXiv:hep-ex/0112011 [hep-ex]}
  \BibitemShut {NoStop}%
\bibitem [{\citenamefont {Heister}\ \emph {et~al.}(2004)\citenamefont {Heister}
  \emph {et~al.}}]{Heister:2003zk}%
  \BibitemOpen
  \bibfield  {author} {\bibinfo {author} {\bibfnamefont {A.}~\bibnamefont
  {Heister}} \emph {et~al.} (\bibinfo {collaboration} {ALEPH Collaboration}),\
  }\href {\doibase 10.1016/j.physletb.2003.12.066} {\bibfield  {journal}
  {\bibinfo  {journal} {Phys.Lett.}\ }\textbf {\bibinfo {volume} {B583}},\
  \bibinfo {pages} {247} (\bibinfo {year} {2004})}\BibitemShut {NoStop}%
\bibitem [{\citenamefont {Abdallah}\ \emph {et~al.}(2003)\citenamefont
  {Abdallah} \emph {et~al.}}]{Abdallah:2003xe}%
  \BibitemOpen
  \bibfield  {author} {\bibinfo {author} {\bibfnamefont {J.}~\bibnamefont
  {Abdallah}} \emph {et~al.} (\bibinfo {collaboration} {DELPHI
  Collaboration}),\ }\href {\doibase 10.1140/epjc/s2003-01355-5} {\bibfield
  {journal} {\bibinfo  {journal} {Eur.Phys.J.}\ }\textbf {\bibinfo {volume}
  {C31}},\ \bibinfo {pages} {421} (\bibinfo {year} {2003})},\ \Eprint
  {http://arxiv.org/abs/hep-ex/0311019} {arXiv:hep-ex/0311019 [hep-ex]}
  \BibitemShut {NoStop}%
\bibitem [{\citenamefont {Abbiendi}\ \emph {et~al.}(2004)\citenamefont
  {Abbiendi} \emph {et~al.}}]{Abbiendi:2004gf}%
  \BibitemOpen
  \bibfield  {author} {\bibinfo {author} {\bibfnamefont {G.}~\bibnamefont
  {Abbiendi}} \emph {et~al.} (\bibinfo {collaboration} {OPAL Collaboration}),\
  }\href {\doibase 10.1016/j.physletb.2004.09.059} {\bibfield  {journal}
  {\bibinfo  {journal} {Phys.Lett.}\ }\textbf {\bibinfo {volume} {B602}},\
  \bibinfo {pages} {167} (\bibinfo {year} {2004})},\ \Eprint
  {http://arxiv.org/abs/hep-ex/0412011} {arXiv:hep-ex/0412011 [hep-ex]}
  \BibitemShut {NoStop}%
\bibitem [{\citenamefont {McDonald}(1994)}]{McDonald:1993ex}%
  \BibitemOpen
  \bibfield  {author} {\bibinfo {author} {\bibfnamefont {J.}~\bibnamefont
  {McDonald}},\ }\href {\doibase 10.1103/PhysRevD.50.3637} {\bibfield
  {journal} {\bibinfo  {journal} {Phys.Rev.}\ }\textbf {\bibinfo {volume}
  {D50}},\ \bibinfo {pages} {3637} (\bibinfo {year} {1994})},\ \Eprint
  {http://arxiv.org/abs/hep-ph/0702143} {arXiv:hep-ph/0702143 [HEP-PH]}
  \BibitemShut {NoStop}%
\bibitem [{\citenamefont {Kim}\ and\ \citenamefont {Lee}(2007)}]{Kim:2006af}%
  \BibitemOpen
  \bibfield  {author} {\bibinfo {author} {\bibfnamefont {Y.~G.}\ \bibnamefont
  {Kim}}\ and\ \bibinfo {author} {\bibfnamefont {K.~Y.}\ \bibnamefont {Lee}},\
  }\href {\doibase 10.1103/PhysRevD.75.115012} {\bibfield  {journal} {\bibinfo
  {journal} {Phys.Rev.}\ }\textbf {\bibinfo {volume} {D75}},\ \bibinfo {pages}
  {115012} (\bibinfo {year} {2007})},\ \Eprint
  {http://arxiv.org/abs/hep-ph/0611069} {arXiv:hep-ph/0611069 [hep-ph]}
  \BibitemShut {NoStop}%
\bibitem [{\citenamefont {Burgess}\ \emph {et~al.}(2001)\citenamefont
  {Burgess}, \citenamefont {Pospelov},\ and\ \citenamefont {ter
  Veldhuis}}]{Burgess:2000yq}%
  \BibitemOpen
  \bibfield  {author} {\bibinfo {author} {\bibfnamefont {C.}~\bibnamefont
  {Burgess}}, \bibinfo {author} {\bibfnamefont {M.}~\bibnamefont {Pospelov}}, \
  and\ \bibinfo {author} {\bibfnamefont {T.}~\bibnamefont {ter Veldhuis}},\
  }\href {\doibase 10.1016/S0550-3213(01)00513-2} {\bibfield  {journal}
  {\bibinfo  {journal} {Nucl.Phys.}\ }\textbf {\bibinfo {volume} {B619}},\
  \bibinfo {pages} {709} (\bibinfo {year} {2001})},\ \Eprint
  {http://arxiv.org/abs/hep-ph/0011335} {arXiv:hep-ph/0011335 [hep-ph]}
  \BibitemShut {NoStop}%
\bibitem [{\citenamefont {Essig}(2008)}]{Essig:2007az}%
  \BibitemOpen
  \bibfield  {author} {\bibinfo {author} {\bibfnamefont {R.}~\bibnamefont
  {Essig}},\ }\href {\doibase 10.1103/PhysRevD.78.015004} {\bibfield  {journal}
  {\bibinfo  {journal} {Phys.Rev.}\ }\textbf {\bibinfo {volume} {D78}},\
  \bibinfo {pages} {015004} (\bibinfo {year} {2008})},\ \Eprint
  {http://arxiv.org/abs/0710.1668} {arXiv:0710.1668 [hep-ph]} \BibitemShut
  {NoStop}%
\bibitem [{\citenamefont {Andreas}\ \emph {et~al.}(2008)\citenamefont
  {Andreas}, \citenamefont {Hambye},\ and\ \citenamefont
  {Tytgat}}]{Andreas:2008xy}%
  \BibitemOpen
  \bibfield  {author} {\bibinfo {author} {\bibfnamefont {S.}~\bibnamefont
  {Andreas}}, \bibinfo {author} {\bibfnamefont {T.}~\bibnamefont {Hambye}}, \
  and\ \bibinfo {author} {\bibfnamefont {M.~H.}\ \bibnamefont {Tytgat}},\
  }\href {\doibase 10.1088/1475-7516/2008/10/034} {\bibfield  {journal}
  {\bibinfo  {journal} {JCAP}\ }\textbf {\bibinfo {volume} {0810}},\ \bibinfo
  {pages} {034} (\bibinfo {year} {2008})},\ \Eprint
  {http://arxiv.org/abs/0808.0255} {arXiv:0808.0255 [hep-ph]} \BibitemShut
  {NoStop}%
\bibitem [{\citenamefont {Agrawal}\ \emph {et~al.}(2010)\citenamefont
  {Agrawal}, \citenamefont {Chacko}, \citenamefont {Kilic},\ and\ \citenamefont
  {Mishra}}]{Agrawal:2010fh}%
  \BibitemOpen
  \bibfield  {author} {\bibinfo {author} {\bibfnamefont {P.}~\bibnamefont
  {Agrawal}}, \bibinfo {author} {\bibfnamefont {Z.}~\bibnamefont {Chacko}},
  \bibinfo {author} {\bibfnamefont {C.}~\bibnamefont {Kilic}}, \ and\ \bibinfo
  {author} {\bibfnamefont {R.~K.}\ \bibnamefont {Mishra}},\ }\href@noop {} {\
  (\bibinfo {year} {2010})},\ \Eprint {http://arxiv.org/abs/1003.1912}
  {arXiv:1003.1912 [hep-ph]} \BibitemShut {NoStop}%
\bibitem [{\citenamefont {Shifman}\ \emph {et~al.}(1978)\citenamefont
  {Shifman}, \citenamefont {Vainshtein},\ and\ \citenamefont
  {Zakharov}}]{Shifman:1978zn}%
  \BibitemOpen
  \bibfield  {author} {\bibinfo {author} {\bibfnamefont {M.~A.}\ \bibnamefont
  {Shifman}}, \bibinfo {author} {\bibfnamefont {A.}~\bibnamefont {Vainshtein}},
  \ and\ \bibinfo {author} {\bibfnamefont {V.~I.}\ \bibnamefont {Zakharov}},\
  }\href {\doibase 10.1016/0370-2693(78)90481-1} {\bibfield  {journal}
  {\bibinfo  {journal} {Phys.Lett.}\ }\textbf {\bibinfo {volume} {B78}},\
  \bibinfo {pages} {443} (\bibinfo {year} {1978})}\BibitemShut {NoStop}%
\bibitem [{\citenamefont {Hill}\ and\ \citenamefont
  {Solon}(2012)}]{Hill:2011be}%
  \BibitemOpen
  \bibfield  {author} {\bibinfo {author} {\bibfnamefont {R.~J.}\ \bibnamefont
  {Hill}}\ and\ \bibinfo {author} {\bibfnamefont {M.~P.}\ \bibnamefont
  {Solon}},\ }\href {\doibase 10.1016/j.physletb.2012.01.013} {\bibfield
  {journal} {\bibinfo  {journal} {Phys.Lett.}\ }\textbf {\bibinfo {volume}
  {B707}},\ \bibinfo {pages} {539} (\bibinfo {year} {2012})},\ \Eprint
  {http://arxiv.org/abs/1111.0016} {arXiv:1111.0016 [hep-ph]} \BibitemShut
  {NoStop}%
\bibitem [{\citenamefont {Lucini}\ and\ \citenamefont
  {Teper}(2001)}]{Lucini:2001ej}%
  \BibitemOpen
  \bibfield  {author} {\bibinfo {author} {\bibfnamefont {B.}~\bibnamefont
  {Lucini}}\ and\ \bibinfo {author} {\bibfnamefont {M.}~\bibnamefont {Teper}},\
  }\href {\doibase 10.1088/1126-6708/2001/06/050} {\bibfield  {journal}
  {\bibinfo  {journal} {JHEP}\ }\textbf {\bibinfo {volume} {0106}},\ \bibinfo
  {pages} {050} (\bibinfo {year} {2001})},\ \Eprint
  {http://arxiv.org/abs/hep-lat/0103027} {arXiv:hep-lat/0103027 [hep-lat]}
  \BibitemShut {NoStop}%
\bibitem [{\citenamefont {Lucini}\ \emph {et~al.}(2010)\citenamefont {Lucini},
  \citenamefont {Rago},\ and\ \citenamefont {Rinaldi}}]{Lucini:2010nv}%
  \BibitemOpen
  \bibfield  {author} {\bibinfo {author} {\bibfnamefont {B.}~\bibnamefont
  {Lucini}}, \bibinfo {author} {\bibfnamefont {A.}~\bibnamefont {Rago}}, \ and\
  \bibinfo {author} {\bibfnamefont {E.}~\bibnamefont {Rinaldi}},\ }\href
  {\doibase 10.1007/JHEP08(2010)119} {\bibfield  {journal} {\bibinfo  {journal}
  {JHEP}\ }\textbf {\bibinfo {volume} {1008}},\ \bibinfo {pages} {119}
  (\bibinfo {year} {2010})},\ \Eprint {http://arxiv.org/abs/1007.3879}
  {arXiv:1007.3879 [hep-lat]} \BibitemShut {NoStop}%
\bibitem [{\citenamefont {Athenodorou}\ \emph {et~al.}(2011)\citenamefont
  {Athenodorou}, \citenamefont {Bringoltz},\ and\ \citenamefont
  {Teper}}]{Athenodorou:2010cs}%
  \BibitemOpen
  \bibfield  {author} {\bibinfo {author} {\bibfnamefont {A.}~\bibnamefont
  {Athenodorou}}, \bibinfo {author} {\bibfnamefont {B.}~\bibnamefont
  {Bringoltz}}, \ and\ \bibinfo {author} {\bibfnamefont {M.}~\bibnamefont
  {Teper}},\ }\href {\doibase 10.1007/JHEP02(2011)030} {\bibfield  {journal}
  {\bibinfo  {journal} {JHEP}\ }\textbf {\bibinfo {volume} {1102}},\ \bibinfo
  {pages} {030} (\bibinfo {year} {2011})},\ \Eprint
  {http://arxiv.org/abs/1007.4720} {arXiv:1007.4720 [hep-lat]} \BibitemShut
  {NoStop}%
\bibitem [{\citenamefont {Del~Debbio}\ \emph {et~al.}(2008)\citenamefont
  {Del~Debbio}, \citenamefont {Lucini}, \citenamefont {Patella},\ and\
  \citenamefont {Pica}}]{DelDebbio:2007wk}%
  \BibitemOpen
  \bibfield  {author} {\bibinfo {author} {\bibfnamefont {L.}~\bibnamefont
  {Del~Debbio}}, \bibinfo {author} {\bibfnamefont {B.}~\bibnamefont {Lucini}},
  \bibinfo {author} {\bibfnamefont {A.}~\bibnamefont {Patella}}, \ and\
  \bibinfo {author} {\bibfnamefont {C.}~\bibnamefont {Pica}},\ }\href {\doibase
  10.1088/1126-6708/2008/03/062} {\bibfield  {journal} {\bibinfo  {journal}
  {JHEP}\ }\textbf {\bibinfo {volume} {0803}},\ \bibinfo {pages} {062}
  (\bibinfo {year} {2008})},\ \Eprint {http://arxiv.org/abs/0712.3036}
  {arXiv:0712.3036 [hep-th]} \BibitemShut {NoStop}%
\bibitem [{\citenamefont {Bali}\ and\ \citenamefont
  {Bursa}(2008)}]{Bali:2008an}%
  \BibitemOpen
  \bibfield  {author} {\bibinfo {author} {\bibfnamefont {G.~S.}\ \bibnamefont
  {Bali}}\ and\ \bibinfo {author} {\bibfnamefont {F.}~\bibnamefont {Bursa}},\
  }\href {\doibase 10.1088/1126-6708/2008/09/110} {\bibfield  {journal}
  {\bibinfo  {journal} {JHEP}\ }\textbf {\bibinfo {volume} {0809}},\ \bibinfo
  {pages} {110} (\bibinfo {year} {2008})},\ \Eprint
  {http://arxiv.org/abs/0806.2278} {arXiv:0806.2278 [hep-lat]} \BibitemShut
  {NoStop}%
\bibitem [{\citenamefont {Bali}\ \emph {et~al.}(2013)\citenamefont {Bali},
  \citenamefont {Bursa}, \citenamefont {Castagnini}, \citenamefont {Collins},
  \citenamefont {Del~Debbio} \emph {et~al.}}]{Bali:2013kia}%
  \BibitemOpen
  \bibfield  {author} {\bibinfo {author} {\bibfnamefont {G.~S.}\ \bibnamefont
  {Bali}}, \bibinfo {author} {\bibfnamefont {F.}~\bibnamefont {Bursa}},
  \bibinfo {author} {\bibfnamefont {L.}~\bibnamefont {Castagnini}}, \bibinfo
  {author} {\bibfnamefont {S.}~\bibnamefont {Collins}}, \bibinfo {author}
  {\bibfnamefont {L.}~\bibnamefont {Del~Debbio}},  \emph {et~al.},\ }\href
  {\doibase 10.1007/JHEP06(2013)071} {\bibfield  {journal} {\bibinfo  {journal}
  {JHEP}\ }\textbf {\bibinfo {volume} {1306}},\ \bibinfo {pages} {071}
  (\bibinfo {year} {2013})},\ \Eprint {http://arxiv.org/abs/1304.4437}
  {arXiv:1304.4437 [hep-lat]} \BibitemShut {NoStop}%
\bibitem [{\citenamefont {Lucini}\ and\ \citenamefont
  {Panero}(2013)}]{Lucini:2012gg}%
  \BibitemOpen
  \bibfield  {author} {\bibinfo {author} {\bibfnamefont {B.}~\bibnamefont
  {Lucini}}\ and\ \bibinfo {author} {\bibfnamefont {M.}~\bibnamefont
  {Panero}},\ }\href {\doibase 10.1016/j.physrep.2013.01.001} {\bibfield
  {journal} {\bibinfo  {journal} {Phys.Rept.}\ }\textbf {\bibinfo {volume}
  {526}},\ \bibinfo {pages} {93} (\bibinfo {year} {2013})},\ \Eprint
  {http://arxiv.org/abs/1210.4997} {arXiv:1210.4997 [hep-th]} \BibitemShut
  {NoStop}%
\bibitem [{\citenamefont {Witten}(1979)}]{Witten:1979kh}%
  \BibitemOpen
  \bibfield  {author} {\bibinfo {author} {\bibfnamefont {E.}~\bibnamefont
  {Witten}},\ }\href {\doibase 10.1016/0550-3213(79)90232-3} {\bibfield
  {journal} {\bibinfo  {journal} {Nucl.Phys.}\ }\textbf {\bibinfo {volume}
  {B160}},\ \bibinfo {pages} {57} (\bibinfo {year} {1979})}\BibitemShut
  {NoStop}%
\bibitem [{\citenamefont {Adkins}\ \emph {et~al.}(1983)\citenamefont {Adkins},
  \citenamefont {Nappi},\ and\ \citenamefont {Witten}}]{Adkins:1983ya}%
  \BibitemOpen
  \bibfield  {author} {\bibinfo {author} {\bibfnamefont {G.~S.}\ \bibnamefont
  {Adkins}}, \bibinfo {author} {\bibfnamefont {C.~R.}\ \bibnamefont {Nappi}}, \
  and\ \bibinfo {author} {\bibfnamefont {E.}~\bibnamefont {Witten}},\ }\href
  {\doibase 10.1016/0550-3213(83)90559-X} {\bibfield  {journal} {\bibinfo
  {journal} {Nucl.Phys.}\ }\textbf {\bibinfo {volume} {B228}},\ \bibinfo
  {pages} {552} (\bibinfo {year} {1983})}\BibitemShut {NoStop}%
\bibitem [{\citenamefont {Jenkins}(1993)}]{Jenkins:1993zu}%
  \BibitemOpen
  \bibfield  {author} {\bibinfo {author} {\bibfnamefont {E.~E.}\ \bibnamefont
  {Jenkins}},\ }\href {\doibase 10.1016/0370-2693(93)91638-4} {\bibfield
  {journal} {\bibinfo  {journal} {Phys.Lett.}\ }\textbf {\bibinfo {volume}
  {B315}},\ \bibinfo {pages} {441} (\bibinfo {year} {1993})},\ \Eprint
  {http://arxiv.org/abs/hep-ph/9307244} {arXiv:hep-ph/9307244 [hep-ph]}
  \BibitemShut {NoStop}%
\bibitem [{\citenamefont {Jenkins}\ and\ \citenamefont
  {Manohar}(1994)}]{Jenkins:1994md}%
  \BibitemOpen
  \bibfield  {author} {\bibinfo {author} {\bibfnamefont {E.~E.}\ \bibnamefont
  {Jenkins}}\ and\ \bibinfo {author} {\bibfnamefont {A.~V.}\ \bibnamefont
  {Manohar}},\ }\href {\doibase 10.1016/0370-2693(94)90377-8} {\bibfield
  {journal} {\bibinfo  {journal} {Phys.Lett.}\ }\textbf {\bibinfo {volume}
  {B335}},\ \bibinfo {pages} {452} (\bibinfo {year} {1994})},\ \Eprint
  {http://arxiv.org/abs/hep-ph/9405431} {arXiv:hep-ph/9405431 [hep-ph]}
  \BibitemShut {NoStop}%
\bibitem [{\citenamefont {Dashen}\ \emph {et~al.}(1995)\citenamefont {Dashen},
  \citenamefont {Jenkins},\ and\ \citenamefont {Manohar}}]{Dashen:1994qi}%
  \BibitemOpen
  \bibfield  {author} {\bibinfo {author} {\bibfnamefont {R.~F.}\ \bibnamefont
  {Dashen}}, \bibinfo {author} {\bibfnamefont {E.~E.}\ \bibnamefont {Jenkins}},
  \ and\ \bibinfo {author} {\bibfnamefont {A.~V.}\ \bibnamefont {Manohar}},\
  }\href {\doibase 10.1103/PhysRevD.51.3697} {\bibfield  {journal} {\bibinfo
  {journal} {Phys.Rev.}\ }\textbf {\bibinfo {volume} {D51}},\ \bibinfo {pages}
  {3697} (\bibinfo {year} {1995})},\ \Eprint
  {http://arxiv.org/abs/hep-ph/9411234} {arXiv:hep-ph/9411234 [hep-ph]}
  \BibitemShut {NoStop}%
\bibitem [{\citenamefont {Jenkins}\ and\ \citenamefont
  {Lebed}(1995)}]{Jenkins:1995td}%
  \BibitemOpen
  \bibfield  {author} {\bibinfo {author} {\bibfnamefont {E.~E.}\ \bibnamefont
  {Jenkins}}\ and\ \bibinfo {author} {\bibfnamefont {R.~F.}\ \bibnamefont
  {Lebed}},\ }\href {\doibase 10.1103/PhysRevD.52.282} {\bibfield  {journal}
  {\bibinfo  {journal} {Phys.Rev.}\ }\textbf {\bibinfo {volume} {D52}},\
  \bibinfo {pages} {282} (\bibinfo {year} {1995})},\ \Eprint
  {http://arxiv.org/abs/hep-ph/9502227} {arXiv:hep-ph/9502227 [hep-ph]}
  \BibitemShut {NoStop}%
\bibitem [{\citenamefont {Cohen}\ and\ \citenamefont
  {Lebed}(2003)}]{Cohen:2003tb}%
  \BibitemOpen
  \bibfield  {author} {\bibinfo {author} {\bibfnamefont {T.~D.}\ \bibnamefont
  {Cohen}}\ and\ \bibinfo {author} {\bibfnamefont {R.~F.}\ \bibnamefont
  {Lebed}},\ }\href {\doibase 10.1103/PhysRevLett.91.012001} {\bibfield
  {journal} {\bibinfo  {journal} {Phys.Rev.Lett.}\ }\textbf {\bibinfo {volume}
  {91}},\ \bibinfo {pages} {012001} (\bibinfo {year} {2003})},\ \Eprint
  {http://arxiv.org/abs/hep-ph/0301167} {arXiv:hep-ph/0301167 [hep-ph]}
  \BibitemShut {NoStop}%
\bibitem [{\citenamefont {Cherman}\ \emph {et~al.}(2009)\citenamefont
  {Cherman}, \citenamefont {Cohen},\ and\ \citenamefont
  {Lebed}}]{Cherman:2009fh}%
  \BibitemOpen
  \bibfield  {author} {\bibinfo {author} {\bibfnamefont {A.}~\bibnamefont
  {Cherman}}, \bibinfo {author} {\bibfnamefont {T.~D.}\ \bibnamefont {Cohen}},
  \ and\ \bibinfo {author} {\bibfnamefont {R.~F.}\ \bibnamefont {Lebed}},\
  }\href {\doibase 10.1103/PhysRevD.80.036002} {\bibfield  {journal} {\bibinfo
  {journal} {Phys.Rev.}\ }\textbf {\bibinfo {volume} {D80}},\ \bibinfo {pages}
  {036002} (\bibinfo {year} {2009})},\ \Eprint {http://arxiv.org/abs/0906.2400}
  {arXiv:0906.2400 [hep-ph]} \BibitemShut {NoStop}%
\bibitem [{\citenamefont {Jenkins}\ \emph {et~al.}(2010)\citenamefont
  {Jenkins}, \citenamefont {Manohar}, \citenamefont {Negele},\ and\
  \citenamefont {Walker-Loud}}]{Jenkins:2009wv}%
  \BibitemOpen
  \bibfield  {author} {\bibinfo {author} {\bibfnamefont {E.~E.}\ \bibnamefont
  {Jenkins}}, \bibinfo {author} {\bibfnamefont {A.~V.}\ \bibnamefont
  {Manohar}}, \bibinfo {author} {\bibfnamefont {J.~W.}\ \bibnamefont {Negele}},
  \ and\ \bibinfo {author} {\bibfnamefont {A.}~\bibnamefont {Walker-Loud}},\
  }\href {\doibase 10.1103/PhysRevD.81.014502} {\bibfield  {journal} {\bibinfo
  {journal} {Phys.Rev.}\ }\textbf {\bibinfo {volume} {D81}},\ \bibinfo {pages}
  {014502} (\bibinfo {year} {2010})},\ \Eprint {http://arxiv.org/abs/0907.0529}
  {arXiv:0907.0529 [hep-lat]} \BibitemShut {NoStop}%
\bibitem [{\citenamefont {Appelquist}\ and\ \citenamefont {et.
  al.}(2014)}]{shortpaper}%
  \BibitemOpen
  \bibfield  {author} {\bibinfo {author} {\bibfnamefont {T.}~\bibnamefont
  {Appelquist}}\ and\ \bibinfo {author} {\bibnamefont {et. al.}},\ }\href@noop
  {} {\  (\bibinfo {year} {(to appear) 2014})}\BibitemShut {NoStop}%
\bibitem [{\citenamefont {Walker-Loud}\ \emph {et~al.}(2009)\citenamefont
  {Walker-Loud}, \citenamefont {Lin}, \citenamefont {Richards}, \citenamefont
  {Edwards}, \citenamefont {Engelhardt} \emph {et~al.}}]{WalkerLoud:2008bp}%
  \BibitemOpen
  \bibfield  {author} {\bibinfo {author} {\bibfnamefont {A.}~\bibnamefont
  {Walker-Loud}}, \bibinfo {author} {\bibfnamefont {H.-W.}\ \bibnamefont
  {Lin}}, \bibinfo {author} {\bibfnamefont {D.}~\bibnamefont {Richards}},
  \bibinfo {author} {\bibfnamefont {R.}~\bibnamefont {Edwards}}, \bibinfo
  {author} {\bibfnamefont {M.}~\bibnamefont {Engelhardt}},  \emph {et~al.},\
  }\href {\doibase 10.1103/PhysRevD.79.054502} {\bibfield  {journal} {\bibinfo
  {journal} {Phys.Rev.}\ }\textbf {\bibinfo {volume} {D79}},\ \bibinfo {pages}
  {054502} (\bibinfo {year} {2009})},\ \Eprint {http://arxiv.org/abs/0806.4549}
  {arXiv:0806.4549 [hep-lat]} \BibitemShut {NoStop}%
\bibitem [{\citenamefont {Edwards}\ and\ \citenamefont
  {Joo}(2005)}]{Edwards:2004sx}%
  \BibitemOpen
  \bibfield  {author} {\bibinfo {author} {\bibfnamefont {R.~G.}\ \bibnamefont
  {Edwards}}\ and\ \bibinfo {author} {\bibfnamefont {B.}~\bibnamefont {Joo}}
  (\bibinfo {collaboration} {SciDAC}),\ }\href {\doibase
  10.1016/j.nuclphysbps.2004.11.254} {\bibfield  {journal} {\bibinfo  {journal}
  {Nucl. Phys. Proc. Suppl.}\ }\textbf {\bibinfo {volume} {140}},\ \bibinfo
  {pages} {832} (\bibinfo {year} {2005})},\ \Eprint
  {http://arxiv.org/abs/hep-lat/0409003} {arXiv:hep-lat/0409003} \BibitemShut
  {NoStop}%
\bibitem [{\citenamefont {Endres}\ \emph {et~al.}(2011)\citenamefont {Endres},
  \citenamefont {Kaplan}, \citenamefont {Lee},\ and\ \citenamefont
  {Nicholson}}]{Endres:2011jm}%
  \BibitemOpen
  \bibfield  {author} {\bibinfo {author} {\bibfnamefont {M.~G.}\ \bibnamefont
  {Endres}}, \bibinfo {author} {\bibfnamefont {D.~B.}\ \bibnamefont {Kaplan}},
  \bibinfo {author} {\bibfnamefont {J.-W.}\ \bibnamefont {Lee}}, \ and\
  \bibinfo {author} {\bibfnamefont {A.~N.}\ \bibnamefont {Nicholson}},\ }\href
  {\doibase 10.1103/PhysRevLett.107.201601} {\bibfield  {journal} {\bibinfo
  {journal} {Phys.Rev.Lett.}\ }\textbf {\bibinfo {volume} {107}},\ \bibinfo
  {pages} {201601} (\bibinfo {year} {2011})},\ \Eprint
  {http://arxiv.org/abs/1106.0073} {arXiv:1106.0073 [hep-lat]} \BibitemShut
  {NoStop}%
\bibitem [{\citenamefont {Young}(2012)}]{Young:2013nn}%
  \BibitemOpen
  \bibfield  {author} {\bibinfo {author} {\bibfnamefont {R.}~\bibnamefont
  {Young}},\ }\href@noop {} {\bibfield  {journal} {\bibinfo  {journal} {PoS}\
  }\textbf {\bibinfo {volume} {LATTICE2012}},\ \bibinfo {pages} {014} (\bibinfo
  {year} {2012})},\ \Eprint {http://arxiv.org/abs/1301.1765} {arXiv:1301.1765
  [hep-lat]} \BibitemShut {NoStop}%
\bibitem [{\citenamefont {Tiburzi}(2009)}]{Tiburzi:2008pa}%
  \BibitemOpen
  \bibfield  {author} {\bibinfo {author} {\bibfnamefont {B.~C.}\ \bibnamefont
  {Tiburzi}},\ }\href {\doibase 10.1016/j.physletb.2009.03.040} {\bibfield
  {journal} {\bibinfo  {journal} {Phys.Lett.}\ }\textbf {\bibinfo {volume}
  {B674}},\ \bibinfo {pages} {336} (\bibinfo {year} {2009})},\ \Eprint
  {http://arxiv.org/abs/0809.1886} {arXiv:0809.1886 [hep-lat]} \BibitemShut
  {NoStop}%
\bibitem [{\citenamefont {Detmold}\ \emph {et~al.}(2010)\citenamefont
  {Detmold}, \citenamefont {Tiburzi},\ and\ \citenamefont
  {Walker-Loud}}]{Detmold:2010ts}%
  \BibitemOpen
  \bibfield  {author} {\bibinfo {author} {\bibfnamefont {W.}~\bibnamefont
  {Detmold}}, \bibinfo {author} {\bibfnamefont {B.}~\bibnamefont {Tiburzi}}, \
  and\ \bibinfo {author} {\bibfnamefont {A.}~\bibnamefont {Walker-Loud}},\
  }\href {\doibase 10.1103/PhysRevD.81.054502} {\bibfield  {journal} {\bibinfo
  {journal} {Phys.Rev.}\ }\textbf {\bibinfo {volume} {D81}},\ \bibinfo {pages}
  {054502} (\bibinfo {year} {2010})},\ \Eprint {http://arxiv.org/abs/1001.1131}
  {arXiv:1001.1131 [hep-lat]} \BibitemShut {NoStop}%
\end{thebibliography}%

\end{document}